\documentclass[11pt,a4paper,english,final]{article}
\usepackage{jheppub}

\usepackage[normalem]{ulem}

\usepackage{ifpdf}
\usepackage{amsmath,amssymb,amsfonts,amsthm}
\usepackage[english]{babel}

\usepackage{enumerate}
\usepackage{enumitem}
\usepackage{epsfig}
\usepackage{latexsym}
\usepackage{xcolor}
\usepackage[colorinlistoftodos, shadow, textsize=small, obeyDraft]{todonotes}
\usepackage{comment}
\usepackage{subfig}
\usepackage{hyperref} 
\usepackage{comment}

\usepackage[backend=bibtex8,url=false,eprint=true,doi=true,sorting=none,backref=true]{biblatex}
\makeatletter

\@ifundefined{textcolor}{}
{
\definecolor{BLACK}{gray}{0}
\definecolor{WHITE}{gray}{1}
\definecolor{RED}{rgb}{1,0,0}
\definecolor{GREEN}{rgb}{0,1,0}
\definecolor{BLUE}{rgb}{0,0,1}
\definecolor{CYAN}{cmyk}{1,0,0,0}
\definecolor{MAGENTA}{cmyk}{0,1,0,0}
\definecolor{YELLOW}{cmyk}{0,0,1,0}
\definecolor{PURPLE}{rgb}{0.5,0,0.5}
}

\newcommand{\bite}{\begin{itemize}}
\newcommand{\eat}{\end{itemize}}
\newcommand{\beq}{\begin{equation}}
\newcommand{\eeq}{\end{equation}}
\newcommand{\rarrow}{\rightarrow}

\newcommand{\ket}{\rangle}
\newcommand{\beqa}{\begin{align}}
\newcommand{\eeqa}{\end{align}}
\newcommand{\beqar}{\begin{eqnarray}}
\newcommand{\eeqar}{\end{eqnarray}}
\newcommand{\barr}{\begin{array}}
\newcommand{\earr}{\end{array}}
\newcommand{\del}{\partial}
\newcommand{\de}{\mathrm{d}}
\renewcommand{\th}{\mathrm{th}}

\newcommand{\mb}[1]{\mathbf{#1}}
\newcommand{\mc}[1]{\mathcal{#1}}
\newcommand{\mbb}[1]{\mathbb{#1}}
\newcommand{\mf}[1]{\mathfrak{#1}}

\newcommand{\abs}[1]{\vert #1 \vert }
\newcommand{\vect}[1]{\boldsymbol{#1}}
\newcommand{\expect}[1]{\langle #1\rangle}
\newcommand{\innerp}[2]{\langle #1 \vert #2 \rangle}

\newcommand{\fullket}[1]{\vert #1 \rangle}
\newcommand{\supersc}[1]{$^{\textrm{#1}}$}

\newcommand{\sltwoc}{\mathfrak{sl}(2,\mathbb{C})}

\newcommand{\ie}{\emph{i.e. }}
\newcommand{\eg}{\emph{e.g. }}
\newcommand{\etc}{\emph{etc. }}

\newtheorem{theorem}{Theorem}[section]

\newcommand{\ignore}[1]{}

\makeatother

\title{LQG for the Bewildered}

\author[a]{Sundance Bilson-Thompson}
\affiliation[a]{School of Chemistry and Physics, University of Adelaide, Adelaide SA, Australia}

\author[b]{Deepak Vaid}
\affiliation[b]{Department of Physics, National Institute of Technology Karnataka (NITK), Surathkal, Karnataka, India}

\emailAdd{sundance.bilson-thompson@adelaide.edu.au}
\emailAdd{deepak@nitk.edu.in}
    
\abstract{We present a pedagogical introduction to the notions underlying the connection formulation of General Relativity - Loop Quantum Gravity (LQG) - with an emphasis on the physical aspects of the framework. We begin by reviewing General Relativity and Quantum Field Theory, to emphasise the similarities between them which establish a foundation upon which to build a theory of quantum gravity. We then explain, in a concise and clear manner, the steps leading from the Einstein-Hilbert action for gravity to the construction of the quantum states of geometry, known as \emph{spin-networks}, which provide the basis for the kinematical Hilbert space of quantum general relativity. Along the way we introduce the various associated concepts of \emph{tetrads}, \emph{spin-connection} and \emph{holonomies} which are a pre-requisite for understanding the LQG formalism. Having provided a minimal introduction to the LQG framework, we discuss its applications to the problems of black hole entropy and of quantum cosmology. A list of the most common criticisms of LQG is presented, which are then tackled one by one in order to convince the reader of the physical viability of the theory.

An extensive set of appendices provide accessible introductions to several key notions such as the \emph{Peter-Weyl theorem}, \emph{duality} of differential forms and \emph{Regge calculus}, among others. The presentation is aimed at graduate students and researchers who have some familiarity with the tools of quantum mechanics and field theory and/or General Relativity, but are intimidated by the seeming technical prowess required to browse through the existing LQG literature. Our hope is to make the formalism appear a little less bewildering to the un-initiated and to help lower the barrier for entry into the field.}

\begin{document}



\maketitle

\tableofcontents

\listoftodos

%
\todo{GLOBALLY: Standardize choice of symbols for various group generators}
\section{Introduction}\label{sec:intro}
The goal of Loop Quantum Gravity (LQG) is to take two extremely well-developed and successful theories, General Relativity and Quantum Field Theory, at ``face value'' and attempt to combine them into a single theory with a minimum of assumptions and deviations from established physics. Our goal, as authors of this paper, is to provide a succinct but clear description of LQG - the main body of concepts in the current formulation of LQG, some of the historical basis underlying these concepts, and a few simple yet interesting results - aimed at the reader who has more curiosity than familiarity with the underlying concepts, and hence desires a broad, pedagogical overview before attempting to read more technical discussions. This paper is inspired by the view that one never truly understands a subject until one tries to explain it to others. Accordingly we have attempted to create a discussion which we would have wanted to read when first encountering LQG. Everyone's learning style is different, and accordingly we make note of several other reviews of this subject \cite{Gaul1999Loop,Ashtekar2004Background,Kiefer2005Quantum,Nicolai2006Loop,
Alexandrov2010Critical,Mercuri2010Introduction,Dona2010Introductory,Esposito2011An-introduction,Rovelli2011Lectures,Ashtekar2012Introduction,Perez2012The-new-spin}, which the reader may refer to in order to gain a broader understanding, and to sample the various points of view held by researchers in the field.

We will begin with a brief review of the history of the field of quantum gravity in the remainder of this section. Following this we review some topics in General Relativity in section \ref{sec:gen_rel} and Quantum Field Theory in section \ref{sec:qft}, which hopefully fall into the ``Goldilocks zone'', providing all the necessary basis for LQG, and nothing more. We may occasionally introduce concepts in greater detail than the reader considers necessary, but we feel that when introducing concepts to a (hopefully) wide audience who find them unfamiliar, insufficient detail is more harmful than excessive detail. We will discuss the Lagrangian and hamiltonian approaches to classical GR in more depth and set the stage for its quantization in section \ref{sec:Expand_GR} then sketch a conceptual outline of the broad program of quantization of the gravitational field in section \ref{sec:firststepsquantgravity}, before moving on to our main discussion of the Loop Quantum Gravity approach in section \ref{sec:hilbert-space}. In section \ref{sec:applications} we cover applications of the ideas and methods of LQG to the counting of microstates of black holes 
and to the problem of quantum cosmology. 
We conclude with criticisms of LQG and rebuttals thereof in section \ref{sec:discussion} along with a discussion of its present status and future prospects.

It is assumed that the reader has a minimal familiarity with the tools and concepts of differential geometry, quantum field theory and general relativity, though we aim to remind the reader of any relevant technical details as necessary.\footnote{Given that we are aiming this paper at a broad audience, we may even hope that some readers will find it helpful with their understanding of GR and/or QFT, quite aside from its intended role explaining quantum gravity.}

Before we begin, it would be helpful to give the reader a historical perspective of the developments in theoretical physics which have led us to the present stage.

We are all familiar with classical geometry consisting of points, lines and surfaces. The framework of Euclidean geometry provided the mathematical foundation for Newton's work on inertia and the laws of motion. In the 19\supersc{th} century Gauss, Riemann and Lobachevsky, among others, developed notions of \emph{curved} geometries in which one or more of Euclid's postulates were loosened. The resulting structures allowed Einstein and Hilbert to formulate the theory of General Relativity which describes the motion of matter through spacetime as a consequence of the curvature of the background geometry. This curvature in turn is induced by the matter content as encoded in Einstein's equations (\ref{eqn:efe}). Just as the parallel postulate was the unstated assumption of Newtonian mechanics, whose rejection led to Riemmanian geometry, the unstated assumption underlying the framework of general relativity is that of the smoothness and continuity of spacetime on all scales.

\ignore{In the 19\supersc{th} century statistical mechanical theories of matter had to be modified to include the effects of the fundamental atomic structure of radiation - in the form of Planck's law - the foundations of which were laid in the early 20\supersc{th} century. Attempts to reconcile classical thermodynamics with the new physics of radiation and photons encountered fatal difficulties - such as James Jeans' ``ultraviolet catastrophe'' - which were resolved only when work by Planck, Einstein and others provided an atomistic description electromagnetic radiation in terms of ``particles'' or ``quanta'' of light known as \emph{photons}.}

Loop quantum gravity and related approaches invite us to consider that our notion of spacetime as a smooth continuum must give way to an atomistic description of geometry in which the classical spacetime we observe around us emerges from the interactions of countless (truly indivisible) \emph{atoms} of spacetime. This idea is grounded in mathematically rigorous results, but is also a natural continuation of the trend that began when 19\supersc{th} century attempts to reconcile classical thermodynamics with the physics of radiation encountered fatal difficulties - such as James Jeans' ``ultraviolet catastrophe''. These difficulties were resolved only when work by Planck, Einstein and others in the early 20\supersc{th} century provided an atomistic description of electromagnetic radiation in terms of particles or ``quanta'' of light known as \emph{photons}. This development spawned quantum mechanics, and in turn quantum field theory, while around the same time the special and general theories of relativity were being developed.

In the latter part of the 20\supersc{th} century physicists attempted, without much success, to unify the two great frameworks of quantum mechanics and general relativity. For the most part it was assumed that gravity was a phenomenon whose ultimate description was to be found in the form of a quantum field theory as had been so dramatically and successfully accomplished for the electromagnetic, weak and strong forces in the framework known as the Standard Model. These three forces could be understood as arising due to interactions between elementary particles mediated by gauge bosons whose symmetries were encoded in the groups $U(1)$, $SU(2)$ and $SU(3)$ for the electromagnetic, weak and strong forces, respectively. The universal presumption was that the final missing piece of this ``grand unified'' picture, gravity, would eventually be found as the QFT of some suitable gauge group. This was the motivation for the various grand unified theories (GUTs) developed by Glashow, Pati-Salam, Weinberg and others where the hope was that it would be possible to embed the gravitational interaction along with the Standard Model in some larger group (such $SO(5)$, $SO(10)$ or $E_8$ depending on the particular scheme). Such schemes could be said to be in conflict with Occam's dictum of simplicity and Einstein and Dirac's notions of beauty and elegance. \emph{More importantly all these models assumed implicitly that spacetime remains continuous at all scales}. As we shall see this assumption lies at the heart of the difficulties encountered in unifying gravity with quantum mechanics.

A significant obstacle to the development of a theory of quantum gravity is the fact that GR is not renormalizable. The gravitational coupling constant $G$ (or equivalently $1/M^2_\mathrm{Planck}$ in dimensionless units where $ G = c = \hbar = 1 $) is not dimensionless, unlike the fine-structure constant $\alpha$ in QED. This means that successive terms in any perturbative series have increasing powers of momenta in the numerator. Rejecting the notion that systems could absorb or transmit energy in arbitrarily small amounts led to the photonic picture of electromagnetic radiation and the discovery of quantum mechanics. Likewise, rejecting the notion that spacetime is arbitrarily smooth at all scales - and replacing it with the idea that geometry at the Planck scale must have a discrete character - leads us to a possible resolution of the ultraviolet infinities encountered in quantum field theory and to a theory of ``quantum gravity''.

Bekenstein's observation \cite{Bekenstein1972Black,Bekenstein1973Black,Bekenstein1973Extraction} of the relationship between the entropy of a black hole and the area of its horizon combined with Hawking's work on black hole thermodynamics led to the realization that there were profound connections between thermodynamics, information theory and black hole physics. These can be succinctly summarized by the famous \emph{area law} relating the entropy of a \emph{macroscopic} black hole $S_{BH}$ to its surface area $A$:
\begin{equation}
\label{eqn:area-law}
S_{BH} = \gamma A
\end{equation}
where $\gamma$ is a universal constant and $A \gg A_{pl}$, with $A_{pl} \propto l_p^2$ being the Planck area. While a more detailed discussion will wait until \ref{subsec:entropy}, we note here that if geometrical observables such as area are quantized, eq.~(\ref{eqn:area-law}) can be seen as arising from the number of ways that one can join together $\mc{N}$ quanta of area to form a horizon. In LQG the quantization of geometry arises naturally - though not all theorists are convinced that geometry should be quantized or that LQG is the right way to do so.

\ignore{Since the mass of a black hole determines the area of the black hole's event horizon, the discrete nature of electromagnetic radiation implies that the area of a black hole must also be discrete.}


With this historical overview in mind, it is now worth summarizing the basic notions of General Relativity and QFT before we attempt to see how these two disciplines may be unified in a single framework.

\subsection{Conventions}\label{sec:conventions}
Before we proceed, a quick description of our conventions for indices will hopefully be useful to the reader;
\bite
\item Greek letters $\mu,\nu,\rho,\lambda,... \in \{0,1,2,3\}$ from the middle of the alphabet are four-dimensional spacetime indices. Other Greek letters, $\alpha, \beta,\ldots$ will be used for general cases in $N$ dimensions.
\item Lowercase letters from the start of the Latin alphabet, $a,b,c,\ldots \in \{1,2,3\}$ are three-dimensional spatial indices. These will often be used when dealing exclusively with the spatial part of a four-dimensional quantity that would otherwise have Greek indices. 
\item Uppercase letters $I,J,K,\ldots \in \{0,1,2,3\}$ are ``internal'' indices which take values in the $\mf{sl}(2,\mathbb{C})$ Lorentz lie-algebra.
\item Lowercase Latin letters $j,k,l,\ldots \in {1,2,3,...,N}$ from the middle of the alphabet are indices for a space of $N$ dimensions. Equations involving these indices are the general cases, which can be applied to Minkowski space, $\mathbb{R}^3$, etc. They will also be used as $\mf{su}(2)$ lie-algebra indices 
\eat
Wherever possible we will attempt to avoid using ``special'' letters (e.g. $\pi$, $i=\sqrt{-1}$, $\gamma$ in the context of the Dirac matrices, $\sigma$ in the context of the Pauli matrices) as indices, unless there is no chance of confusion.

\section{Classical GR}\label{sec:gen_rel}
General Relativity (GR) is an extension of Einstein's Special Theory of Relativity (SR), which was required in order to include observers in non-trivial gravitational backgrounds. SR applies in the absence of gravity, and in essence it describes the behavior of vector quantities in a four-dimensional Galilean space, with the Minkowski metric\footnote{Of course the choice $\mathrm{diag}(+1,-1,-1,-1)$ is equally valid but we will have occasion later to restrict our attention to the spacial part of the metric, in which case a positive (spatial) line-element is cleaner to work with.}
\beq
\eta_{\mu\nu} = \mathrm{diag}(-1,+1,+1,+1),
\eeq
leading to a 4D line-element
\beq
ds^2 = -c^2 dt^2 + dx^2 + dy^2 + dz^2\,.
\eeq
The speed of a light signal, measured by any inertial observer, is a constant, denoted $c$. If we denote the components of a vector in four-dimensional spacetime with Greek indices \mbox{(e.g. $v^\mu$)} the Minkowski metric\footnote{Strictly speaking it is a pseudo-metric, as the distance it measures between two distinct points can be zero.} divides vectors into three categories; \emph{timelike} (those vectors for which $\eta_{\mu\nu}v^\mu v^\nu<0$), \emph{null} or \emph{light-like} (those vectors for which $\eta_{\mu\nu}v^\mu v^\nu=0$), and \emph{spacelike} (those vectors for which $\eta_{\mu\nu}v^\mu v^\nu>0$). Any point, with coordinates $(ct,x,y,z)$, is referred to as an \emph{event}, and the set of all null vectors having their origin at any event define the future light-cone and past light-cone of that event. Events having time-like or null displacement from a given event $E_0$ (\ie lying inside or on $E_0$'s lightcones) are causally connected to $E_0$. Those in/on the past light-cone can influence
$E_0$, those in/on the future lightcone can be influenced by $E_0$.

General Relativity extends these concepts to non-Euclidean spacetime. The metric of this (possibly curved) spacetime is denoted $g_{\mu\nu}$. Around each event it is possible to consider a sufficiently small region that the curvature of spacetime within this region is negligible, and hence the central concepts of Special Relativity apply locally. Rather than developing the idea that the curvature of spacetime gives rise to gravitational effects, we shall treat this as assumed knowledge, and discuss how the curvature of spacetime may be investigated. Since spacetime is not assumed to be flat (we'll define ``flat'' and ``curved'' rigorously below) and Euclidean, in general one cannot usefully extend the coordinate system from the region of one point in spacetime (one event) to the region of another arbitrary point. This can be seen from the fact that a Cartesian coordinate system which defined ``up'' to be the $z$-axis at one point on the surface of the Earth, would have to define ``up'' not to be parallel to the $z$-axis at most other points. In short, a freely-falling reference frame cannot be extended to each point in the vicinity of the surface of the Earth - or any other gravitating body. We are thus forced to work with local coordinate systems which vary from region to region. We shall refer to the basis vectors of these local coordinate systems by the symbols $e_i$. A set of four such basis vectors at any point is called a \emph{tetrad} or \emph{vierbein}. The metric is related to the dot product of basis vectors by 
${\mbox g_{ij}=e_i\cdot e_j}$. As the basis vectors are not necessarily orthonormal, we also may define a set of dual basis vectors $e^i$, where $e^i \cdot e_j = \delta^i_j$.  

\begin{figure}[htbp]
\begin{center}
{\includegraphics[height=50mm,angle=0]{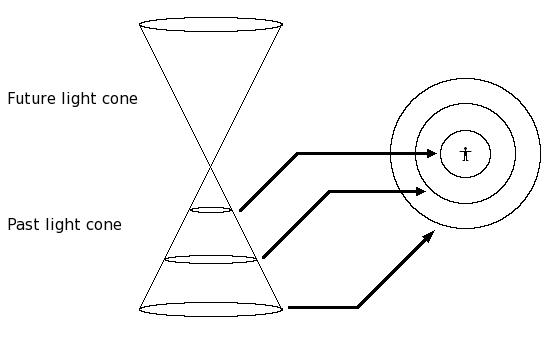}}
\caption{The future-pointing and past-pointing null vectors at a point define the future and past light cones of that point. Slices (at constant time) through the past light cone of an observer are two-spheres centred on the observer, and hence map directly to that observer's celestial sphere.}
\label{fig:lightcone}
\end{center}
\end{figure}

\subsection{Parallel Transport and Curvature}
Given the basis vectors $e_i$ of a local coordinate system, an arbitrary vector is written in terms of its components $v^i$ as $\vec{V}=v^i e_i$. It is of course also possible to define vectors with respect to the dual basis. These dual vectors will have components with lowered indices, for example $v_i$, and take the general form $v_i e^i$. The metric is used to switch between components referred to the basis or dual basis, e.g. $v_j=g_{ij}v^i$. Vectors defined with raised indices on their components are called `contravariant vectors' or simply `vectors'. Those with lowered indices are called `covariant vectors, `covectors' or `1-forms'. Note that $e_i$, having lowered indices, are basis vectors, while the $e^i$, having raised indices, are basis 1-forms. We will return to the distinction between vectors and 1-forms in section~\ref{sec:k-vectors-k-forms}.

When we differentiate a vector along a curve parametrised by the coordinate $x^k$ we must apply the product rule, as the vector itself can change direction and length, and the local basis will in general also change along the curve, hence
\beq
\frac{\de \vec{V}}{\de x^k} = \frac{\del v^j}{\del x^k}e_j + v^j\frac{\del e_j}{\del x^k}.
\label{eqn:chain_rule_deriv}
\eeq
We extract the $i^\th$ component by taking the dot product with the dual basis vector (basis 1-form) $e^i$, since $e^i \cdot e_j = \delta^i_j$. Hence we obtain
\beq
\frac{\de v^i}{\de x^k} = \frac{\del v^i}{\del x^k} + v^j\frac{\del e_j}{\del x^k}\cdot e^i\, ,
\label{eqn:chain_rule_deriv_component}
\eeq
which by a suitable choice of \emph{notation} is usually rewritten in the form
\beq\label{eqn:cov_deriv}
\nabla_k v^i = \del_k v^i + v^j\Gamma^i_{jk}.
\eeq
The derivative written on the left-hand-side is termed the \emph{covariant derivative}, and consists of a partial derivative due to changes in the vector, and a term $\Gamma^i_{jk}$ called the \emph{connection} due to changes in the local coordinate basis from one place to another. If a vector is parallel-transported along a path, its covariant derivative will be zero. In consequence any change in the components of the vector is due to (and hence equal and opposite to) the change in local basis, so that
\beq
\frac{\del v^i}{\del x^k} = - v^j\frac{\del e_j}{\del x^k}\cdot e^i\, .
\label{eqn:paral_transp_equal_opp}
\eeq
\begin{figure}[bp]
\centering
\includegraphics[scale=0.39]{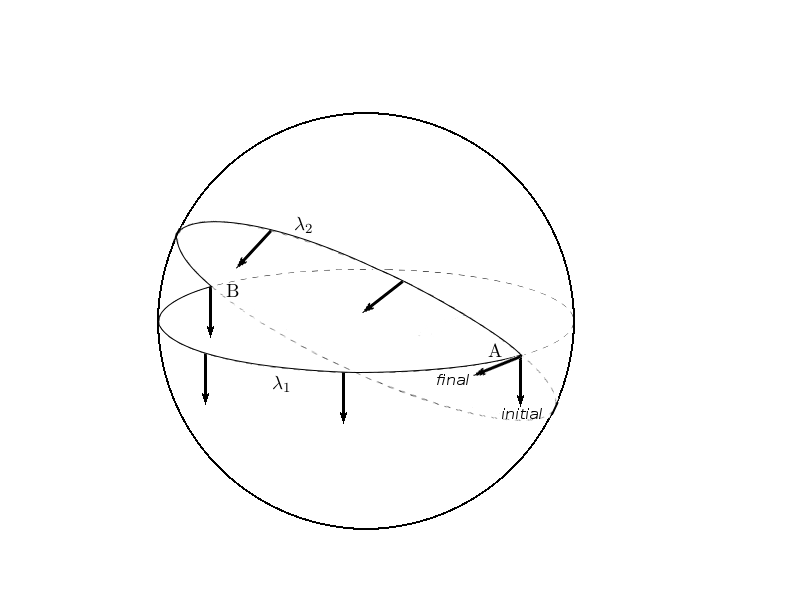}
\caption{The parallel transport of a vector around a closed path tells us about the curvature of a region bounded by that path. Here a vector is parallel transported along curve $\lambda_1$ from A to B, and back from B to A along $\lambda_2$. Both $\lambda_1$ and $\lambda_2$ are sections of great circles, and so we can see that the vector maintains a constant angle to the tangent to the curve between A and B, but this angle changes abruptly at B when the vector switches from $\lambda_1$ to $\lambda_2$. The difference in initial and final orientation of the vector at A tells us that the surface (a sphere in this case) is curved. Just as an arbitrarily curved path in $\mathbb{R}^2$ can be built up from straight line segments, an arbitrary path in a curved manifold can be built up from sections of geodesics (of which great circles are an example).}
\label{fig:holonomy}
\end{figure}
The transport of a vector along a single path between two distinct points does not reveal any curvature of the space (or spacetime) through which the vector is carried. To detect curvature it is necessary to carry a vector all the way around a closed path and back to its starting point, and compare its initial and final orientations. If they are the same, for an arbitrary path, the space (or spacetime) is \emph{flat}. If they differ, the space is \emph{curved}, and the amount by which the initial and final orientations of the vector differ provides a measure of how much curvature is enclosed within the path. Alternatively, one may transport two copies of a vector from the same starting point, A, along different paths, $\lambda_1$ and $\lambda_2$ to a common end-point, B. Comparing the orientations of the vectors after they have been transported along these two different paths reveals whether the space is flat or curved. It should be obvious that this is equivalent to following a closed path (moving along $\lambda_1$ from A to B, and then along $\lambda_2$ from B to A, c.f. figure~\ref{fig:holonomy}). The measure of how much this closed path (loop) differs from a loop in flat space (that is, how much the two transported vectors at B differ from each other) is called the \emph{holonomy} of the loop.

In light of the preceding discussion, suppose a vector $\vec{V}$ is transported from point A some distance in the $\mu$-direction. The effect of this transport upon the components of $\vec{V}$ is given by the covariant derivative $\nabla_\mu$ of $\vec{V}$. The vector is then transported in the $\nu$-direction to arrive at point B. An identical copy of the vector is carried first from A in the $\nu$-direction, and then in the $\mu$-direction to B. The difference between the two resulting (transported) vectors, when they arrive at B is given by
\beq
(\nabla_\mu \nabla_\nu - \nabla_\nu \nabla_\mu)\vec{V}.
\eeq
This commutator defines the Riemann curvature tensor,
\beq
R^\lambda{}_{\rho\mu\nu}v^\rho = [\nabla_\mu,\,\nabla_\nu]v^\lambda.
\eeq
If and only if the space is flat, all the components of $R^\lambda{}_{\rho\mu\nu}$ will be zero, otherwise the space is curved. 

Since the terms in the commutator of covariant derivatives differ only in the ordering of the indices, it is common to place the commutator brackets around the indices only, rather than the operators, hence we can write
\beq
 \nabla_{[\mu}\nabla_{\nu]} = [\nabla_\mu,\,\nabla_\nu] = \nabla_\mu \nabla_\nu - \nabla_\nu \nabla_\mu\, .
\eeq

\subsection{Einstein's Field Equations}
\label{subsec:efe}
Einstein's equations relate the curvature of spacetime with the energy density of the matter and fields present in the spacetime. Defining the Ricci tensor $R_{\rho\nu}=R^\mu{}_{\rho\mu\nu}$ and the Ricci scalar $R = R^\nu{}_\nu$ (\ie it is the trace of the Ricci tensor, taken after raising an index using the metric $g^{\mu\nu}$), the relationship between energy density and spacetime curvature is then given by
\beq
R^{\mu\nu} - \frac{1}{2}R g^{\mu\nu} + \Lambda g^{\mu\nu} = 8\pi \mc{G} T^{\mu\nu},
\label{eqn:efe}
\eeq
where $\mc{G}$ is Newton's constant, and the coefficient $\Lambda$ is the cosmological constant, which prior to the 1990s was believed to be identically zero. The tensor $T^{\mu\nu}$ is the energy-momentum tensor (also referred to as the stress-energy tensor). We will not discuss it in great detail, but its components describe the flux of energy and momentum (\ie 4-momentum) across various timelike and spacelike surfaces\footnote{The presence of the energy-momentum tensor is related to the fact that it is not merely the mass of matter that creates gravity, but its momentum, as required to maintain consistency when transforming between various Lorentz-boosted frames}. The component $T^{\mu\nu}$ describes the flux of the $\mu^\th$ component of 4-momentum across a surface of constant $x^\nu$. For instance, the zeroeth component of 4-momentum is energy, and hence $T^{00}$ is the amount of energy crossing a surface of constant $time$ (\ie energy that is moving into the future but stationary in space, hence it is the energy density).  

It should be noted that we can write $\Gamma^\rho_{\mu\nu}$ in terms of the metric $g_{\mu\nu}$ (see for e.g. \cite{Wald1984General}),
\begin{equation}
\Gamma^\rho_{\mu\nu} = \frac{1}{2} g^{\rho\delta}\left( \partial_\mu g_{\delta\nu} + \partial_\nu g_{\delta\mu} - \partial_\delta g_{\mu\nu} \right)\, .
\label{eqn:gamma_from_g}
\end{equation}
Since the Riemann tensor is defined from the covariant derivative, and the covariant derivative is defined by the connection, the metric $g^{\mu\nu}$ should be interpreted as a solution of the Einstein field equations, eq.~(\ref{eqn:efe}).

It is sometimes preferable to write equation~(\ref{eqn:efe}) in the form
\beq\label{eqn:einstein-eqn}
G^{\mu\nu} = 8\pi \mc{G} T^{\mu\nu} - \Lambda g^{\mu\nu}
\eeq
where the \emph{Einstein tensor} $G^{\mu\nu} = R^{\mu\nu} - R g^{\mu\nu}/2$ is the divergence-free part of the Ricci tensor. The explicit form of equation~(\ref{eqn:efe}) emphasises the relationship between mass-energy and spacetime curvature. All the quantities related to the structure of the spacetime (\ie $R^{\mu\nu}$, $R$, $g^{\mu\nu}$) are on the left-hand side. The quantity related to the presence of matter and energy, $T^{\mu\nu}$, is on the right-hand side. For now it remains a question of interpretation whether this means that mass-energy is equivalent to spacetime curvature, or identical to it. Perhaps more importantly the form of the Einstein Field Equations makes it clear that GR is a theory of dynamical spacetime. As matter and energy move, so the curvature of the spacetime in their vicinity changes.

It is worth noting (without proof, see for instance \cite{Wald1984General}) that the gravitational field in the simplest case of a static, spherically-symmetric field around a mass $M$, defines a line element of the form derived by Schwarzschild,
\beq
\de s^2 = -c^2\left(1-\frac{2\mc{G}M}{c^2 r}\right)\de t^2 + \left(1-\frac{2\mc{G}M}{c^2 r}\right)^{-1}\!\!\!\de r^2 + r^2(\de \theta^2 + \sin^2\theta \de \phi^2).
\eeq
For weak gravitational fields, and test masses moving at low velocities ($v\ll c$) the majority of the deviation from the line element in empty space is caused by the coefficent of the $\de t^2$ term on the right. This situation also coincides with the limit in which Newtonian gravity becomes a good description of the mechanics. In the Newtonian picture the force of gravity can be written as the gradient of a potential,
\beq
\vec{F} = \nabla V.
\eeq
It can be shown that
\beq
\del g_{00} \propto \nabla V,
\eeq
implying that gravity in the Newtonian or weak-field limit can be understood, primarily, as the amount of distortion in the local ``speed'' of time caused by the presence of matter.

\subsection{Changes of Coordinates and Diffeomorphism Invariance}
\label{sec:diffeo_inv}
General relativity embodies a principle called \emph{diffeomorphism invariance}. This principle states, in essence, that the laws of physics should be invariant under different choices of coordinates. In fact, one may say that coordinates have no meaning in the formulation of physical laws, and in principle we could do without them.

In a practical sense, however, when performing calculations it is often necessary to work with a particular choice of coordinates. When translating between different points we may find that that basis vectors are defined differently at different points (giving rise to a connection, as we saw above). However if we restrict our attention to a particular point we find that the coordinate basis may be changed by performing a transformation on the basis, leading to new coordinates derived from the old coordinates. Transformations of coordinates take a well-known form, which we will briefly recap. Suppose the two coordinate systems have basis vectors $x^1,\ldots, x^n$ and $y^1,\ldots, y^n$. Then for a given vector $\vec{V}$ with components $a_k$ and $b_l$ in the two coordinate systems it must be true that $a_k x^k = \vec{V} = b_l y^l$. Differentiating with respect to $y$, the relationship between coordinate systems is given by
\beq
b_l = a_k \frac{\del x^k}{\del y^l}\, .
\label{eqn:coord_transf}
\eeq
This tells us how to find the components of a vector in a ``new'' coordinate system (the $y$-basis), given the components in the ``old'' coordinate system (the $x$-basis). Let us write $J_l^k={\del x^k}/{\del y^l}$, and then since a summation is implied over $k$ the transformation of coordinates can be written in terms of a matrix acting upon the components of vectors, $b_l = J_l^k a_k$. Such a matrix, relating two coordinate systems is called a {\em Jacobian matrix}. While one transformation matrix is needed to act upon vectors (which have only a single index), one transformation matrix per index is needed for more complex objects, e.g.
\beq
 b_{jkl} = J^m_j J^n_k J^p_l a_{mnp}\, .
\eeq 

Since the metric defines angles and lengths (and hence areas and volumes) calculations involving the volume of a region of spacetime (e.g. integration of a lagrangian) must introduce a supplementary factor
of $\sqrt{-g}$ (where $g$ is the determinant of the metric $g^{\mu\nu}$) in order to remain invariant under arbitrary coordinate transformations. Hence instead of $d^n x \rightarrow d^n y$ we have
\beq
d^n x \sqrt{-g(x)} \rightarrow d^n y \sqrt{-g(y)}\, .
\label{eqn:change_of_coords}
\eeq
The square root of the determinant of the metric is an important factor in defining areas, as we can see by considering a parallelogram whose sides are defined by two vectors, $\vec{x}$ and $\vec{y}$. The area of this parallelogram is given by the magnitude of the cross product of these vectors, hence
\beq
 A_\mathrm{parallel.} = \sqrt{(\vec{x}\times \vec{y})\cdot(\vec{x}\times \vec{y})} = \sqrt{(\vec{x}\cdot\vec{x})(\vec{y}\cdot\vec{y})\sin^2\theta}
 = \sqrt{(\vec{x}\cdot\vec{x})(\vec{y}\cdot\vec{y})(1-\cos^2\theta)}\, .
 \label{eqn:area_parallel}
\eeq 
Suppose that $\vec{x}$ and $\vec{y}$ are basis vectors lying in a plane. Then the metric in this plane will be 
\beq
	 m_{ab} = \left( \begin{array}{cc}
						 \vec{x}\cdot\vec{x}\,\,\, & \vec{x}\cdot\vec{y} \\
						 \vec{y}\cdot\vec{x}\,\,\, & \vec{y}\cdot\vec{y}
						\end{array}
				 \right)
 \label{eqn:planar_metric}
\eeq 
where $a,\,b \in \{\vec{x},\,\vec{y}\}$. Comparing equations~(\ref{eqn:area_parallel}) and (\ref{eqn:planar_metric}), we see that 
\beq
 A_\mathrm{parallel.} = \sqrt{(\vec{x}\cdot\vec{x})(\vec{y}\cdot\vec{y}) - (\vec{x}\cdot\vec{y})^2} = \sqrt{\det m_{ab}}\, .
 \label{eqn:area_det_metric}
\eeq 
It is therefore reasonable to expect an analogous function of the metric to play a role in changes of coordinates. Furthermore we would expect the total area of some two-dimensional surface, which can be broken up into many small parallelograms, to be given by integrating the areas of such parallelograms together (this point will be taken up again is sec.~\ref{subsec:areaop}). 

To see why eq.~(\ref{eqn:change_of_coords}) applies in the case of 
coordinate transformations\footnote{This argument is taken from chapter 8 of \cite{Koks}, where a more detailed discussion can be found.}, consider an infinitessimal region of a space. Let this region be a parallelipiped in some coordinate system $x^1,\ldots, x^n$. Now suppose we want to change to a different set of coordinates, $y^1,\ldots, y^n$, which are functions of the first set (e.g. we want to change from polar coordinates to cartesian).
The Jacobian of this transformation is
\beq
J = \frac{\del(x^1,\ldots, x^n)}{\del(y^1,\ldots, y^n)} =
\begin{pmatrix}
\frac{\del x^1}{\del y^1} & \cdots & \frac{\del x^1}{\del y^n} \\
\vdots & & \vdots \\
\frac{\del x^n}{\del y^1} & \cdots & \frac{\del x^n}{\del y^n}\, .
\end{pmatrix}
\eeq
The entries in the Jacobian matrix are the elements of the vectors defining the sides of the infinitessimal region we began with, referred to the new basis. Each row corresponds with one vector, and the absolute value of the determinant of such a matrix, multiplied by $d^ny = dy^1\ldots dy^n$ gives the volume of the infinitessimal region. An integral referred to these new coordinates must include a factor of this volume, to ensure that the coordinates have been transformed correctly and the integral doesn't over-count the infinitessimal regions of which it is composed, hence
\beq
\int f(x^1,\ldots,\,x^n) dx^1\ldots dx^n = \int f(y^1,\ldots,\,y^n) |{\rm det}J| dy^1\ldots dy^n\, .
\label{eqn:change coord}
\eeq

The Jacobian matrix defines the transformation between coordinate systems. To be specific, we will choose the Minkowski metric $\eta_{\mu\nu}$ for the first coordinate system. The metric of the second coordinate system remains unspecified, hence
\beq
g_{\alpha\beta} = \frac{\del x^\mu}{\del y^\alpha}\frac{\del x^\nu}{\del y^\beta} \eta_{\mu\nu}\, . 
\label{eqn:g_from_eta}
\eeq
We can treat this expression as a product of matrices. If we do so, we must be careful about the ordering of terms, since matrix multiplication is non-commutative, and it is useful to replace one of the Jacobian matrices by its transpose. However this extra complication can be avoided since we are interested in the determinants of the matrices, and ${\rm det}(AB)={\rm det}A\,{\rm det}B={\rm det}B\,{\rm det}A$, and also ${\rm det}A^T={\rm det}A$ so the ordering of terms is ultimately unimportant. Taking the absolute value of the determinant of eq.~(\ref{eqn:g_from_eta}),
\beq
|J| = \sqrt{\frac{g}{\eta}} = \sqrt{-g}
\eeq
since $\eta=\rm{det}\eta_{\mu\nu}=-1$. From this and eq.~(\ref{eqn:change coord}) the use of a factor $\sqrt{-g}$ follows immediately.

The transformations described above, where a new coordinate basis is derived from an old one is called a {\em passive} transformation. By contrast, it is possible to leave the coordinate basis unchanged and instead change the positions of objects, whose coordinates will consequently change as measured in this basis. This is called an {\em active} coordinate transformation. With this distinction in mind, we will elaborate on the concept of diffeomorphism invariance in GR.   

A diffeomorphism is a mapping of coordinates $f:x\rightarrow f(x)$ from a manifold $U$ to a manifold $V$ that is smooth, invertible, one-to-one, and onto. As a special case we can take $U$ and $V$ to be the same manifold, and define a diffeomorphism from a spacetime manifold to itself. A passive diffeomorphism will change the coordinates, but leave objects based on them unchanged, so that for instance the metric before a passive diffeomorphism is $g_{\mu\nu}(x)$ and after it is $g_{\mu\nu}(f(x))$. Invariance under passive diffeomorphisms is nothing special, as any physical theory can be made to yield the same results under a change of coordinates. An active diffeomorphism, on the other hand, would yield a new metric $g'_{\mu\nu}(x)$, which would in general measure different distances between any two points than does $g_{\mu\nu}(x)$. General relativity is significant for being invariant under active diffeomorphisms. This invariance requires that if $g_{\mu\nu}(x)$ is any solution of the Einstein field equations, an active diffeomorphism yields  $g'_{\mu\nu}(x)$ which must be another valid solution of the EFEs. We require that any theory of quantum gravity should also embody a notion of diffeomorphism invariance, or at the very least, should exhibit a suitable notion of diffeomorphism invariance in the classical limit.

An understanding of classical General Relativity helps us to better understand transformations between locally-defined coordinate systems. We will now proceed to a discussion of Quantum Field Theory, where these local coordinate systems are abstracted to ``internal" coordinates. And just as the discussion of GR provides us with tools to more easily visualise the concepts at the heart of QFT, the quantisation of field theories discussed in the next section will lay the foundations for our attempts to extend classical GR into a quantum theory of gravity.

\section{Quantum Field Theory}\label{sec:qft}

Quantum Field Theory should be familiar to most (if not all) modern physicists, however we feel it is worth mentioning the basic details here, in order to emphasize the similarities between QFT and GR, and hence illustrate how GR can be written as a gauge theory. In short, we will see that a local change of phase of the wavefunction is equivalent to the position-dependent change of basis we considered in the case of GR. Just as the partial derivative of a vector gave (via the product rule) a derivative term corresponding to the change in basis, we will see that a derivative term arises corresponding to the change in phase of the quantum field. This introduces a connection and a covariant derivative defined in terms of the connection.

\subsection{Covariant Derivative and Curvature}
\label{subsec:covariantderiv}
We may write the wavefunction of a particle as a product of wavefunctions $\phi(x)$ and $u(x)$ corresponding respectively to the external and internal degrees of freedom\footnote{A more thorough discussion of the material in this subsection can be found in chapter 3 of \cite{Moriyasu}},
\beq
\psi(x) = \phi(x)_a u(x)_a
\eeq
where there is an obvious analogy to the definition of a vector, with the $u_a$ playing the role of basis vectors, the $\phi(x)_a$ playing the role of the components, and summation implied over the repeated index $a$. 
In complete analogy with
eq.~(\ref{eqn:chain_rule_deriv}), by applying the chain rule we find that
\beq
\frac{d\psi}{dx^\mu} = \frac{\del\phi_a}{\del x^\mu} u_a + \phi_a \frac{\del u_a}{\del x^\mu}
\label{eqn:chain_rule_wavefunction}
\eeq
For illustrative purposes, let us consider a fairly simple choice of basis, where we have only one $u$ and so we drop the index $a$. We will write $u = e^{ig\theta(x)}$. Then the derivative of $\psi$ will take the form
\beqar
\frac{d\psi}{dx^\mu} & = & \frac{\del\phi}{\del x^\mu} e^{ig\theta(x)} + ig e^{ig\theta(x)}\phi \frac{\del \theta(x)}{\del x^\mu} \nonumber \\
& = & e^{ig\theta(x)}\left(\frac{\del}{\del x^\mu} + ig \frac{\del \theta(x)}{\del x^\mu}\right) \phi
\label{eqn:chain_rule_deriv_wavefunction}
\eeqar
Next we can pre-multiply the whole expression by $e^{-ig\theta(x)}$ to eliminate the exponential term on the right hand side. This is equivalent to eq.~(\ref{eqn:chain_rule_deriv_component}) where we extracted an expression for the derivative of the components using $e^i\cdot e_j = \delta^i_j$. Lastly we switch notation slightly to more closely resemble eq.~(\ref{eqn:cov_deriv}), and define the term in brackets to be a covariant derivative
\beq
D_\mu = \partial_\mu + ig A_\mu
\label{eqn:cov_deriv_gauge}
\eeq
where $A_\mu = \del_\mu \theta$, and $D_\mu$ satisfies all the properties required of a derivative operator (linearity, Leibniz's rule, etc.).

A transformation $\theta\rightarrow\theta'=\theta+\lambda$ will result in a transformation of the wavefunction $\psi\rightarrow\psi'=e^{ig\lambda}\psi$, and a transformation of the connection $A_\mu\rightarrow A'_\mu$. For brevity, let us write $G=e^{ig\lambda}$. We can find the transformation of $A_\mu$ from the requirement that $D'_\mu\psi'=D'_\mu G\psi = G D_\mu\psi$, which means that
\beqar
(\partial_\mu + ig A'_\mu) G \psi & = & G(\partial_\mu + ig A_\mu) \psi \nonumber \\
\therefore (\partial_\mu G)\psi + G \partial_\mu\psi + ig A'_\mu G \psi & = & G \partial_\mu\psi + ig G A_\mu \psi \nonumber \\
\therefore (\partial_\mu G)\psi + ig A'_\mu G \psi & = & ig G A_\mu \psi \nonumber \\
\therefore ig A'_\mu G & = & ig G A_\mu - (\partial_\mu G) \nonumber \\
\therefore A'_\mu & = & G A_\mu G^{-1} +\frac{i}{g} (\partial_\mu G) G^{-1}\, .
\label{eqn:Dmu_transf}
\eeqar
Substituting in $G=e^{ig\lambda}$ we deduce that $A_\mu$ transforms as
\beq
A'_\mu = A_\mu - \partial_\mu \lambda\, .
\label{eqn:A_transf}
\eeq
Since we defined $A_\mu = \del_\mu \theta$ above, the presence of a minus sign might be a bit surprising. Surely from the definition of $A_\mu$ we expect that $\partial\theta'=\partial\theta+\partial\lambda$. However what eq.~(\ref{eqn:A_transf}) is telling us is 
simply that when we locally change the basis of a wavefunction but leave the overall physics unchanged, the 
connection
must change in an equal and opposite manner to compensate.
This is akin to the concept of diffeomorphism invariance discussed in section~\ref{sec:diffeo_inv}. In both GR and QFT there are two ways to change the local coordinate basis. The first is by moving from an initial position to a new position where the basis is defined differently. The second is by staying at one point and performing a transformation (a diffeomorphism in GR, a gauge transformation in QFT) to change the coordinate basis. In each case, we want the laws of physics to remain the same, despite any change to the chosen coordinate basis. We can see how this condition is enforced by the transformation of the connection, eq.~(\ref{eqn:A_transf}), and the role of the covariant derivative in the action for a Dirac field $ \psi $ of mass $m$;
\beq \label{eqn:diracfield}
S = \int d^4 x \, \bar{\psi}(i\hbar c \gamma^\mu \partial_\mu - m c ^2)\psi\, .
\eeq
A \emph{global} gauge transformation corresponds to rotating $\psi$ by a \emph{constant} phase $ \psi \rightarrow e^{ig\lambda}\psi $. Under this change we can see that the value of the action
\beq
S \rightarrow \int d^4 x \, \bar{\psi}e^{-ig\lambda}(i\hbar c \gamma^\mu \partial_\mu - m c ^2)e^{ig\lambda}\psi
\eeq
does not change because the factor of $e^{ig\lambda}$ acting on $\psi$ and the corresponding factor of $e^{-ig\lambda}$ acting on $\bar{\psi}$ pass through the partial derivative unaffected, and cancel out. However if we allow $\lambda$ to become a function of position $\lambda(x)$, then the global gauge transformation is promoted to a \emph{local} gauge transformation, due to which the partial derivative becomes
\beq
\partial_\mu \left(e^{ig\lambda(x)}\psi\right) = e^{ig\lambda(x)}\left( \del_\mu + ig(\del_\mu \lambda(x)) \right) \psi
\label{eqn:chain_rule_deriv2}
\eeq
leading to a modification of the action $ S \rightarrow S - \int d^4 x \hbar c \gamma^\mu (\partial_\mu \lambda) \bar{\psi} \psi $.
The covariant derivative, however, compensates for the $x$-dependence of $\lambda$, since as we saw in eq.~(\ref{eqn:Dmu_transf}) it has the property that
\beq
D_\mu \psi \rightarrow D_\mu\left(e^{ig\lambda(x)}\psi\right) = e^{ig\lambda(x)} D_\mu\psi
\eeq
and so the phase factor passes through the covariant derivative as desired. It is now trivial to show that the Dirac action defined in terms of the covariant derivative,
\beq\label{eqn:diracaction}
S_\mathrm{Dirac} = \int d^4 x \, \bar{\psi}(i\hbar c \gamma^\mu D_\mu - m c ^2)\psi
\eeq
is invariant under local phase transformations of the form $\psi \rightarrow e^{ig\lambda(x)}\psi$, $\bar{\psi} \rightarrow \bar{\psi} e^{-ig\lambda(x)}$, so long as $A_\mu(x)$ transforms as per eq.~(\ref{eqn:A_transf}). The connection $A_\mu$ tells us how the phase of the wavefunction at each point corresponds to the phase at a different point, in analogy to the connection in GR which told us how coordinate bases varied from point to point, but additionally the requirement that the action be invariant under \emph{local} gauge transformations necessitates that it is not simply the wavefunction, but also the connection that changes under a gauge transformation. 

The discussion above has been restricted to the case of a simple rotation of the phase (that is, $e^{ig\lambda}\in U(1)$, the rotation group of the plane). In GR, by contrast, the local bases at different points may be rotated in three dimensions relative to each other (that is, the basis vectors are acted upon by elements of $SO(3)$). We can accordingly generalise the discussion above to include phase rotations arising from more elaborate groups. For instance, in the case of $SU(2)$ we replace the wavefunction $\psi$ by a Dirac doublet
\beq
\psi \rightarrow \psi = \begin{pmatrix} \psi_1(x) \\ \psi_2(x) \end{pmatrix}
\eeq
and act upon this with transformations of the form
\beq
U(x) = \exp(i\lambda^I(x)t^I).
\eeq
Here $t^I = \sigma^I/2$, (with $\sigma^I$ the $I^\th$ Pauli matrix)\footnote{In general the $t^I$ will be the appropriate generators of the symmetry group, where $I=1,\,2,\,\ldots \,N$.}. In this case the covariant derivative becomes
\beq
D_\mu = \del_\mu + igA^I_\mu t^I
\eeq
(summation on the repeated index is implied). In analogy to the case discussed above for GR, we can form the commutator of covariant derivatives. In this case, we obtain the field strength tensor $F_{\mu\nu}$, the analogue of the Riemann curvature tensor,
\beq
[D_\mu,\,D_\nu] = igF^I_{\mu\nu}t^I
\label{eqn:commut_Fmunu}
\eeq
where we can see (by applying the standard commutation relations for the Pauli matrices, namely $[\sigma^I,\,\sigma^J] = 2i \epsilon^{IJK}\sigma^K$, and relabelling some dummy indices) that
\beq\label{eqn:nonabeliancurvature}
F^I_{\mu\nu} = \del_\mu A^I_\nu-\del_\nu A^I_\mu - g\epsilon^{IJK}A^J_\mu A^K_\nu .
\eeq

When our gauge group is abelian (as in QED) all the generators of the corresponding Lie algebra commute with each other and thus the structure constants of the group ($ \epsilon^{IJK} $ in the $SU(2)$ example of eq.~(\ref{eqn:nonabeliancurvature})) vanish. In this event the field strength simplifies to
\beq\label{eqn:abeliancurvature}
F^I_{\mu\nu} = \del_\mu A^I_\nu-\del_\nu A^I_\mu
\eeq

The field strength $F_{\mu\nu}^I$ itself is gauge \emph{covariant} but not gauge \emph{invariant}. Under an infinitesimal gauge transformation $ A_0 \rightarrow A_0 + \delta A$ the field strength also changes by $F[A_0] \rightarrow F[A_0 + \delta A] = F_0 + \delta F$ where the variation in field strength is given by $\delta F = D_\mu [A_0]$ as the reader can easily verify by substituting and expanding in eq.~(\ref{eqn:nonabeliancurvature}) or eq.~(\ref{eqn:abeliancurvature}). Here $D_{\mu} [A_0]$ denotes that the covariant derivative is taken with respect to the original connection $A_0$.

The basic statement of Einstein's gravitational theory, often expressed in the saying
\begin{quote}
``Matter tells geometry how to curve and geometry tells matter how to move.''
\end{quote}
has a parallel statement in the language of gauge theory. In a gauge theory, matter is represented by the fields $\psi$ whereas the ``geometry'' (not of the background spacetime, but of the interactions between the particles) is determined by the configurations of the gauge field. The core idea of GR can then be generalised to an equivalent idea in field theoretic terms,
\begin{quote}
``Gauge charges tell gauge fields how to curve and gauge fields tell gauge charges how to move.''
\end{quote}

Now, what we have so far is an action, eq.~(\ref{eqn:diracaction}) which describes the dynamics of spinorial fields, interactions between which are mediated by the gauge field. The gauge field itself is not yet a dynamic quantity. In any gauge theory, consistency demands that the final action should also include terms which describe the dynamics of the gauge field alone. We know this to be true from our experience with QED where the gauge field becomes a particle called the photon. From classical electrodynamics Maxwell's equations possess propagating solutions of the gauge field - or more simply \emph{electromagnetic waves}. The term giving the dynamics of the gauge field can be uniquely determined from the requirement of gauge invariance. We need to construct out of the field strength an expression with no indices. This can be achieved by contracting $F_{\mu\nu}^I$ with itself and then taking the trace over the Lie algebra indices. Doing this we get the term
\beq\label{eqn:gaugefieldaction}
S_{gauge} = - \frac{1}{4}\int d^4 x \mathrm{Tr} \left[ F^{\mu\nu} F_{\mu\nu} \right]
\eeq
which in combination with (\ref{eqn:diracaction}) gives us the complete action for a gauge field interacting with matter
\beq\label{eqn:fullaction}
S = S_{gauge} + S_{Dirac} = \int d^4 x \left\{ - \frac{1}{4}\mathrm{Tr} \left[ F^{\mu\nu} F_{\mu\nu}\right] + \bar{\psi}(i\hbar c \gamma^\mu D_\mu - m c ^2)\psi \right\}
\eeq

\subsection{Dual tensors, bivectors and $k$-forms}
\label{sec:k-vectors-k-forms}
The field strength is usually first encountered in the case of electromagnetism, where the relevant gauge group is $U(1)$ which has only one group generator and so we can drop the index $I$ in eq.~(\ref{eqn:abeliancurvature}). The electromagnetic field strength $F_{\mu\nu}$ combines the electric and magnetic fields into a single entity,
\beq
F_{\mu\nu} = \del_\mu A_\nu-\del_\nu A_\mu =
\begin{pmatrix}
   0  &  E_1 &  E_2 &  E_3 \\
 -E_1 &   0  & -B_3 &  B_2 \\
 -E_2 &  B_3 &   0  & -B_1 \\
 -E_3 & -B_2 &  B_1 &   0  \\
\end{pmatrix}
\label{eqn:fmunu_components}
\eeq
Since each component of $F_{\mu\nu}$ is associated with two index values, we can think of the components as ``bivectors'' (oriented areas lying in the $\mu$-$\nu$ plane), in analogy with vectors which carry only a single index (and are oriented lengths lying along a single axis). For the reader unfamiliar with bivectors we will very quickly review them.

\begin{figure}[b]
\centering
\includegraphics[scale=0.3]{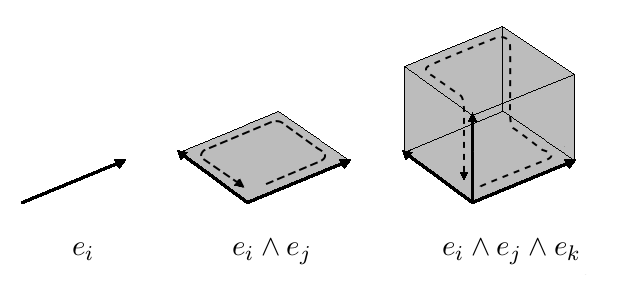}
\includegraphics[scale=0.3]{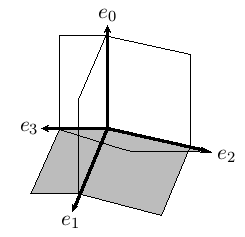}
\caption{Wedge products of basis vectors define basis bivectors, basis trivectors, and so on. While a vector's magnitude is its length, a bivector's magnitude is its area, and the magnitude of a trivector is its volume. The orientation of the unit bivector and unit trivector are shown here by the dashed arrows. The field strength $F_{\mu\nu}$ can be represented as a set of bivectors oriented between pairs of timelike and spacelike axes in four dimensions (shown here by distorting the angles between axes, as is done in a two-dimensional drawing of  a cube). Shaded (unshaded) bivectors are the magnetic (electric) field components.}
\label{fig:multivector}
\end{figure}

A unit basis vector $e_i$ can be visualised as a line segment with a ``tail'' and a ``head'', and an orientation given by traversing the vector from its tail to its head. A general vector is a linear combination of basis vectors, $\vec{v}=v^1 e_1 + v^2 e_2 +v^3 e_3 + \ldots$ Similarly a unit basis bivector can be visualised as an area bounded by the vectors $e_i$ and $e_j$, written as the wedge product $e_i\wedge e_j$, and with an orientation defined by traversing the boundary of this area along the first side, in the same direction as $e_i$, then along the second side parallel to $e_j$, and continuing anti-parallel to $e_i$ and $e_j$ to arrive back at the origin (this concept can be extended arbitrarily to define trivectors, etc. as illustrated in Figure.~\ref{fig:multivector}). A general bivector is a linear combination of basis bivectors. Writing the field strength as a general bivector we find that it takes the form 
\beq
  F_{\mu\nu} = E_1 (e_1\wedge e_0) + E_2 (e_2\wedge e_0) + E_3 (e_3\wedge e_0) + B_1 (e_2\wedge e_3) + B_2 (e_3\wedge e_1) + B_3 (e_1\wedge e_2)
\eeq
Electric fields are those parts of $F_{\mu\nu}$ lying in a plane defined by one space axis and the time axis, while magnetic fields are those lying in a plane defined by two space axes (Figure~\ref{fig:multivector}). 
Reversing the orientation of a bivector is equivalent to traversing its boundary ``backwards'', so we may write $e_j\wedge e_i = -e_i\wedge e_j$. This is consistent with the fact that the field strength is antisymmetric, \ie $F_{\mu\nu} = - F_{\nu\mu}$.

We can also combine the electric and magnetic fields into a single entity by defining the dual field strength,
\beq
 \star\!F^{\mu\nu} = \frac{1}{2}\epsilon^{\lambda\rho\mu\nu}F_{\lambda\rho} =
 \begin{pmatrix}
   0  & -B_1 & -B_2 & -B_3 \\
  B_1 &   0  &  E_3 & -E_2 \\
  B_2 & -E_3 &   0  &  E_1 \\
  B_3 &  E_2 & -E_1 &   0  \\
\end{pmatrix}
\label{eqn:dualfmunu_components}
\eeq 
We can see that the mapping between field strength and dual field strength\footnote{The notation ${\tilde F}$ is also used for the dual field strength.} associates a given electric field component with a corresponding magnetic field component, such that $E_j \leftrightarrow -B_j$. Thinking in terms of bivectors, the quantity defined on the plane between any pair of spacetime axes is associated to the quantity defined on the plane between the other two spacetime axes. The field strength is said to be {\em self-dual} if $\star F = +F$, and {\em anti-self-dual} if $\star F = - F$. Although we will not be concerned with (anti-)self-dual field strengths in the rest of this paper, we will be dealing with (anti-)self-dual gauge connections from section~\ref{subsec:ashtekar-connection} onwards. The EM field strength as presented here is merely the simplest example to use to introduce the concept of self-duality, and illustrate its physical meaning. Further discussion of duality, for the reader requiring a deeper understanding, is presented in appendix~\ref{app-sec:duality}. Some readers will also no doubt have noticed the similarity between bivectors $e_i\wedge e_j$, and differential 2-forms $dx^i\wedge dx^j$. The two are indeed very similar. 
\begin{figure}[t] 
\begin{center}
{\includegraphics[height=39mm,angle=0]{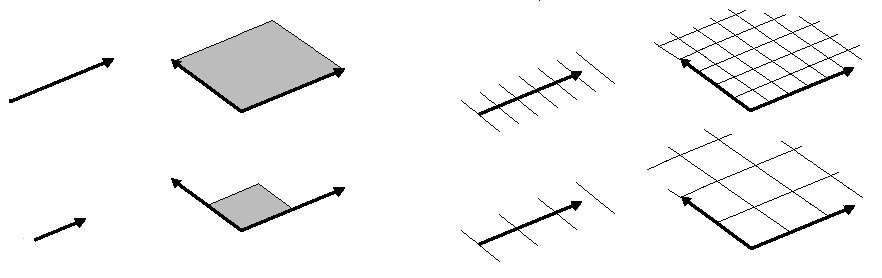}}
\caption{Two vectors (far left) with the same direction and different magnitude differ in their length, while two bivectors (left) differ in their area. The magnitude of $k$-forms is a density, and can be represented by interval lines. A 1-form (right) has a direction, just like a vector, but the spacing of interval lines represents its magnitude. A 2-form (far right) defines a plane, just like a bivector does, and once again the magnitude is represented by the spacing of interval lines. In all cases, the greater magnitude object is on the top row.}
\label{fig:vec_form_magn}
\end{center}
\end{figure}
A bivector defined by the wedge product of two vectors $a\wedge b$ can be imagined as a parallelogram with two sides parallel to $a$, and the other sides parallel to $b$. The magnitude of this bivector is the area of the enclosed parallelogram. Differential forms, on the other hand, have a magnitude which is thought of as a density. This is often drawn as a series of lines (similar to the contour lines on a topographical map or the isobars on a weather map) with smaller spacing between lines indicating higher density (Figure~\ref{fig:vec_form_magn}). Hence a 1-form can be thought of as a density of contour lines or contour surfaces perpendicular to the direction of the 1-form. The inner product of a vector with a 1-form is a scalar - the number of lines that the vector crosses. Similarly a 2-form can be thought of as a series of contours spreading out through a plane (this plane being defined by the directions of the two 1-forms wedged together to make the 2-form). Clearly there is a one-to-one mapping between vectors and 1-forms, and between bivectors and 2-forms, which involves changing one's choice of magnitude, (length or area) $\leftrightarrow $ (density). It is certainly more common to see 1-forms, 2-forms, and higher-dimensional forms used throughout physics, but bivectors and higher-dimensional multivectors can be very useful too (see appendix~\ref{app-subsec:diff-forms} for a further discussion of multivectors) and are often easier to visualise.    


\subsection{Wilson Loops and Holonomies}
\label{subsec:holonomies}
In section~\ref{sec:gen_rel} we defined a holonomy as a measure of how much the initial and final values of a vector transported around a closed loop differ. The discussion in the previous section demonstrates that the internal degrees of freedom of a spinor can also be position-dependant, and hence it should be possible to define a holonomy by the difference between the initial and final values of a spinor transported around a closed loop\footnote{The name holonomy is also used within the LQG community to refer to a closed loop itself. We feel this is unnecessarily confusing, and hence we shall avoid using the term ``holonomy'' for a closed loop or closed path. The reader should be aware that this terminology does, however, exist within the wider literature.}. As a first step to constructing such a definition, let us consider what happens when we compare the values of a field at different points, separated by a displacement $dx^\mu$. We begin by using eq.~(\ref{eqn:chain_rule_wavefunction}), eq.~(\ref{eqn:chain_rule_deriv_wavefunction}) and eq.~(\ref{eqn:cov_deriv_gauge}) to write
\beq
\frac{d\psi}{dx^\mu} = \frac{\del\phi}{\del x^\mu} u + \phi \frac{\del u}{\del x^\mu} = u\left(\partial_\mu + ig A_\mu\right)\phi 
\eeq
from which we readily see that $ig A_\mu u = \del_\mu u$, or equivalently $ig A_\mu u dx^\mu = du$. The internal components of the fields will be related by a gauge rotation which we will call $U(dx^\mu)$. The action of this rotation can be expanded as 
\beq
 U(dx^\mu)u = u + du = u + ig A_\mu u dx^\mu = (1 + ig A_\mu dx^\mu) u
\eeq
and we immediately see that 
\beq
 U(dx^\mu) = \exp\{ig A_\mu dx^\mu\}
\eeq
$U(dx^\mu)$ is the parallel transport operator that allows us to bring two field values at different positions together so that they may be compared. Remembering that the effect of parallel transport is path-dependant, this operator can be readily generalised to finite separations along an arbitrary path $\lambda$ and connections valued in arbitrary gauge groups, in which case we find
\beq\label{eqn:schwinger_line_integral}
U(x,\,y) = \mc{P} \exp \left\{ \int_\lambda igA_\mu{}^I(x) t^I dx^\mu \right\} 
\eeq
where the $\mc{P}$ tells us that the integral must be \emph{path ordered}\footnote{See Appendix \ref{app:pathordered} for the definition of a ``path ordered'' exponential.}, $t^I$ are gauge group generators as before, and $x$ and $y$ are the two endpoints of the path $\lambda$ we are parallel transporting along. If the gauge connection vanishes along this path then the gauge rotation is simply the identity matrix and $\psi$ is unchanged by being parallel transported along the path. In general, however, the connection will \emph{not} vanish. 

\begin{figure}[htbp]
\centering
\includegraphics[scale=0.3]{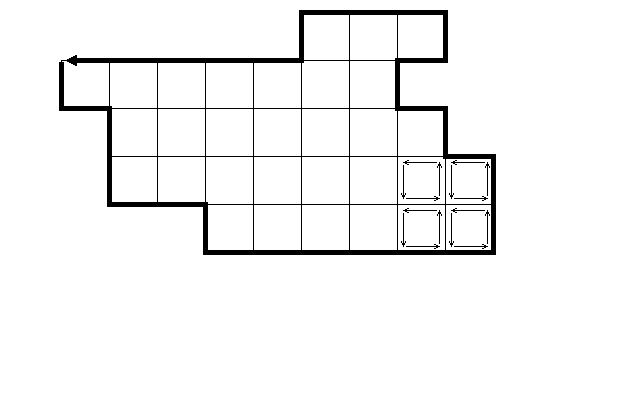}
\caption{An arbitrary closed path in the plane can be approximated by tilings of plaquettes. Since each plaquette is traversed anti-clockwise, adjacent edges make cancelling contributions to the parallel transport of a spinor, leaving only the contribution at the boundary of the tiling (as illustrated for the plaquettes in the lower-right corner).}
\label{fig:plaq_tiling}
\end{figure}

Now consider the situation when the path $\lambda$ is a closed loop, \ie its beginning and end-point coincide. Analogously to the situation for a curved manifold, where the parallel transport of a vector along a closed path gives us a measure of the curvature of the spacetime bounded by that path, the parallel transport of a spinor around a closed path yields a measure of the \emph{gauge} curvature living on a surface bounded by this path. We can see this simply in the case of a small square ``plaquette'' in the $\mu$-$\nu$ plane, with side length $a$. The gauge rotation in this case is a product of the rotation induced by parallel-transporting a spinor along each of the four sides of the plaquette in order. The parallel transport operators for each side of the plaquette are found from eq.~(\ref{eqn:schwinger_line_integral}), and explicitly, their product around a plaquette is 
\beq
W = e^{igaA^\dag_\nu(x+a\nu)}e^{igaA^\dag_\mu(x+a\mu+a\nu)}e^{igaA_\nu(x+a\mu)}e^{igaA_\mu(x)}
\eeq
Assuming that we are dealing with a non-Abelian field theory, this product of exponentials can be converted to a single exponential by use of the Baker-Cambell-Haussdorf rule, which for the product of four terms takes the form 
\beq
e^Ae^Be^Ce^D=\exp\{A+B+C+D+[A,B]+[A,C]+[A,D]+[B,C]+[B,D]+[C,D]+...\}
\eeq 
After a bit of algebra we find that this simplifies to 
\beq
 W = \exp\{iga^2F_{\mu\nu}+...\}
\eeq
where the $...$ represent higher-order terms. An arbitrary loop can be approximated by a tiling of small plaquettes, to yield a result proportional to the total tiled area, multiplied by $F_{\mu\nu}$. Since the common edges of adjacent plaquettes are traversed in opposite directions, the contributions along these edges are cancelled, and the entire tiling results in a path around the outside of the tiled area (Fig.~\ref{fig:plaq_tiling}). Such an arbitrary loop is called a Wilson loop, and the holonomy associated to it is called the {\em Wilson loop variable}, denoted $W_\lambda$. To obtain a single variable from the parallel transport around a loop, we take the trace of the parallel transport operator, hence 
\beq
W_\lambda={\mathrm {Tr}}\mc{P} \exp \left\{ \oint_\lambda igA_\mu{}^I(x) t^I dx^\mu \right\} 
\label{eqn:holonomy_wilsonloop}
\eeq 
The Wilson loop is gauge-invariant, since each line segment of which the loop is composed transforms as 
\beq
  U(x,y) \rightarrow G(y)U(x,y)G^{-1}(x)
\eeq 
under a gauge transformation like that in eq.~(\ref{eqn:Dmu_transf}), and so the product of several line segments forming a closed loop transforms as
\beq
  W \rightarrow W' = G(x_1)U(x_1,\,x_2)G^{-1}(x_2) \ldots G(x_n)U(x_n,\,x_1)G^{-1}(x_1)
\eeq 
Different gauge transformations therefore correspond with different choices of starting point for the loop. However the trace is invariant under cyclic permutations, Tr{\em ABC} = Tr{\em BCA} = Tr{\em CAB}, and so the Wilson loop variable is independent of choice of gauge transformation \cite{PeskinSchroeder}. 

This discussion shows that $F_{\mu\nu}$ is a measure of the gauge curvature within a surface, as well as a measure of the holonomy of the loop enclosing the surface (that is, the gauge rotation induced on a spinor when it is parallel-transported around a closed loop). Hence when the connection does not vanish the associated holonomy will in general not be trivial.\todo{Insert some sentences about Yang Mills theory (1st mention of it in this paper). Show how it can be built from loop variables?}



\subsection{Dynamics of Quantum Fields}

We will conclude this section with a discussion of two approaches to the dynamics of quantum fields. These are well-established in the case of theories like QED and QCD, and so it will be natural later on to consider equivalent approaches when we wish to quantise spacetime, which is the dynamical field in GR. These two approaches are based on lagrangian and hamiltonian dynamics.

\subsubsection{Lagrangian (or Path Integral) Approach}\label{subsec:path-integrals}
As shown in eq.~(\ref{eqn:gaugefieldaction}), starting with the curvature of a gauge field it is possible to define an action which governs the dynamics of the gauge field. In the path-integral approach to quantum field theory the basic element is the propagator (or the partition function when $\mc{M}$ is a Euclidean manifold) which allows us to calculate the probability amplitudes between pairs of initial and final states of our Hilbert space. Although we will be concerned with fields throughout the majority of the following discussion, the prototypical example is that of the non-relativistic point particle in flat space moving under the influence of an external potential $V(x)$ for which the action is given by
\beq\label{eqn:nonrel-particle-action}
S_{pp}[\gamma] = \int_\gamma d^3x dt \, \left(\frac{1}{2}m \dot{x}^2 - V(x)\right)
\eeq
Note that the potential term must be replaced by a gauge field $A_\mu$ in the relativistic case, in which case the action takes the form
\beq\label{eqn:rel-particle-action}
S_{Rel}[\gamma] = \int_\gamma d^3x dt \,\frac{(p^\mu + A^\mu)(p_\mu + A_\mu)}{m_0}
\eeq
where $p^\mu$ is the energy-momentum 4-vector of the particle and $m_0$ is its rest mass. This is the familiar action for a charged point particle moving under the influence of an external potential encoded in the abelian gauge potential $A_\mu$. The action integral \emph{depends} on the choice of the path $\gamma$ taken by the system as it evolves from the initial to final states in question. The action can be evaluated for \emph{any} such path and not just the ones which extremize the variation of the action. This allows us to assign a complex amplitude (or real probability in the Euclidean case) to any path $\gamma$ by:
\beq\label{eqn:path-amplitude}
\exp \left\{ i S[\gamma] \right\}
\eeq\todo{imaginary and real. Wick rotations, and so on}
Using this complex amplitude as a weighting function we can calculate matrix elements for transitions between an arbitrary pair of initial $\Psi_i(t)$ and final $\Psi_f(t')$ states by summing all paths or \emph{histories} which interpolate between the two states,
\beq\label{eqn:matrix-element}
\left\langle \Psi_i(t) \mid \Psi_f(t') \right\rangle = \int \mc{D}[\psi] \exp \left\{ i S[\gamma] \right\},
\eeq
in contrast to the classical view of dynamics, in which a system moves from an initial state to a final state in exactly one way. Here $\mc{D}[\psi]$ is an appropriate measure on the space of allowed field configurations.

For the point-particle $\fullket{q,t}$ represents a state where the particle is localized at position $q$ at time $t$. The matrix-element between states at two different times then takes the form
\beq\label{eqn:nonrel-matrix-element}
\left\langle q,t \mid q,t' \right\rangle = \int \mc{D}[\psi] \exp \left\{ i S_{pp}[\gamma] \right\}\, .
\eeq
The weighting factor gives higher value to the contribution from those paths which have an associated action close to the minimum. It is this which results in classical behaviour, in the appropriate limit. However the contributions of all possible paths must still be taken into account to accurately calculate the transitions between states. 

\subsubsection{Hamiltonian approach: Canonical quantisation}
The alternative to the lagrangian or path-integral approach is to study the dynamics of a system through its Hamiltonian. This leads to Dirac's procedure for canonical (or ``second'') quantisation\footnote{The quantisation of the motion of a particle in a classical potential is sometimes referred to as ``first quantisation''. This is the basis for the somewhat un-intuitive name ``second quantisation'' for quantisation extended to the potential as well.}. The Hamiltonian $H$ for a dynamical system can be constructed from the Lagrangian $L$ by performing a Legendre transformation. Given a configuration variable $q$, which we can think of as a generalised position, and a corresponding generalised momentum $p$ defined by 
\beq
  p = \frac{\del L}{\del \dot{q}}\, ,
\eeq
then the Hamiltonian is given by
\begin{equation}
H[p,\,q] = p\dot{q} - {L}[q,\,\dot{q}] 
\end{equation}
in the case of a point particle, and generalisations of this equation for other systems. If we define the Poisson bracket of two functions by
\beq
  \left\{f,g\right\} = \sum_{i=1}^n
  \left(\frac{\del f}{\del q_i}\frac{\del g}{\del p_i} - 
  \frac{\del f}{\del p_i}\frac{\del g}{\del q_i}\right)
\eeq
where $f=f(q,\,p,\,t)$ and $g=g(q,\,p,\,t)$, then Hamilton's equations can be written in the form
\beq
\dot{q} = \frac{\partial H}{\partial p} = \left\{H,p\right\} \hspace{5mm} \mathrm{and} \hspace{5mm} \dot{p}= -\frac{\partial H}{\partial q} = \left\{H,q\right\}
\eeq 
and give the time evolution of the system. Hence, leaving the second spot in the brackets empty, time evolution is generated by the operator $\{ H,\,\}$ which acts upon the generalised coordinates and momenta. 

In quantum mechanics and quantum field theory observables are replacd by operators, \ie $x \rightarrow \hat{x}$. While operators do not necessarily commute, classical observables do. However the Poisson bracket of two observables will not necessarily be zero, and Dirac was led to postulate that in the transition from classical to quantum mechanics, Poisson brackets between observables should be replaced by commutation relations, where the scalar value of the commutator is $i\hbar$ times the scalar value of the equivalent Poisson bracket, \ie
\beq
  \left\{f,g\right\} = 1 \hspace{5mm} \mathrm{implies} \hspace{5mm} \left[\hat{f},\hat{g}\right] = i\hbar 
\eeq 
This prescription will be central to our attempts to quantise spacetime in later sections. 

This completes the necessary background discussion of quantum field theories. 

\section{Expanding on classical GR}
\label{sec:Expand_GR}
We now return to the discussion of General Relativity. Equipped with the preceding discussions of both the quantisation of field theories, and the geometrical interpretations of gauge transformations, it is time to set about formulating what will eventually become a theory of dynamical spacetime obeying rules adapted from quantum field theory. But before we get there we must cast classical GR into a form amenable to quantisation.

From classical mechanics we know that dynamics can be described either in the Hamiltonian or the Lagrangian frameworks. The benefits of a Lagrangian framework are that it provides us with a covariant perspective on the dynamics and connects with the path-integral approach to the quantum field theory of the given system. The Hamiltonian approach, on the other hand, provides us with a phase space picture and access to the Schrodinger method for quantization. Each has its advantages and difficulties and thus it is prudent to be familiar with both frameworks. We will begin with discussing these approaches in a classical framework, and move to quantisation in section \ref{sec:firststepsquantgravity}.

\subsection{Lagrangian approach: The Einstein-Hilbert Action}
\label{subsec:eh-action}

The form of the Lagrangian, and hence the action, can be determined by requirements of covariance and simplicity. Out of the dynamical elements of geometry - the metric and the connection - we can construct a limited number of quantities which are invariant under coordinate transformations, hence they should have no uncontracted indices. These quantities must be constructed out of the Riemann curvature tensor or its derivatives. These possibilities are of the form: $ \{ R, R_{\mu\nu} R^{\mu\nu}, R^2, \nabla_\mu R \nabla^{\mu} R, \ldots \} $. The simplest of these is the Ricci scalar $R = R_{\mu\nu\alpha\beta} g^{\mu\alpha} g^{\nu\beta} $. As it turns out this term is sufficient to fully describe Einstein's general relativity, yielding a lagrangian that is simply $\sqrt{-g}R$, where as noted in subsection~\ref{sec:diffeo_inv}, $g=\mathrm{det}(g^{\mu\nu})$.

This allows us to construct the simplest lagrangian which describes the coupling of geometry to matter:
\beq\label{eqn:ehm-action}
S_\mathrm{EH+M} = \frac{1}{\kappa}\int d^4 x \sqrt{-g}R + \int d^4 x \sqrt{-g} \mc{L}_\mathrm{matter}
\eeq
where $\mc{L}_\mathrm{matter}$ is the lagrangian for the matter fields that may be present and $\kappa$ is a constant, to be determined. If the matter lagrangian is omitted, one obtains the usual vacuum field equations of GR. This action (omitting the matter term) is known as the Einstein-Hilbert action, $S_\mathrm{EH}$.

It is worth digressing to prove (at least in outline form) that the Einstein field equations (EFEs) can be found from $S_\mathrm{EH+M}$. The variation of the action (\ref{eqn:ehm-action}) yields a classical solution which, by the action principle, is chosen to be zero,
\beq\label{eqn:gmunuvariation}
\delta S=0=\int d^4x\left[\frac{1}{\kappa}\frac{\delta\sqrt{-g}}{\delta g^{\mu\nu}}R+\frac{1}{\kappa}\sqrt{-g}\frac{\delta R}{\delta g^{\mu\nu}}+\frac{\delta\sqrt{-g}{\mc{L}_\mathrm{matter}}}{\delta g^{\mu\nu}}\right]\delta g^{\mu\nu}
\eeq
which implies that
\beq
\frac{1}{\sqrt{-g}}\frac{\delta\sqrt{-g}}{\delta g^{\mu\nu}}R+\frac{\delta R}{\delta g^{\mu\nu}} = -\kappa\frac{1}{\sqrt{-g}}\frac{\delta\sqrt{-g}{\mc{L}_\mathrm{matter}}}{\delta g^{\mu\nu}}\,.
\eeq
The energy-momentum tensor can be defined as
\beq
 T^{\mu\nu} = -\frac{2}{\sqrt{-g}} \frac{\delta\sqrt{-g}{{\cal L}_\mathrm{matter}}}{\delta g^{\mu\nu}}
 \label{eqn:tmunu}
\eeq 
where $g=\mathrm{det}(g^{\mu\nu})$, and ${\cal L}_\mathrm{matter}$ is a lagrangian encoding the presence of matter\footnote{This definition of the energy-momentum tensor may seem to come out of thin air, and in many texts it is simply presented as such. To save space we will follow suit, but the reader who wishes to delve deeper should consult \cite{HobsonEfstathiouLasenby}, in which $T_{\mu\nu}$ is referred to as the {\em dynamical} energy-momentum tensor, and it is proven that it obeys the conservation law $\nabla_\mu T^{\mu\nu}=0$ (as one would hope, since energy and momentum are conserved quantities), as well as being consistent with the form of the electromagnetic energy-momentum tensor.}. From equation~(\ref{eqn:tmunu}) we can immediately see that
\beq
\frac{1}{\sqrt{-g}}\frac{\delta\sqrt{-g}}{\delta g^{\mu\nu}}R+\frac{\delta R}{\delta g^{\mu\nu}} = \frac{\kappa}{2}T^{\mu\nu}\,.
\label{eqn:tmunuvariation}
\eeq
We now need to work out the variation of the terms on the left-hand-side. Omitting the details, which can be found elsewhere (see e.g. the appendix of \cite{Wald1984General}), we find that
\beq
\delta\sqrt{-g} = -\frac{1}{2\sqrt{-g}}\delta \sqrt{g} = \frac{1}{2}\sqrt{-g}(g^{\mu\nu}\delta g_{\mu\nu}) = - \frac{1}{2}\sqrt{-g}(g_{\mu\nu}\delta g^{\mu\nu})
\eeq
thanks to Jacobi's formula for the derivative of a determinant. The variation of the Ricci scalar can be found by differentiating the Riemann tensor, and contracting on two indices to find the variation of the Ricci tensor. Then, since the Ricci scalar is given by $R=g^{\mu\nu} R_{\mu\nu}$ we find that
\beq
\delta R = R_{\mu\nu}\delta g^{\mu\nu} + g^{\mu\nu} \delta R_{\mu\nu}\,.
\eeq
The second term on the right may be neglected when the variation of the metric vanishes at infinity, and we obtain $\delta R /\delta g^{\mu\nu}= R_{\mu\nu}$. Plugging these results into eq.~(\ref{eqn:tmunuvariation}) we find that
\beq
-\frac{1}{2}g_{\mu\nu} R + R_{\mu\nu} = \frac{\kappa}{2}T^{\mu\nu}
\eeq
which yields the Einstein equations if we set $\kappa=16\pi \mc{G}$.

As noted in eq.~(\ref{eqn:gamma_from_g}), we can write $\Gamma^\rho_{\mu\nu}$ in terms of the metric $g_{\mu\nu}$,
\begin{equation}
\Gamma^\rho_{\mu\nu} = \frac{1}{2} g^{\rho\delta}\left( \partial_\mu g_{\delta\nu} + \partial_\nu g_{\delta\mu} - \partial_\delta g_{\mu\nu} \right) \nonumber
\end{equation}
and since the covariant derivative $\nabla_\mu$ is a function of $\Gamma^\rho_{\mu\nu}$, and the Riemann tensor is defined in terms of the covariant derivative, the Einstein-Hilbert action is ultimately a function of the metric $g_{\mu\nu}$ and its derivatives.

\begin{figure}[t]
    \centering{{\includegraphics[scale=0.2]{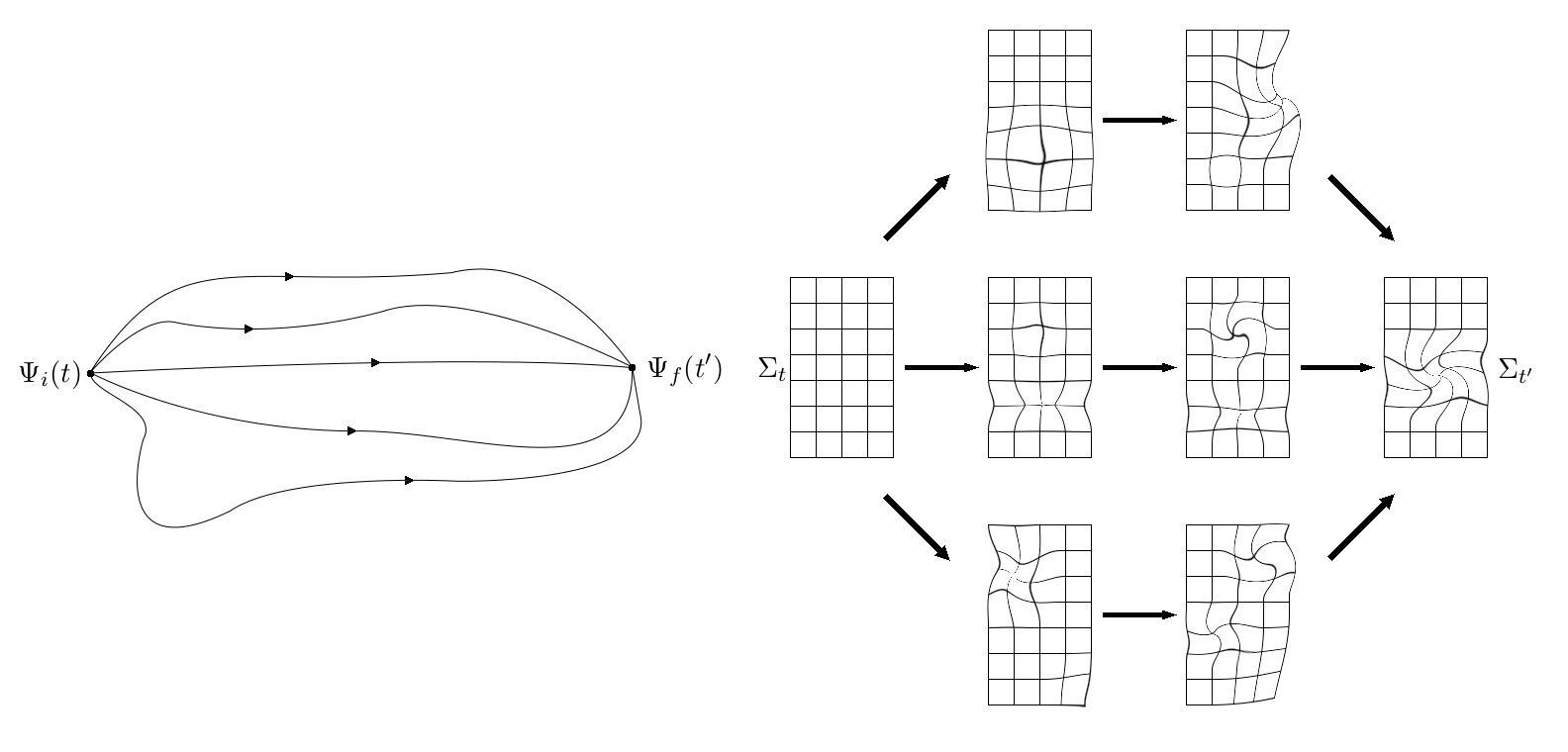}}}
    \caption{Weighted sums of transitions between different configurations of spacelike hypersurfaces may be used to calculate the transition amplitude between an initial and final state of geometry (right), see eq.~(\ref{eqn:propagator}). This is analogous to the path-integral approach used in quantum field theory (left), see eq.~(\ref{eqn:matrix-element}).}
    \label{fig:geometrypath}
\end{figure}

As a further aside, we will briefly describe how the Lagrangian formulation allows us to make contact with the path-integral or sum-over-histories approach outlined in sec.~\ref{subsec:path-integrals}, and apply it to the behaviour of spacetime as a dynamical field. In general, this approach involves calculating transition amplitudes with each path between the initial and final states being weighted by an exponential function of the action associated with that path. In the case of gravity we may think of four-dimensional spacetime as a series of spacelike hypersurfaces, $\Sigma_t$, corresponding to different times. Each complete 4-dimensional geometry consisting of a series of 3-dimensional hypersurfaces that interpolate between the initial and final states may be thought of as the generalisation of a ``path''. This 3+1 splitting of spacetime into foliated three-dimensional hypersurfaces will be covered in more detail in the next subsection. To calculate the matrix-elements (as in eq.~(\ref{eqn:matrix-element})) for transition amplitudes between initial and final states of geometry, $\Sigma_t$ and $\Sigma_{t'}$ (see Fig. \ref{fig:geometrypath}) we use the Einstein-Hilbert action for GR on a manifold $\mc{M}$ without matter
\beq\label{eqn:eh-action}
S_{EH} = \frac{1}{\kappa}\int d^4 x \sqrt{-g}\,R
\eeq
Let us represent the states corresponding to the initial and final hypersurfaces as $\fullket{h_{ab},t}$ and $\fullket{h'_{ab},t'}$, where $h_{ab}$ is the intrinsic metric of a given spatial hypersurface, and $a,b\in\{1,2,3\}$. Then the probability that evolving the geometry will lead to a transition between these two states is given by
\beq\label{eqn:propagator}
\innerp{h_{ab},t}{h'_{ab},t'} = \int \mc{D}[g_{\mu\nu}] \exp \left\{ iS_{EH}(g_{\mu\nu})\right\}
\eeq
where the action is evaluated over all 4-metrics $g_{\mu\nu}$ interpolating between the initial and final hypersurfaces. $\mc{D}[g_{\mu\nu}]$ is the appropriate measure on the space of 4-metrics. While this approach is noteworthy, and ultimately leads to a very successful {\em computational} approach to quantising gravity \cite{CDT2005}, it is not the path we follow to formulate Loop Quantum Gravity. Instead, as mentioned above, the lagrangian formulation of General Relativity is used as a stepping-stone to the hamiltonian formulation. 

\subsection{Hamiltonian approach: The ADM splitting}
\label{sec:ADMsplit}

Since General Relativity is a theory of dynamical spacetime, we will want to describe the dynamics of spacetime in terms of some variables which make computations as tractable as possible. The Hamiltonian formulation is well suited to a wide range of physical systems, and the ADM (Arnowitt-Deser-Misner) formalism, described below, allows us to apply it to General Relativity. We can think of the action (\ref{eqn:ehm-action}), which is clearly written in the form of an integral of a lagrangian, as a stepping-stone to this hamiltonian approach. This hamiltonian formulation of GR takes us to the close of our discussion of classical gravity, and will be used as the jumping-off point for the quantisation of gravity, to be undertaken in Section~\ref{sec:firststepsquantgravity}.

The ADM formalism involves foliating spacetime into a set of three-dimensional spacelike hypersurfaces, and picking an ordering for these hypersurfaces which plays the role of time, so that the hypersurfaces are level surfaces of the parameter $t$. This is a necessary feature of the hamiltonian formulation of a dynamical system, although it seems at odds with the way GR treats space and time as interchangable parts of spacetime. However this time direction is actually a ``fiducial time''\footnote{The term ``fiducial'' refers to a standard of reference, as used in surveying, or a standard established on a basis of faith or trust.} and will turn out not to affect the dynamics. It is essentially a parameter used as a scaffold, which in the absence of a metric is not directly related to the passage of time as measured by a clock.

\begin{figure}[htbp]
      \begin{center}
      {\includegraphics[height=50mm,angle=0]{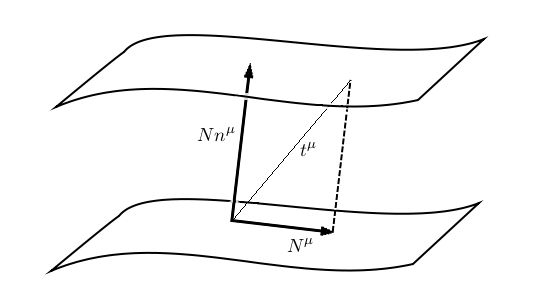}}
      \caption{When performing the ADM splitting, the lapse function $N$ and shift vector $N^\mu$ define how points on successive hypersurfaces are mapped together.}
      \label{fig:foliation}
\end{center}
\end{figure}

To begin, we will suppose that the 4-dimensional spacetime is embedded within a manifold $\mc{M}$ (which may be $\mathbb{R}^4$ or any other suitable manifold). Next we choose a local foliation\footnote{Generally one assumes that our 4 manifolds can always be foliated by a set of spacelike 3 manifolds. For a general theory of quantum gravity the assumption of trivial topologies must be dropped. In the presence of topological defects in the 4 manifold, in general, there will exist inequivalent foliations in the vicinity of a given defect. This distinction can be disregarded in the following discussion for the time being.} $ \{ \Sigma_t,t\} $ of $ \mc{M} $ into spacelike 3-manifolds, where $ \Sigma_t $ is the 3-manifold corresponding to a given value of the parameter $t$. We will refer to such a manifold as a ``leaf of foliation''. The topology of the original four-dimensional spacetime is then $\Sigma\otimes\mathbb{R}$, while $t$ 
is a parametrization of the set of geodesics orthogonal to $ \Sigma_t $, 
c.f. (Fig.~\ref{fig:foliation}). In addition at each point of a leaf we have a unit time-like vector $n^{\mu}$ (with $ n^\mu n_\mu = -1 $) which defines the normal at each point on the leaf.

Given the full four-metric $g_{\mu\nu}$ on $ \mc{M} $ and the vector field $n^{\mu}$ the foliation is completely determined by the requirement that the surfaces $\Sigma_t$ of constant ``time'' are normal to $ n^{\mu} $.

The diffeomorphism invariance of general relativity implies that there is no canonical choice of the time-like vector field $t^\mu$ which maps a point $x^{\mu}$ on a leaf $\Sigma_t$ to the point $x'^{\mu}$ on the leaf $\Sigma_{t+\delta t}$, \ie which generates time evolution of the geometry. This property is in fact the gauge symmetry of general relativity. It implies that we can choose any vector field $t^{\mu}$ as long as it is time-like. Such a vector field can be projected onto the three-manifold to obtain the {\em shift vector} $N^a = t_{\parallel}$ which is the part tangent to the surface, while the component of $t^{\mu}$ normal to the three-manifold is then identified as the ``distance between hypersufaces'' and is called the {\em lapse function} $N=t_\perp$. Therefore $t^{\mu}$ can be written as
\begin{equation}\label{eqn:time-evolution}
t^{\mu} = N n^{\mu} + N^{\mu}
\end{equation}
where, though we have written the shift as a four-vector to keep our choice of indices consistent, it is understood that $N^0 = 0$ in a local basis of coordinates adapted to the splitting.

Now we can determine the components of the four-metric in a basis adapated to the splitting as follows:
\begin{align}
  g_{00} = & g_{\mu\nu} t^\mu t^\nu \nonumber \\
	 = & g_{\mu\nu} \left(N n^\mu + N^\mu \right)\left(N n^\nu + N^\nu \right) \nonumber \\
	 = & N^2 n^\mu n_\mu + N^\mu N_\mu + 2 N (N^\mu n_\mu) \nonumber \\
	 = & -N^2 + N^\mu N_\mu
\end{align}
where we have used $n^\mu n_\mu = -1$ and $N^\mu n_\mu = 0$ in the third line. Working in a coordinate basis where $N^\mu = (0,N^a)$, we have $g_{00} = -N^2 + N^a N_a$\footnote{From this expression we can also see that \ignore{the norm $\abs{{\cal N}} = -N^2 + N^a N_a$ of the ``embedding vector field'' ${\cal N}^\mu = (N,N^x,N^y,N^z)$} $g_{00} = -N^2 + N^a N_a$ is a measure of the \emph{local} speed of time evolution and hence is a measure of the \emph{local} gravitational energy density.}. Similarly to obtain the other components of the metric we project along the time-space and the space-space directions:
\begin{equation}
  g_{\mu\nu} t^\mu N^\nu = N^\mu N_\mu \equiv N^a N_a
\end{equation}
Since, by definition $g_{0\nu} \equiv g_{\mu\nu} t^\mu$, this implies that $g_{0a} = N_a$. The space-space components of $g_{\mu\nu}$ are simply given by selecting values of the indices $\mu,\,\nu \in \{1,2,3\}$. Thus the full metric $g_{\mu\nu}$ can be written schematically as
\beq\label{eqn:spatial_metric}
	g_{\mu\nu} = \left( \begin{array}{cc}
						-N^2 + N^a N_a & \vect{N} \\
						\vect{N}^T & g_{ab}
						\end{array}
				 \right)
\eeq
where $a,\,b \in \{1,2,3\}$ and $\vect{N}\equiv \{N^a\}$. The 4D line-element can then be read off from the above expression
\begin{equation}\label{eqn:line-element}
 ds^2 = g_{\mu\nu} dx^\mu dx^\nu = (-N(t)^2 + N^a N_a) dt^2 + 2 N^a dt \, dx_a + g_{ab} dx^a dx^b
\end{equation}
where again $a,\,b \in \{1,2,3\}$ are spatial indices on $\Sigma_t$.
\begin{figure}[htbp]
	\begin{center}
	\subfloat{
      		\includegraphics[height=40mm]{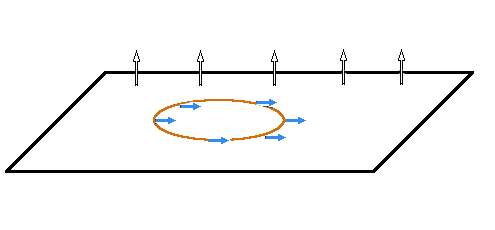}\label{fig:intrinsic-curv}
	}
	\subfloat{
		\includegraphics[height=50mm]{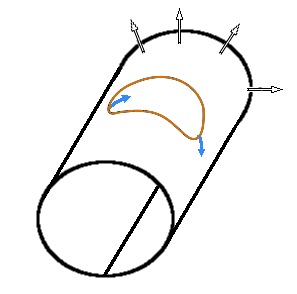}\label{fig:extrinsic-curv}
	}
	\caption{Intrinsic curvature measured by parallel transport (\emph{left}), and extrinsic curvature measured by changes in the normal vectors (\emph{right}).}
    \label{fig:curvature}
    \end{center}
\end{figure}

The components $g_{ab}$ of the metric restricted to a leaf of foliation are not the same as the intrinsic metric in a leaf of foliation. The intrinsic metric is related to the projection operator that takes any object $T_{\mu,\ldots,\nu}$ defined in the full four-dimensional manifold and projects out its component normal to the leaf $\Sigma_t$. To understand how to decompose $T_{\mu,\ldots,\nu}$ into a part $T_\parallel$, which lies only in the hypersurface $\Sigma_t$ and a part $T_\perp$, orthogonal to $\Sigma_t$, we may consider a vector $v^{\mu}$. The orthogonal component is given by $v_\perp = v^{\mu} n_{\mu}$. Similarly the component lying in $\Sigma_t$ is obtained by projecting the vector along the direction of the shift, so $v_\parallel = v^\mu N_\mu$. Writing a general four-vector as $v^{\mu} = v_\perp n^\mu + v_\parallel \frac{N^\mu}{\abs{N}}$ (where $\abs{N} = N^{\mu} N_{\mu}$ is the norm of the shift vector) and acting on it with $g_{\mu\nu} + n_\mu n_\nu$ we have
\begin{equation}
(g_{\mu\nu} + n_\mu n_\nu) \left(v_\perp n^\nu + v_\parallel \frac{N^\nu}{\abs{N}}\right) = v_\perp n_\mu (1 + n^\nu n_\nu) + \frac{v_\parallel}{\abs{N}}(N_\mu + n^\nu N_\nu) \nonumber \\
= v_\parallel \frac{N_\mu}{\abs{N}}
\end{equation}
Since $n^\mu n_\mu = -1$, and $n^\nu N_\nu=0$ by definition, we are left with only the component of $v^\mu$ parallel to $\Sigma_t$. We see that $h_{\mu\nu} = g_{\mu\nu} + n_\mu n_\nu$ is the required projection operator. This tensor also happens to correspond to the intrinsic three-metric on $\Sigma_t$, \emph{induced by its embedding in $\mc{M}$}:
\begin{equation}\label{eqn:intrinsic_curv}
	h_{ab} = g_{ab} + n_a n_b
\end{equation}
where as above $a,\,b \in \{1,2,3\}$. The reader might wonder how a rank 3 tensor $h_{ab}$ can be written in terms of a rank 4 object $g_{\mu\nu}$. To understand this, note that the spatial metric can also be written as a rank 4 tensor:
\[ h_{\mu\nu} = g_{\mu\nu} + n_\mu n_\nu \]
However, by construction, the time-time ($h_{tt}$) \& space-time ($h_{tx}, h_{ty}, h_{tz}$) components vanish and we are left with a rank 3 object. There is no contradiction in writing the spatial metric with either spatial indices ($a,b,\ldots$) or with spacetime indices ($\mu,\nu,\ldots$) as its contraction with another object is non-zero if and only if that object has a purely spatial character.

We have already seen how the Einstein-Hilbert action can be written in terms of the metric $g_{\mu\nu}$ and its derivatives. It makes sense, therefore, that in the case of General Relativity, where we have foliated the spacetime into spacelike hypersurfaces, we should take the intrinsic metric on $\Sigma$ (from now on we drop the $t$ superscript as we will deal with only one, representative, leaf of the foliation) as our configuration or ``position'' variable. To find the relevant hamiltonian density we proceed in a manner that parallels the approach in classical mechanics or field theory - namely we perform a Legendre transform to obtain the Hamiltonian function from the Lagrangian. In the case of classical mechanics, given a Lagrangian $L$ dependent on some coordinates $q$, we see that
\begin{equation}
H[p,\,q] = p\dot{q} - {L}[q,\,\dot{q}] \hspace{5mm} \mathrm{where} \hspace{5mm} p = \frac{\del L}{\del \dot{q}}\, ,
\label{eqn:hamiltonian_defn}
\end{equation}
where $p$ is the generalised momentum conjugate to $q$. Similarly, in the case of scalar field theory, we find that
\beq
H[\pi,\,\phi] = \int d^4x\,\pi\dot{\phi} - {L}[\phi,\,\dot{\phi}].
\eeq
In the case of GR we find that 
\beq
H[\pi^{\mu\nu},\,h_{\mu\nu}] = \int d^3x\,\pi^{ab}\dot{h}_{ab} - L[h_{ab},\,\dot{h}_{ab}]
\label{eqn:GRhamilt}
\eeq


In addition to the intrinsic metric $h_{ab}$, the hypersurfaces ${}^3 \Sigma$ \footnote{The notation ${}^3\Sigma$ is sometimes used to denote that these are three-dimensional hypersurfaces, however this is redundant in our present discussion.} also have a tensor which describes their embedding in $\mc{M}$, as shown in Fig.~\ref{fig:curvature}. This object is known as the extrinsic curvature, and is measured by taking the spatial projection of the gradient of the normal vectors to the hypersurface:
\beq
\label{eqn:extrinsic_curvature}
k_{ab} = h_a{}^c h_b{}^d \nabla_c n_d \equiv D_a n_b
\eeq
where $D_a$ is now the covariant derivative operator which acts only on purely spatial objects. As shown in Appendix~\ref{app-subsec:spatial-deriv}, that as is true in the case of the intrinsic metric, contracting the extrinsic curvature with any object with a time-like component gives zero: $k_{\mu\nu} n^\mu = 0$, implying that the extrinsic curvature is a quantity with only spacelike indices: $k_{ab}$. Moreover $k_{ab} = k_{[ab]}$ is a symmetric object by virtue of its construction \ref{eqn:extrinsic_curvature} (Appendix~\ref{app-subsec:extrinsic-curv}).

Due to the properties of the Lie derivative and the purely spatial character of the extrinsic curvature one can show (see Appendix \ref{app-sec:lie-derivative})that $ k_{ab} = {\pounds}_n h_{ab} $, \ie the \emph{ extrinsic curvature is the Lie derivative of the intrinsic metric w.r.t. the unit normal vector field $n^a$}. Now the Lie derivative $\pounds_{\vec{v}}X$ of an object $X$ w.r.t. a vector field $v^a$ can be interpreted as the \emph{rate of change of $X$ along the integral curves generated by $v^a$}. By analogy with the definition of $p$ in eq.~(\ref{eqn:hamiltonian_defn}) we might be tempted to identify the extrinsic curvature with the ``momentum variable'' conjugate to the ``position variable'' (namely the intrinsic metric). This is not far off the mark. As we will see the conjugate momentum will, indeed, turn out to be a function of $k_{ab}$.

The Einstein-Hilbert action can be re-written in terms of quantities defined on the spatial hypersurfaces, by making two substitutions. Firstly, and analogously to $g$, we write $h$ for the determinant of $h^{ab}$ and recognise that the four-dimensional volume form $\sqrt{-g}$ is equal to $N \sqrt{h}$ (that is, the three-dimensional volume form multiplied by the distance between hypersurfaces). Secondly, using the Gauss-Codazzi equation\footnote{A derivation of which can be found in Appendix 1.3 of \cite{Dona2010Introductory}},
\beq
  {}^{(3)}R^\mu_{\nu\rho\sigma} = h^\mu_\alpha h^\beta_\nu h^\gamma_\rho h^\delta_\sigma R^\alpha_{\beta\gamma\delta} - k_{\nu\sigma}k^\mu_\rho - k_{\nu\rho} k^\mu_\sigma
\eeq 
the four-dimensional Ricci curvature scalar $R$ can be re-written in terms of the three-dimensional Ricci scalar ${}^{(3)}R$ (that is, the Ricci scalar restricted to a hypersurface $\Sigma$), and the extrinsic curvature of $ \Sigma $ as
\beq
	R = {}^{(3)}R + k^{ab}k_{ab} - k^2
	\label{eqn:gauss-codazzi}
\eeq
where $ k$ is the trace of the extrinsic curvature taken with respect to the 3-metric
\beq
k := k^{ab}h_{ab}\,.
\eeq
The Gauss-Codazzi relation is a very general result which is true in an arbitrary number of dimensions. The reader with time on their hands may wish to derive it for themselves by using the definition of the Ricci scalar in terms of the Christoffel connection and using the 3-metric $ h^\mu_\nu$ to project quantities in $3+1$ dimensions down to the three dimensions of $ \Sigma $. By repeating this process with objects living in $n$ \& $n+1$ dimensions, one can obtain the version which applies for manifolds of any dimensionality $n$.

Using these substitutions, the Einstein-Hilbert action can be rewritten in a form that is convenient for identifying the parts which depend only on $\Sigma$
\beq
S_\mathrm{EH} = \int dt\,d^3x\,N\sqrt{h}\left({}^{(3)}R + k^{ab}k_{ab} - k^2\right) = \int dt\,{L_\mathrm{EH}}
\eeq
We next need to find $\dot{h}_{ab}$, which is obtained by taking the Lie derivative (Appendix \ref{app-sec:lie-derivative}) with respect to the vector field $t^\mu$ which generates time-translations (for a detailed derivation, see Appendix \ref{app-subsec:canonical-momentum} ):
\beq
	\dot{h}_{ab} = \pounds_{\vec{t}} h_{ab} = 2 N k_{ab} + \pounds_{\vec{N}} h_{ab}
    \label{eqn:dot_h_ab}
\eeq
The conjugate momentum is then found to be
\beq
\label{eqn:canonical-momentum}
	\pi^{ab} = \frac{\delta  L}{\delta \dot{h}_{ab}} = \sqrt{h}(k^{ab} - k \, h^{ab})
\eeq
Substituting these results into eq.~(\ref{eqn:GRhamilt}) we obtain
\begin{subequations}
\begin{eqnarray}
 H[\pi^{ab},\,h_{ab}] & = & \int d^3x\,\pi^{ab}\dot{h}_{ab} - L[h_{ab},\,\dot{h}_{ab}] \\
                                      & = & \int d^3x\,N\left(-\sqrt{h}{}^{(3)}R + \frac{1}{\sqrt{h}}(\pi^{ab}\pi_{ab}-\frac{1}{2}\pi^2) \right)- 2N_a D_b \pi^{ab} \\
                                      & = & \int d^3x\,N{\cal H} - N_a {\cal C}^a \label{eqn:HamiltConstraintForm}
\end{eqnarray}
\end{subequations}
where for brevity we have adopted the notation
\begin{subequations}
\begin{eqnarray}
\label{eqn:metric-constraints}
 & {\cal H} = \left(-\sqrt{h}{}^{(3)}R + \frac{1}{\sqrt{h}}(\pi^{ab}\pi_{ab}-\frac{1}{2}\pi^2) \right) & \textrm{(Hamiltonian constraint)} \label{eqn:metric-hamiltonian-constraint} \\
 & {\cal C}^a = 2 D_b \pi^{ab} & \textrm{(Diffeomorphism constraint)} \label{eqn:metric-diffeo-constraint}
\end{eqnarray}
\end{subequations}
where $\pi$ is the trace of $\pi^{ab}$, and $D$ is the covariant derivative with respect to the 3-metric $h_{ab}$.

We can reverse the Legendre transform to rewrite the action for GR as
\begin{subequations}
\begin{eqnarray}
	S_{EH} = \int dt L_{EH} & = & \int dt d^3 x \left( \pi^{ab} \dot h_{ab} - H[\pi^{ab},h_{ab}] \right)\\
	& = & \int dt d^3 x \left( \pi^{ab} \dot h_{ab} - N{\cal H} + N_a {\cal C}^a \right)
\end{eqnarray}
\end{subequations}

It is now apparent that the action written in this form is a function of the lapse and shift but \emph{not} their time derivatives. Consequently the Euler-Lagrange equations of motion obtained by varying $S_{EH}$ with respect to the lapse and shift are
\begin{subequations}
\begin{eqnarray}
 	\frac{\delta S_{EH}}{\delta N} & = & - \mc{H} = 0 \\
	\frac{\delta S_{EH}}{\delta N_a} & = & \mc{C}^{a} = 0
\end{eqnarray}
\end{subequations}
implying that $\mc{H}$ and $\mc{C}^a$ are identically zero and are thus to be interpreted as constraints on the phase space! This is nothing more than the usual prescription of Lagrange multipliers - when an action depends only on a configuration variable $q$ but not on the corresponding momentum $p$, the terms multiplying the configuration variable are constraints on the phase space.

$\mc{C}^a$ and $\mc{H}$ are referred to as the vector (or diffeomorphism) constraint and the scalar (or Hamiltonian) constraint, respectively. The diffeomorphism constraint generates diffeomorphisms within the spatial hypersurfaces $\Sigma_t$. The Hamiltonian constraint generates the time evolution which takes the geometry of $\Sigma_t$ to $\Sigma_{t+1}$. A little later, when we cast GR in the first order formulation we will encounter a third constraint, referred to as the Gauss constraint. We shall discuss the interpretation of the constraints once the Gauss constraint has been properly introduced, but note here that the Hamiltonian constraint is relevant to the time evolution of the spacelike hypersurfaces, while other two constraints act spatially (\ie within the hypersurfaces).

We see that the Hamiltonian density $H_{EH}$ in eq.~(\ref{eqn:HamiltConstraintForm}), obtained after performing the $3+1$ split of the Einstein-Hilbert action via the ADM procedure \cite{Romano1993Geometrodynamics}, is a sum of constraints, \ie $H_{EH} = N {\mc H} -N_a \mc{C}^a = 0$. This is a generic feature of diffeomorphism invariant theories.





%

\subsection{Physical Interpretation of Constraints}\label{subsec:poisson-brackets}

Here we briefly describe the form of the Poisson brackets between the various constraints and their physical interpretation\footnote{For what follows, it will be helpful to recall some aspects of symplectic geometry. In the symplectic formulation of classical mechanics a system consists of a phase space in the form of an even-dimensional manifold $\Gamma$ equipped with a symplectic structure (anti-symmetric tensor) $\Omega_{\mu\nu}$. Given any function $f:\Gamma \rarrow \mbb{R}$ on the phase space, and a derivative operator $\nabla$, there exists a vector field associated with $f$, given by $X_f^\alpha = \Omega^{\alpha \beta} \nabla_\beta f$. Given two functions $f,g$ on $\Gamma$, the Poisson brackets between the two can be written as $\{f,g\} = \Omega^{\alpha\beta}\nabla_\alpha f \nabla_\beta g $ which can also be identified with $-\mc{L}_{X_f} g = \mc{L}_{X_g} f$ - the Lie derivative of $g$ along the vector field generated by $f$ or vice-versa. Thus in this picture, the Poisson bracket between two functions tells us the change in one function when it is Lie-dragged along the vector field generated by the other function (or vice-versa). For more details see \cite[Appendix B]{Ashtekar1991Lectures}}. The Poisson brackets between two functions $f$ and $g$ defined on the phase space is given by
\begin{equation}\label{eqn:poisson-brackets}
	\{ f, g \} = \int d^3 x \frac{\delta f}{\delta h_{ab}} \, \frac{\delta g}{\delta \pi^{ab}} - \frac{\delta f}{\delta \pi^{ab}} \, \frac{\delta g}{\delta h_{ab}} 
\end{equation}
where $h_{ab}, \pi^{ab}$ are the canonical coordinates and momenta respectively. Since these variables are fields defined over the three-dimensional manifold $\Sigma$, it is necessary to integrate over $\Sigma$ to obtain a number. Since the diffeomorphism constraint $\mc{C}^a = 2 D_{b} \pi^{ab}$ is a function of momenta only, the Poisson bracket of this constraint with the canonical coordinate is given by
\begin{align}
	\{ h_{cd}(x'), \xi_a \mc{C}^a (x'') \} & = - \int d^3 x \frac{\delta h_{cd}(x')}{\delta h_{ef}(x)}\, \frac{\delta \left[ 2 \xi_a D_{b} \pi^{ab}(x'') \right]}{\delta \pi^{ef}(x)} \nonumber \\
	& = - \int d^3 x \, 2 \, \delta_c^e \, \delta_d^f \, \delta^a_e \, \delta^b_f \, \delta (x-x') \, \delta (x'' - x) \, D_{b} \xi_a (x'') \nonumber \\
	& = - \delta(x' - x'') \, 2 \, D_d \xi_c 
\end{align}
where $\xi_a$ is a vector field defined on $\Sigma$, which serves to ``smear out'' the constraint $\mc{C}^a$ over the manifold so that we get a function defined over the entire phase space, rather than just being defined at each point of $\Sigma$. To go from the first line to the second we have integrated by parts and dropped the term which is a pure divergence. This is justified if the field $\xi_a$ has support only on a compact subset of $\Sigma$. The constraint $\mc{C}^a$ takes the metric $h_{ab}$ to a neighboring point on the phase space: $ h_{ab} \rarrow h_{ab} - 2 D_b \xi_a $. Using the properties of the Lie derivative, the second term can also be written as $ \mc{L}_{\xi} h_{ab} = D_{a} \xi_b + D_{b} \xi_a $ implying that $ h_{ab} \rarrow h_{ab} - \mc{L}_{\xi} h_{ab} $, and that therefore \ie $\xi_a \mc{C}^a $ is the generator of spatial diffeomorphisms along the vector field $\xi_a$ on the spatial manifold $\Sigma$. This is the reason for calling it the ``diffeomorphism constraint'' in the first place.

Similarly a much more involved calculation along the lines of the one above yields for the Poisson bracket between a function $f$ on the phase space and the ``Hamiltonian constraint'' $\mc{H}$ \cite[Sec. I.1.1]{Thiemann2001Introduction}

\begin{equation}
		\{ N \mc{H}, f \} = \mc{L}_{N \vec{n}} f
\end{equation}

\ie $\mc{H}$ generates diffeomorphisms along the vector field $N \vec{n}$ orthogonal to the hypersurface $\Sigma$. In other words $\mc{H}$ maps functions defined on the hypersurface $\Sigma_t$ at a given time $t$ to functions on a hypersurface $\Sigma_{t'}$ at a later time $t'$. This is the reason for referring to $\mc{H}$ as the ``Hamiltonian constraint''; because it generates time evolution of functions on the phase space, the same way the Hamiltonian in classical or quantum mechanics does.

We do not wish to provide more details of the ADM procedure than are strictly necessary. Further details about the ADM splitting and canonical quantization can be found in \cite{Wald1984General} in the metric formulation, and \cite{Romano1993Geometrodynamics} in the connection formulation\footnote{The terms ``metric formulation'' and ``connection formulation''will be defined in sec.~\ref{sec:GR_in_connect_vars}}.

\subsection{Seeking a path to canonical quantum gravity}
\label{sec:can_quant_gravity}
In the Hamiltonian formulation one works with a phase space spanned by a set of generalized coordinates $\mathbf{q}_i$, and a set of generalized momenta $\mathbf{p}_i$. For the case of General Relativity, the generalised coordinate is the intrinsic metric $h_{ab}$ of the spatial 3-manifold $\Sigma$ and the extrinsic curvature $k_{ab}$ induced by its embedding in $\mc{M}$ determines the corresponding generalized momentum, as per (\ref{eqn:canonical-momentum}). For comparison the phase spaces of various classical systems are listed in the following table
\begin{center}\label{table:phase_spaces}
	\begin{tabular}{|l|c|c|}
		\hline \textbf{System} & \textbf{Coordinate} & \textbf{Momentum} \\
		\hline Simple Harmonic Oscillator & $x$ & $p$ \\ \hline
		Ideal Rotor & $\theta$ & $L_\theta$ \\ \hline
		Scalar Field & $\phi(x,t)$ & $\pi(x,t)$ \\ \hline
		Geometrodynamics & $h_{ab}$ & $\pi^{ab} = \sqrt{h}(k^{ab}k_{ab} - k^2)$ \\ \hline
		Connection- dynamics & $A_a{}^i$ & $E^a{}_i$ \\ \hline
	\end{tabular}
\end{center}
%
Now, given our phase space co-ordinatized by $\{h_{ab},\pi^{ab}\}$ and the explicit form of the Hamiltonian of GR in terms of the Hamiltonian eq.~(\ref{eqn:metric-hamiltonian-constraint}) and diffeomorphism eq.~(\ref{eqn:metric-diffeo-constraint}) constraints, we may expect that we can proceed directly to quantization by promoting the Poisson brackets on the classical phase space to commutation relations between the operators acting on a Hilbert space $\mathrm{H}_{GR}$:
\begin{subequations}
 \begin{eqnarray}
  \label{eqn:quantum-phase-space}
	h_{ab} \rightarrow \hat h_{ab} & \qquad & \pi^{ab} \rightarrow i\hbar\frac{\delta}{\delta h_{ab}} \\
	\left\{h_{ab}(x),\pi^{a'b'}(x')\right\} = \delta(x-x') \delta^{a'}{}_a \delta^{b'}{}_b & \rightarrow & \left[ \hat h_{ab} , i\hbar\frac{\delta}{\delta h_{ab}} \right] = i \hbar \delta^{a'}{}_a \delta^{b'}{}_b \\
	f[h_{ab}] & \rightarrow & \fullket{\Psi_{h_{ab}}}
 \end{eqnarray}
\end{subequations}
It should then remain to write the constraints $\mc{H}$ and $\mc{C}^\mu$ in operator form 
\begin{eqnarray}
\label{eqn:quantum-constraints}
\mc{H}, \mc{C}^a & \rightarrow & \hat{\mc{H}}, \hat{\mc{C}}^a 
\end{eqnarray}
which act upon states $\fullket{\Psi_q}$ which 
would then be identified with the \emph{physical} states of quantum gravity. The physical Hilbert space is a subset of the kinematic Hilbert space which consists of all functionals of the 3-metrics: $ |\Psi_{q'}\rangle \in \mathrm{H}_{phys} \subset \mathrm{H}_{kin}$.

Unfortunately the above prescription is only formal in nature and we run into severe difficulties when we try to implement this recipe. The primary obstacle is the fact that the Hamiltonian constraint stated in eq.~(\ref{eqn:metric-hamiltonian-constraint}) has a \emph{non-polynomial} dependence on the 3-metric via the Ricci curvature ${}^3 R$. We can see this schematically by noting that ${}^3 R$ is a function of the Christoffel connection $\Gamma$ which in turn is a complicated function of $h_{ab}$:
\beq
\label{eqn:non-polynomial}
{}^3 \mathrm{R} \sim (\partial \Gamma)^2 + (\Gamma)^2; \qquad \Gamma \sim q \partial q \Rightarrow \partial \Gamma \sim \partial q \partial q + q \partial^2 q
\eeq
This complicated form of the constraints raises questions about operator ordering and is also very non-trivial to quantize. Therefore, in this form, the constraints of general relativity are not amenable to quantization.

This is in contrast to the situation with the Maxwell and Yang-Mills fields, which being gauge fields can be quantized in terms of holonomies (see section \ref{subsec:holonomies}), which form a complete set of gauge-invariant variables. An optimist might believe that were we able to cast General Relativity as a theory of a gauge field, we could make considerably more progress towards quantization than in the metric formulation. This does turn out to be the case as we see in the following sections.

\subsection{Connection Formulation}
\label{sec:GR_in_connect_vars}
Our ultimate goal is to cast general relativity in the mould of gauge field theories such as Maxwell or Yang-Mills. The parallel between covariant derivaties and connections in GR and QFT suggests that gravity may be treated as a gauge field theory with $\Gamma^\rho_{\mu\nu}$ as the gauge connection. 
However, though the Christoffel connection is an affine connection it does not transform as a tensor under arbitrary coordinate transformations (c.f. \cite[chapter 4]{Wald1984General}) and thus cannot play the role of a gauge connection which should be a covariant quantity. 

$\Gamma^\rho_{\mu\nu}$ allows us to parallel transport vectors $v^\mu$ and, in general, arbitrary tensors (vectors are of course a special case of tensors) 
\ie it allows us to map the tangent space $T_p$ at point $p$ to the tangent space $T_{p'}$ at the point $p'$. The map depends on the path connecting $p$ and $p'$ and it is this fact that allows us to measure \emph{local} geometric properties of a manifold. However, in order to allow the parallel transport of spinors the Christoffel connection is not sufficient.

The Christoffel connection does not ``know'' about 
spinor fields of the form $\psi_\mu{}^I$ (where $I$ is a lie-algebra index). A theory of quantum gravity which does not know about fermions would not be very useful. Thus we need an alternative to the Christoffel connection which has both these properties: covariance with respect to coordinate transformations and coupling with spinors.


Up until now we have worked with GR in \emph{second-order} form, \ie with the metric $g_{\mu\nu}$ as the only configuration variable (hence this is also called the metric formulation). The Christoffel connection $\Gamma^\rho{}_{\mu\nu}$ is determined by the metric compatibility condition,
\beq
	\nabla g_{\mu\nu} = 0
	\label{eqn:metric-compat}
\eeq
The passage to the quantum theory is facilitated by switching to a \emph{first-order} formulation of GR (also called the connection formulation), in which \emph{both} the metric and the connection are treated as independent configuration variables. However due to the problems with the Christoffel connection noted above, we shall choose a first-order formulation in terms of a tetrad or ``frame-field'' (which we will see shortly takes the role of the metric) and a gauge connection (the ``spin connection''), both of which take values in the Lie algebra of the Lorentz group. 
In the following subsections we will describe the tetrads and the spin connection in some detail, before proceeding to our first example of a first-order formulation of gravity, the Palatini formulation. 

%

The connection formulation exposes a hidden symmetry of geometry as illustrated by the following analogy. The introduction of spinors in quantum mechanics (and the corresponding Dirac equation) allows us to express a scalar field $\phi(x)$ as the ``square'' of a spinor $ \phi = \Psi^i \Psi_i $. In a similar manner the use of the tetrads allows us to write the metric as a square $ g_{\mu\nu} = e_\mu^I e_\nu^J \eta_{IJ} $. The transition from the metric to connection variables in GR is analogous to the transition from the Klein-Gordon equation
\beq
  (-\partial_t^2 + \partial^i\partial_i - m^2)\psi = 0
\eeq
to the Dirac equation
\beq
  (i\gamma^\mu\partial_\mu -m)\psi = 0
\eeq
in field theory (where here we have used $c=\hbar=1$).

The connection is a Lie-algebra valued one-form $A_\mu{}^{IJ} \tau_{IJ}$ where $\tau_{IJ}$ are the generators of the Lorentz group. Our configuration space is then spanned by a tetrad and a connection pair: $ \{e_\mu^I, A^\mu_{IJ}\}$. The tetrads are naturally identified as 
mappings between the Lie algebra $ \sltwoc $, and the Lie algebra $\mathfrak{so}(3,1)$ of 4-vectors.


\subsubsection{Tetrads}
\label{subsubsec:tetrads}

We begin by considering the four dimensional manifold ${\mc M}$, introduced in section~\ref{sec:ADMsplit}, above. As we know, any sufficiently small region of a curved manifold will look flat\footnote{So long as the manifold is continuous, not discrete. This is an important point to keep in mind for later.} and so we may define a tangent space to any point $P$ in ${\mc M}$. Such a tangent space will be a flat Minkowski spacetime, and the point $P$ may be regarded as part of the worldline of an observer, without loss of generality. This tangent space will be spanned by four vectors, $e_\mu$. Each basis vector will have four components, $e^I_\mu$ where $I\in\{0,\,1,\,2,\,3\}$, referred to the locally-defined reference frame (the ``laboratory frame'' of the observer who's worldline passes through $P$, with lengths and angles measured using the Minkowski metric). As noted back in section~\ref{sec:intro}, such a set of four basis vectors is referred to as a tetrad or vierbein (German for ``four legs'')\footnote{The similar word \emph{vielbein} (``any legs'') is used for the generalisation of this concept to an arbitrary number of dimensions (e.g. triads, pentads).}. Since the tetrads live in Minkowski space, their dot product is taken using the Minkowski metric. But the dot product of basis vectors is just the metric itself, so the metric of ${\mc M}$ at any point is just given by
\beq
 g_{\mu\nu} = e_\mu^I e_\nu^J \eta_{IJ}
 \label{eqn:tetrad}
\eeq
where $\eta_{IJ} = \mathrm{diag}(-1,+1,+1,+1)$ is the Minkowski metric. Taking the determinant of both sides we find that
\begin{subequations}
\beq
\mathrm{det}(g_{\mu\nu}) = \mathrm{det(\eta_{IJ})}\mathrm{det}(e^I_\mu)^2 = -\mathrm{det}(e^I_\mu)^2
\eeq
\beq
\therefore e = \sqrt{-g}
\eeq
\end{subequations}
where $g \equiv \mathrm{det}(g_{\mu\nu})$ and $e \equiv \mathrm{det}(e^I_\mu)$. Due to this fact the tetrad can be thought of as the ``square-root'' of the metric.

Tetrads can thus be interpreted as the transformation matrices that map between two sets of coordinates, as can be seen by comparing eq.~(\ref{eqn:tetrad}) with the standard form for a coordinate transformation, eq.~(\ref{eqn:coord_transf}). 
It is this fact which makes the tetrads a useful tool in modern formulations of GR. Since the components of spinors are defined relative to the flat ``laboratory frame'' of the tangent space, and tetrads map the metric of this tangent space to the metric of the full four-dimensional spacetime, they serve the role we mentioned above, of allowing us to construct a connection that knows about spinor quantities as well as vectors and tensors. The construction of such a connection will be described in the following subsection. 

As an aside, we note that any vector $v^\mu$ can be written as an $\sltwoc$ spinor $v_{ab}$ as
\beq
v_{ab} := v_\mu e^\mu{}_I \sigma^I{}_{ab}
\label{eqn:v_from_e_and_sigma}
\eeq
where $\sigma^I = \{ \mb{1}, \sigma^x, \sigma^y, \sigma^z \}$ is a basis of the lie-algebra $\sltwoc$ and $a,b$ are the spinorial matrix indices shown explicitly for clarity.


\subsubsection{Spin Connection}
\label{subsubsec:connection}
It is a truth universally acknowledged, that a student in possession of a basic familiarity with Loop Quantum Gravity will be in want of an explanation of the significance of $SL(2,\mathbb{C})$. If we wish to construct a theory that encompasses GR under the framework of gauge field theories we should anticipate that the local symmetries of spacetime will define the gauge group of our quantum gravity theory. As noted in sec.~\ref{sec:gen_rel} the causal structure of spacetime defines a future light-cone and past light-cone at each event. The past light-cone of an observer at any given value of time is the celestial sphere at a fixed distance from the observer. The celestial sphere can be parametrised by the angles $\theta$, $\phi$, and any point on a sphere can be stereographically projected onto a plane. For our purposes, this shall be taken to be the complex plane, so that any point on the celestial sphere corresponds with a complex number $\zeta = X + iY$. We can write this as the ratio of two complex numbers $\zeta = \alpha/\beta$, which can (if we so desire) be written as functions of $\theta$, $\phi$. A change of the complex coordinates (which is equivalent to a coordinate transformation of the real angles $\theta$, $\phi$) can
be effected by acting on the 2-vector with components $\alpha$, $\beta$ with a linear transformation, written in the form of a $2\times 2$ matrix with complex components. If we take the determinant of this matrix to be +1 (which we can do, without loss of generality) this is an $SL(2,\mathbb{C})$ transformation. Thus the Lorentz group, $SL(2,\mathbb{C})$, is the local gauge group of special relativity.

While dynamics on a flat spacetime can be described by the Poincare group, in a general curved spacetime such as we would expect in GR, translational symmetry is broken and only local Lorentz invariance remains as an unbroken symmetry. The mapping between local coordinate bases is encoded in the connection. As noted above, the Christoffel connection does not allow for the parallel transport of spinors. It is therefore not suitable to be used in constructing a theory of quantum gravity. The simplest candidate that allows for parallel transport of spinors is an $\sltwoc$ valued connection $A_\mu{}^{IJ}$. Such a choice of connection is a logical candidate for casting GR as a gauge theory, and will be referred to as a \emph{spin connection}.  

In order to be able to parallel transport objects with spinorial indices we need a suitable extension of the notion of a covariant derivative which acts on vectors to one which acts on spinors (we follow \cite[Appendix B]{Peldan1993Actions}). The condition for parallel transport of a vector is that its covariant derivative with respect to the Christoffel connection should vanish, \ie
\beq
\nabla_k v^i = \del_k v^i + v^j\Gamma^i_{jk} = 0
\label{eqn:gr_paral_tr}
\eeq
Similarly the condition for parallel transport of a spinor requires that its covariant derivative with respect to the gauge connection should vanish
\beq
D_\mu \psi = \del_\mu \psi + igA_\mu \psi = 0
\label{eqn:paralleltransport}
\eeq
where $A_\mu \equiv A_\mu^I t^I\,$ is the gauge connection. Analogously, given the tetrad $e^I_\mu$ and the Christoffel connection $\Gamma^\gamma_{\alpha\beta}$ we define an $\sltwoc$ valued \emph{spin connection} $\omega_{\alpha}^{IJ} $ and use these to construct the \textit{generalised} derivative operator on $\mc{M}$ which annihilates the tetrad
\begin{equation}\label{eqn:spin-connection}
\mc{D}_\alpha e_\beta^I = \partial_\alpha e_\beta^I - \Gamma^\gamma_
{\alpha\beta} e_\gamma^I + \omega_{\alpha}^{I}{}_J e_\beta^J = 0
\end{equation}

The term ``spin connection'' may cause some confusion, by tricking newcomers into thinking they have to learn a new concept, when it fact this is nothing more than the notion of parallel transport of a particle along a Wilson line.


Now one would expect that this derivative operator should also annihilate the (internal) Minkowski metric $\eta_{IJ} = e_{\alpha I} e^\alpha_I$ and the spacetime metric $g_{\mu\nu} = e_\mu^I e_\nu^J \eta_{IJ}$. One can check that requiring this to be the case yields that the spin-connection is anti-symmetric $\omega_\alpha^{\{IJ\}} = 0$ and the Christoffel connection is symmetric $\Gamma^\alpha_{[\beta\gamma]} = 0$.

We can solve for $\Gamma^\alpha_{\beta\gamma}$ in the usual manner (see for e.g. \cite{Wald1984General}) to obtain
\begin{equation}
\Gamma^\gamma_{\alpha\beta} = \frac{1}{2} g^{\gamma\delta}\left( \partial_\alpha g_{\delta\beta} + \partial_\beta g_{\delta\alpha} - \partial_\delta g_{\alpha \beta} \right)
\end{equation}
Inserting the above into eq.~(\ref{eqn:spin-connection}) we can solve for $\omega$ to obtain
\begin{equation}\label{eqn:default-spin-connection}
\omega_{\alpha}^{IJ} = \frac{1}{2} e^{\delta [I} \left( \partial_{[\alpha} e_{\delta]}^{J]} + e^{|\beta|J]} e_\alpha^K \partial_\beta e_{\delta K} \right)
\end{equation}
Note that in the above expression the Christoffel connection does not occur.

In the definition of $\mc{D}$ we have included the Christoffel connection. Ideally, in a gauge theory of gravity, we would not want any dependence on the spacetime connection. That this is the case can be seen by noting that all derivatives that appear in the Lagrangian or in expressions for physical observables are \textit{exterior} derivatives, \ie of the form $\mc{D}_{[\alpha}e_{\beta]}^I$. The anti-symmetrization in the spacetime indices and the symmetry of the Christoffel connection $\Gamma^\gamma_{[\alpha\beta]} = 0$ implies that the exterior derivative of the tetrad can be written without any reference to $\Gamma$:
\begin{equation}\label{eqn:exterior-derivative}
\mc{D}_{[\alpha}e_{\beta]}^I = \partial_{[\alpha} e_{\beta]}^I + \omega_{[\alpha}{}^{IL}e_{\beta]L} = 0
\end{equation}
We can solve for $\omega$ by a trick similar to one used in solving for the Christoffel connection. Following \cite[Appendix B]{Peldan1993Actions}, first contract the above expression with $e^\alpha_J e^\beta_K$ to obtain
\begin{equation}
	e^\alpha_J e^\beta_K  \left(\partial_{[\alpha} e_{\beta]}^I + \omega_{[\alpha}{}^{IL}e_{\beta]L} \right) = 0
\end{equation}
Now let us define $\Omega_{IJK} = e^\alpha_I e^\beta_J \partial_{[\alpha} e_{\beta]K} $. Performing a cyclic permutation of the indices $I, J, K$ in the above expression, adding the first two terms thus obtained and subtracting the third term we are left with
\begin{equation}
	\Omega_{JKI} + \Omega_{IJK} - \Omega_{KIJ} + 2 e^\alpha_J \omega_{\alpha I K} = 0
\end{equation}
This can be solved for $\omega$ to yield
\begin{equation}
	\omega_{\alpha IJ} = \frac{1}{2} e^K_\alpha [\Omega_{KIJ} + \Omega_{JKI} - \Omega_{IJK}]
\end{equation}
which is equivalent to the previous expression, eq.~(\ref{eqn:default-spin-connection}), for $\omega$.

Next we consider the curvature tensors for the Christoffel and the spin connections and show the fundamental identity that allows us to write the Einstein-Hilbert action solely in terms of the tetrad and the spin-connection. The Riemann tensor for the spacetime and the spin connections, respectively are defined as:
\begin{equation}
 \mc{D}_{[\alpha} \mc{D}_{\beta]} v_\gamma = R_{\alpha \beta \gamma}{}^\delta v_\delta; \qquad
 \mc{D}_{[\alpha} \mc{D}_{\beta]} v_I = R_{\alpha \beta I}{}^J v_J
\end{equation}
Writing $v_\gamma = e_\gamma^I v_I$ and inserting into the first expression we obtain
\begin{equation}
 R_{\alpha \beta \gamma}{}^\delta v_\delta = \mc{D}_{[\alpha} \mc{D}_{\beta]} v_\gamma = \mc{D}_{[\alpha} \mc{D}_{\beta]} e_\gamma^I v_I = e_\gamma^I R_{\alpha \beta I}{}^J v_J = e_\gamma^I R_{\alpha \beta I}{}^J e^\delta_J v_\delta
\end{equation}
where we have used the fact that $\mc{D}_\mu e_\nu^I = 0$. Since the above is true for all $v_\delta$, we obtain
\begin{equation}\label{eqn:curvature-identity}
 R_{\alpha \beta \gamma}{}^\delta = R_{\alpha \beta I}{}^J e_\gamma^I e^\delta_J
\end{equation}
The Ricci scalar is given by $R = g^{\mu\nu} R_{\mu\nu} = g^{\mu\nu} R_{\mu\delta\nu}{}^\delta$. Using the previous expression we find
\begin{equation}
 R_{\mu\delta\nu}{}^\delta = R_{\mu\delta I}{}^J e^I_\nu e^\delta_J
\end{equation}
Contracting over the remaining two spacetime indices then allows us to write the Ricci scalar in terms of the curvature of the spin-connection and the tetrads,
\begin{equation}\label{eqn:ricci-tetrad}
 R = R_{\mu\nu}{}^{IJ} e^\mu_I e^\nu_J\,.
\end{equation}

\subsubsection{Palatini Action}\label{subsec:palatini_action}
 
 The Einstein-Hilbert action, from the discussion in sec.~\ref{subsec:eh-action}, can be written in the form
\beq
 S_{EH} = \frac{1}{\kappa}\int d^4 x \sqrt{-g} g^{\mu\nu}R_{\mu\nu} \,.
 \label{eqn:action_EH_gmunu_explicit}
\eeq 
The Palatini approach to GR starts with this action and treats the metric and the connection as independent dynamical variables. Variation of the action with respect to the metric yields the vacuum field equations ($R_{\mu\nu}=0$), while variation with respect to the connection implies that the connection is the Christoffel connection. Discussion of the Palatini approach in terms of the metric and connection can be found in many textbooks (e.g. \cite{DInverno}, chapter 11). 

Having gone to the effort of defining tetrads and the spin connection we now wish to write the action for GR in terms of these variables. We saw in sec.~\ref{subsec:eh-action} that requirements of covariance and simplicity dictated the form of the action for GR. Similarly our construction of an action based on tetrads and the spin connection is guided by physical considerations. Firstly we want the action to be diffeomorphism invariant. We also require the lagrangian density to be a four-form, which we can integrate over a four-dimensional spacetime to give a scalar (thus this action is valid only in four dimensions). The curvature of the connection is already a two-form, so (suppressing spacetime indices for simplicity) we include $ e^I \wedge e^J \equiv e_{[\mu}{}^I e_{\nu]}{}^J $, which is a two-form\footnote{If we
use two copies of the curvature tensor then we get Yang-Mills theory ($F \wedge F$). But that doesn't include the tetrad.}. This yields the Palatini action, the simplest diffeomorphism-invariant action one can construct using tetrads and the
curvature of the gauge connection. 
We emphasise that this is not simply $S_{EH}$ rewritten with a change of variables, but a parallel construction. The discussion above is intended to describe the physical intuition behind this construction. It is conventional to 
{{use}}
the notation $F_{\mu\nu}^{IJ}$ for the curvature of the spin connection, to yield 
\begin{align}\label{eqn:connection-action}
	S_P\left[e,\omega\right] & = \frac{1}{2\kappa}\int d^4 x\, \star (e^I \wedge e^J) \wedge F^{KL} \, \epsilon_{IJKL} \nonumber \\
	& =  \frac{1}{4\kappa} \int d^4 x\, \epsilon^{\mu\nu\alpha\beta} \epsilon_{IJKL} \, e_{\mu}{}^I e_{\nu}{}^J F_{\alpha\beta}{}^{KL}\,,
\end{align}
where 
\begin{equation}\label{eqn:gauge-curvature}
	F^{KL}{}_{\gamma\delta} = \partial_{[\gamma}\omega_{\delta]}{}^{KL} + \frac{1}{2} \left[ \omega_{\gamma}{}^{KM}, \omega_{\delta\,M}{}^L \right]\,.
\end{equation}
\ignore{and $\tilde{e}^I{}_\mu = \sqrt{e}\, e^I{}_\mu$ is a \emph{densitized} tetrad. We have suppressed space-time indices for simplicity but note that: $ e^I \wedge e^J \equiv e_{[\mu}{}^I e_{\nu]}{}^J $, which is a two-form.} 

The similarity between eqs.~(\ref{eqn:action_EH_gmunu_explicit}) and (\ref{eqn:connection-action}) should be clear, especially when we remember that $g_{\mu\nu} = e_\mu^I e_\nu^J \eta_{IJ}$ (eq.~(\ref{eqn:tetrad})). At this point $F_{\mu\nu}{}^{IJ}$ is the curvature of $\omega$, but it remains to be shown that it satisfies the identity of eq.~(\ref{eqn:ricci-tetrad}). The equations of motion obtained by varying the Palatini action are
\begin{subequations}
\begin{eqnarray}\label{eqn:palatini_eom}
 \frac{\delta S_P}{\delta \omega_\nu{}^{IJ}} & = & \epsilon^{\mu\nu\alpha\beta} \epsilon_{IJKL} \, D_{\nu} \left( e_{\alpha}{}^I e_{\beta}{}^J \right)  = 0 \,, \label{eqn:connection_eom}\\
 \frac{\delta S}{\delta e_I{}^\mu} & = & \epsilon^{\mu\nu\alpha\beta} \epsilon_{IJKL}\, e_{\nu}{}^J F_{\alpha\beta}{}^{KL} = 0\,. \label{eqn:tetrad_eom}
\end{eqnarray}
\end{subequations}
One can see that eq.~(\ref{eqn:palatini_eom}) is equivalent to the statement that 
\beq
\frac{\delta S[g,\Gamma]}{\delta \Gamma} = 0 \Rightarrow \nabla g = 0
\eeq
therefore in this approach the metric compatibility condition eq.~(\ref{eqn:metric-compat}) arises as the equation of motion obtained by varying the action with respect to the connection.

In deriving eq.~(\ref{eqn:palatini_eom}) we have utilized the fact that $ F[\omega + \delta \omega] = F[\omega] + \mc{D}_{[\omega]} (\delta \omega) $, where $\mc{D}_{[\omega]}$ is the covariant derivative defined with respect to the unperturbed connection $\omega$ as in eq.~(\ref{eqn:exterior-derivative}). The resulting equation of motion, eq.~(\ref{eqn:connection_eom}), is then the \emph{torsion-free} or \emph{metric-compatibility} condition which tells us that the tetrad is parallel transported by the connection $\omega$. This then implies that eq.~(\ref{eqn:ricci-tetrad}) holds, \ie $F_{\mu\nu}{}^{IJ} \equiv R_{\mu\nu}{}^{IJ}$. The second equation of motion can be obtained by inspection, since $F$ does not depend on the tetrad. Already we see dramatic technical simplification compared to when we had to vary the Einstein-Hilbert action with respect to the metric as in eq.~(\ref{eqn:gmunuvariation}).

We will digress at this point, much as we did in sec.~\ref{subsec:eh-action}, in order to show that eq.~(\ref{eqn:tetrad_eom}) is equivalent to Einstein's vacuum equations. We first note that the volume form can be written as
\begin{equation}\label{eqn:volume-form}
\epsilon_{\mu\nu\alpha\beta} = \frac{1}{4!} \epsilon_{PQRS} \, e_{[\mu}{}^P e_{\nu}{}^Q e_{\alpha}{}^R e_{\beta]}{}^S
\end{equation}
Contracting both sides with $e^{\nu}{}_J$ we find that
\begin{align}\label{eqn:contraction-one}
\epsilon_{\mu\nu\alpha\beta} \, e^{\nu}{}_J & = \frac{1}{4!} \epsilon_{PQRS}\, e_{[\mu}{}^P e_{\nu}{}^Q e_{\alpha}{}^R e_{\beta]}{}^S e^{\nu}{}_J \nonumber \\
			& = - \frac{1}{3!} \epsilon_{JPQR} \, e_{[\mu}{}^P e_{\alpha}{}^Q e_{\beta]}{}^R
\end{align}
where in the second line we have switched some dummy indices and relabelled others. Inserting the right hand side of the above in eq.~(\ref{eqn:tetrad_eom}) and using the fact that eq.~(\ref{eqn:ricci-tetrad}) implies $F_{\mu\nu}{}^{IJ} \equiv R_{\mu\nu}{}^{IJ}$, we find:
\begin{align}\label{eqn:efe-vacuum}
	\frac{\delta S}{\delta e_I{}^\mu} & = \epsilon^{\mu\nu\alpha\beta} \,e_{\nu}{}^J \epsilon_{IJKL}\,  R_{\alpha\beta}{}^{KL} \nonumber\\
	& = - \frac{1}{3!} \epsilon^{JPQR} \, \epsilon_{IJKL} \, e^{[\mu}_P \, e^{\alpha}_Q \, e^{\beta]}_R \, R_{\alpha\beta}{}^{KL} \nonumber \\
	& = \delta^P_{[I} \, \delta^Q_{K} \, \delta^R_{L]} \, e^{\mu}_P \, e^{\alpha}_Q \, e^{\beta}_R \, R_{\alpha\beta}{}^{KL} \nonumber \\
	& = e^{\mu}_{[I} \, e^{\alpha}_K \, e^{\beta}_{L]} \, R_{\alpha\beta}{}^{KL} \nonumber \\
	& = \left(e^{\mu}_{I} e^{\alpha}_K  e^{\beta}_{L} + e^{\mu}_{K} e^{\alpha}_L e^{\beta}_{I} + e^{\mu}_L e^{\alpha}_I e^{\beta}_K \right) \, R_{\alpha\beta}{}^{KL} \nonumber \\
	& = e^{\mu}_{I} R + e^{\beta}_{I} R_{\alpha\beta}{}^{\mu\alpha} + e^{\alpha}_{I} R_{\alpha\beta}{}^{\beta\mu} \nonumber \\
	& = e^{\mu}_{I} R - 2 e^{\beta}_{I} R_\beta{}^\mu = 0
\end{align}
In the first step we have used the result in eq.~(\ref{eqn:contraction-one}). In the second step we have used the fact that the contraction of two epsilon tensors can be written in terms of anti-symmetrized products of Kronecker deltas. In the third and fourth steps we have simply contracted some indices using the Kronecker deltas and expanded the anti-symmetrized product explicitly. In the fifth and sixth steps we have made use of eq.~(\ref{eqn:curvature-identity}) and the definition of the Ricci tensor as the trace of the Riemann tensor: $R_\beta{}^\mu = \sum_{\alpha} R_{\alpha\beta}{}^{\alpha \mu}$. Contracting the last line of the above with $e^{\nu I}$ and using the fact that $g_{\mu\nu} = e_{\mu}^I e_{\nu}^J \eta_{IJ}$ we find
\begin{equation}\label{eqn:einstein-fe}
	 R_{\mu\nu} -\frac{1}{2} g_{\mu\nu}R = 0\,.
\end{equation}
Thus the tetradic action in the first-order formulation - where the connection and tetrad are independent variables - is completely equivalent to classical general relativity.

\subsubsection{Palatini Hamiltonian \& Constraints}
Up to this point we have been discussing classical approaches to GR. The Palatini and ADM approaches reproduce Einstein's original formulation of GR, but as mentioned in sec~\ref{sec:can_quant_gravity}, one would hope that they provide a formulation amenable to canonical quantisation. We can perform a $3+1$ split of the Palatini action, eq.~(\ref{eqn:connection-action}) and obtain a hamiltonian which, once again, is a sum of constraints. However, while the resulting formulation appears simpler than that in terms of the metric variables, there are some second class constraints which when solved \cite[Section 2.4]{Peldan1993Actions} yield the same set of constraints as obtained in the ADM framework. Thus, the Palatini approach does not appear to yield any substantial improvements over the ADM version as far as canonical quantization is concerned. To proceed to a quantum theory, we must transition to a description of gravity in terms of the Ashtekar variables.

\section{First steps to a theory of Quantum Gravity}\label{sec:firststepsquantgravity}
As discussed in the previous section, we wish to attempt to canonically quantise GR, which means turning the Hamiltonian, diffeomorphism and Gauss constraints into operators and replacing Poisson brackets with commutation relations.
This procedure is easier said than done, however. In a practical sense one must be careful with the ordering of operators, and hence constructing appropriate commutation relations is not as easy as one might at first hope. We shall discuss the way forward in outline, before turning to a more detailed discussion of each step. Firstly we simplify the constraints by adopting a complex-valued form for the connection and tetrad variables. These are the Ashtekar variables. Next one performs a $3+1$ decomposition to obtain the Einstein-Hilbert-Ashtekar Hamiltonian $\mc{H}_{eha}$ which turns out to be a sum of constraints. We have already seen that these constraints all equal zero, and so when treated as operators they should act upon a state of quantum spacetime, $\fullket{\Psi}$ to yield $ \mc{H}_{eha} \fullket{\Psi} = 0 $. This condition does not force a particular choice of basis for $\fullket{\Psi}$ upon us, but it does admit a choice built from objects we are already familiar with - Wilson loops. These loops are then allowed to intersect, to yield area and volume operators of the spacetime. As a result, the states of quantum spacetime come to be represented by graphs whose edges are labelled by representations of the gauge group (for GR this is $SU(2)$). Throughout, the notion of \emph{background independence}\footnote{It is important to mention one aspect of background independence that is \emph{not} implemented, \emph{a priori}, in the LQG framework. This is the question of the topological degrees of freedom of geometry. On general grounds, one would expect any four dimensional theory of quantum gravity to contain non-trivial topological excitations at the quantum level. Classically, these excitations would correspond to defects which would lead to deviations from smoothness of any coarse-grained geometry.}, which is central to General Relativity, is considered sacrosanct.

The reader interested in the history behind the canonical quantization program, with further mathematical details, is referred to \cite{Thiemann2001Introduction}.

\subsection{Ashtekar Formulation: ``New Variables" for General Relativity}
\label{subsec:ashtekar-connection}
We have already discussed the first-order form of GR above. Now let us turn our attention to Ashtekar's complex-valued version of this formalism. We begin with tetradic GR whose action is written \ignore{in terms of a tetrad and connection} in the Palatini form\ignore{ as
$ S_{EH}\left[e,\omega\right] = \frac{1}{2\kappa}\int d^4 x\, \tilde{e}_I{}^\mu \tilde{e}_J{}^\nu F[\omega]_{\mu\nu}{}^{IJ}$}.
This action is equivalent to the usual Einstein-Hilbert action \emph{on-shell}, \ie for configurations which satisfy Einstein's field equations, as shown in subsection~\ref{subsec:palatini_action}. For dealing with spinors, a formalism defined in terms of connections and tetrads is more useful than one defined in terms of the metric, as shown above.  
When we perform the ADM splitting of the Palatini action, we switch from variables defined in the full four-dimensional spacetime to the three-dimensional hypersurfaces $\Sigma_t$. Hence the tetrads at each point become ``triads'', $e^I_\mu \rightarrow e_a^i$ where $\mu\rightarrow a\in\{1,2,3\}$, $I\rightarrow i\in\{1,2,3\}$, and the spin connection is likewise restricted, to become $\Gamma_a^i = \omega_{ajk}\epsilon^{jki}$. 
The phase space variables of the Palatini picture $(e_a^i, \Gamma_a^i)$ are the intrinsic metric of the spacelike manifold $\Sigma$ and a function of its extrinsic curvature respectively, similarly to the situation we noted in sec.~\ref{sec:ADMsplit}. Unfortunately in this case the Hamiltonian constraint is still a complicated non-polynomial function and canonical quantization does not appear to be any easier in this formalism.\todo{Show what the H constraint looks like in this case}  

Ashtekar made the remarkable observation that the form of the constraints simplifies dramatically\footnote{for the detailed derivation of these constraints starting with the self-dual Lagrangian see for e.g. \cite[Section 6.2]{Romano1993Geometrodynamics}} if instead of the \emph{real} connection $\omega_\mu{}^{IJ}$ one works with a \emph{complex}, self-/anti-self-dual connection 
(this means that the connection is equal to $\pm 1$ times the dual connection, which is defined in an analogous manner to the dual field strength of eq.~(\ref{eqn:dualfmunu_components})).
At the heart of the formulation of general relativity as a gauge theory lies a canonical transformation from the triad and connection to the ``new" or Ashtekar variables;
\begin{eqnarray}
e_a^i\rightarrow\frac{1}{\gamma}e_a^i\, , & \,\,\, & \Gamma_a^i\rightarrow A^i_a = \Gamma_a^i + \gamma K^i_a
\end{eqnarray}
where $\gamma$ is the so-called {\em Immirzi parameter}, $A^i_a$ is the Ashtekar-Barbero connection, and $K^i_a = \kappa_{ab}e^{bi}$ with $\kappa_{ab}$ the extrinsic curvature of $\Sigma$.

It's standard practice to rewrite the triads as ``densitised triads'', denoted $\tilde{e}$, and defined by
\beq
\tilde{e}^i_a = e^i_a \sqrt{h} 
\eeq
so that the new variables we actually work with are $(\frac{1}{\gamma} \tilde{e}^i_a, A_a^i)$. Both $A^i_a$ and $\tilde{e}^i_a$ admit $SU(2)$ rotations with respect to the internal indices (and hence the choice of densitised triads is non-unique). We can therefore treat the Ashtekar formulation of gravity as an $SU(2)$ gauge theory. This is consistent with our previous discussion about the choice of gauge group for gravity (sec.~\ref{subsubsec:connection}), as $SU(2)$ is a subgroup of $SL(2,\mathbb{C})$. 

Given this choice of variables, the constraints simplify to
\begin{subequations}
\begin{eqnarray}\label{eqn:palatini_constraints}
{\cal H} = & \hspace{-11mm}\epsilon^{ij}_k \tilde{e}^a_i \tilde{e}^b_j F^k{}_{ab} 
\label{eqn:palatini-hamiltonian} \qquad\qquad & \textrm{(Hamiltonian constraint)} \\
{\cal G}_i = & \hspace{-15mm}D_a \tilde{e}^a_i 
\label{eqn:palatini-gauss} \qquad\qquad\qquad & \textrm{(Gauss constraint)} \\
{\cal C}_a = & \hspace{1mm}\tilde{e}^b_i F^i{}_{ab}-A^i_a {\cal G}_i 
\label{eqn:palatini-diffeo} \qquad\qquad\qquad & \textrm{(Diffeomorphism constraint)} 
\end{eqnarray}
\end{subequations}
if we choose $\gamma = -i$. This makes quantisation {\em feasible}, although not necessarily {\em easy}. 
The phase space configuration and momentum variables are the three dimensional triad $\tilde{e}_a{}^i$ and the spatial connection $A^i{}_a$. 
The second class constraints which were present in the Palatini framework must now vanish due to the Bianchi identity \todo{Show the 2nd class constraints, and the B ident} and the diffeomorphism constraint becomes a polynomial quadratic function of the momentum variables - in this case the triad. We thereby obtain a form for the constraints which is polynomial in the coordinates and momenta and thus amenable to methods of quantization used for quantizing gauge theories such as Yang-Mills. The resulting expression for the Einstein-Hilbert-Ashtekar hamiltonian of GR is
\begin{equation}
\mc{H}_{eha} = N^a \mc{C}_a + N \mc{H} + T^i \mc{G}_i = 0
\end{equation}
where $ \mc{C}_a $, $\mc{H}$ and $\mc{G}_i$ are the vector, scalar and Gauss constraints respectively. The terms $N_i^a$ and $N$ are the shift and lapse, while $T^i$ is a lie-algebra valued function over our spatial surface which encodes the freedom we have in choosing the gauge for the gauge connection. As in \ref{subsec:poisson-brackets} we can calculate the Poisson brackets between these constraints and the canonical variables. Doing so verifies the intuition gained from \ref{subsec:poisson-brackets}. The Poisson brackets of a function $f$ with the Hamiltonian and diffeomorphism constraints gives:
\begin{equation}
	\{ f,\mc{H} \} = \pounds_{N\vec{n}} f \, , \qquad \{ f, \xi^a \mc{C}_a \} = \pounds_{\vec{\xi}} f\, ,
\end{equation}
implying that as expected $\mc{H}$ and $\mc{C}_a$ generate time-evolution and spatial diffeomorphism respectively. Introducing the gauge degrees of freedom has also led to the introduction of a third constraint $\mc{G}_i$, for whose Poisson bracket we have
\begin{equation}
	\{ f, T^i \mc{G}_i \} = - \tilde e^a_i D_a T^i\, ,
\end{equation}
implying that $\mc{G}_i$ is the generators of gauge rotations.

It is instructive to compare the above form of the constraints to their metric counterparts in eq.~(\ref{eqn:metric-constraints}) which are reproduced below for the reader's convenience:
\begin{eqnarray*}
{\cal H} & = & \left(-\sqrt{h}{}^{(3)}R + \frac{1}{\sqrt{h}}(\pi^{ab}\pi_{ab}-\frac{1}{2}\pi^2) \right)\\
{\cal C}^a & = & 2 D_b \pi^{ab}
\end{eqnarray*}
The price to be paid for this simplification is that the theory we are left with is no longer the theory we started with - general relativity with a manifestly real metric geometry. The connection is now a complex connection. However the new concoction is also not too far from the original theory and can be derived from an action. That this is the case was shown independently by Jacobson and Smolin \cite{Jacobson1988Covariant} and by Samuel \cite{Samuel1987A-lagrangian}. They completed the analysis by writing down the Lagrangian from which Ashtekar's form of the constraints would result:
\begin{equation}
S_\pm\left[ e,A\right] = \frac{1}{4\kappa} \int d^4 x \, {}^\pm\Sigma^{\mu\nu}{}_{IJ} {}^\pm F_{\mu\nu}{}^{IJ}
\end{equation}
Here $ {}^{\pm} F$ is the curvature of a \emph{self-dual} (anti-self-dual) four-dimensional connection ${}^{\pm} A$ one-form, which we will discuss more in the next subsection. The field ${}^\pm\Sigma$ is the self-dual (anti-self-dual) portion of the two-form $ \tilde e^I \wedge \tilde e^J $. The Palatini action is then simply given by the real part of the self-dual (or anti-self-dual) action,
\begin{equation}\label{eqn:realpart}
S_{P} = \mb{Re}[S_\pm]\,.
\end{equation}

\subsection{Loop Quantization}
As noted above, the program of Loop Quantum Gravity involves the following steps; 
\begin{enumerate}
  \item Write GR in connection and tetrad variables (in first order form).
  \item Perform a $3+1$ decomposition to obtain the Einstein-Hilbert-Ashtekar Hamiltonian $\mc{H}_{eha}$ which turns out to be a sum of constraints.
  \item Obtain a quantized version of the Hamiltonian whose action on elements of the physical space of states yields $ \mc{H}_{eha} \fullket{\Psi} = 0 $. 
  \item Identify an appropriate basis for the physical states of spacetime.
\end{enumerate}
The first two steps have been thoroughly covered. So now, after a fairly lengthy digression, we are ready to return to the task mentioned in sec.~\ref{sec:can_quant_gravity}, rewriting the constraints in operator form, and identifying the physical states of quantum gravity.  The first part of this process was completed in eqs.~(\ref{eqn:palatini-hamiltonian})-(\ref{eqn:palatini-diffeo}).

The following exposition only gives us a bird's eye view of the process of canonical quantization. The reader interested in the mathematical  details of and the history behind the canonical quantization program is  referred to \cite{Thiemann2001Introduction}.

\subsection{Canonical Quantization}
\label{subsec:canonical-quantization}
To find solutions of the equations of motion we want to find states $\Psi[A]$ such that they are acted upon appropriately by the constraints. This means that they satisfy
\begin{eqnarray}
{\hat{\cal H}}|\Psi\ket & = & 0 \nonumber \\
{\hat{\cal C}}_a|\Psi\ket & = & 0 \nonumber \\
{\hat{\cal G}}_i|\Psi\ket & = & 0 \nonumber
\end{eqnarray}
The Gauss constraint tells us that $\Psi[A]$ should be gauge-invariant functions of the connection. The diffeomorphism constraint is telling us that $\Psi[A]$ should be invariant under diffeomorphisms of the paths along which the connection lies. These constraints taken together do not impose a particular choice of $\Psi[A]$ upon us, but they do admit Wilson loops as one possible, and particularly convenient, choice.

Let us consider solutions of the form $\Psi[A] = \sum_\lambda \Psi[\lambda]W_\lambda[A]$. A given state will therefore be a sum of loops. These loops may in general be knotted, and hence topologically distinct from each other. Such states will satisfy the Gauss constraint, as Wilson loops are gauge-invariant. They will also satisfy the diffeomorphism constraint. In fact, diffeomorphism invariance actually helps us reduce the number of basis states, thereby avoiding a potentially awkward problem. In a theory with a fixed background and a well-defined metric any tiny change in the shape of a Wilson loop will lead to a different holonomy, since parallel transport is path-dependent. If different loops are taken to be the orthonormal basis states, this means that each deformation of a loop results in a new state, orthonormal to every other loop. But in a diffeomorphism-invariant theory it is not possible to distinguish between any two loops that may be smoothly deformed into each other, and hence the space of loops consists of only a single member of each topological equivalence class. 

Now we must ask whether Wilson loops satisfy the hamiltonian constraint. Firstly we observe that the triads (or tetrads when we are working in four dimensions) are the conjugate momenta to the connection. In usual quantum mechanics the operator for the momentum corresponds to derivation with respect to the position coordinate, $p \rightarrow \hat p = -i\hbar \frac{\partial}{\partial q}$. Similarly the quantum operator for the triad (or tetrad!) is given by the derivative with respect to the connection, hence $ e_a{}^i \rightarrow -i\hbar \frac{\partial}{\partial A_a{}^i} $. The action of ${\hat{\cal H}}$ on a Wilson loop is therefore
\beq
\label{eqn:WilsonHamConstraint}
{\hat{\cal H}}W_\lambda[A] = \epsilon^{ij}{}_k \frac{\delta}{\delta A_a^i} \frac{\delta}{\delta A_b^j} F^k{}_{ab} W_\lambda[A]\, .
\eeq
The derivatives pull out factors of $\dot{\lambda}$. Then since $F^k{}_{ab} = -F^k{}_{ba}$ it follows that the summation over indices of the curvature yields zero, hence $F^k{}_{ab}\dot{\lambda}^a\dot{\lambda}^b = 0$

This loop basis gives us a picture of spacetime at the smallest scale, consisting of closed paths carrying representations of $SU(2)$. It now remains to interpret the loop basis in terms of physical observables.

\section{Kinematical Hilbert Space}\label{sec:hilbert-space}
%
The allowed loop states that spacetime is composed of can take several forms. They may consist of simple closed loops. These loops may be linked through each other.
\begin{figure}[t]
\begin{center}
\includegraphics[scale=0.4]{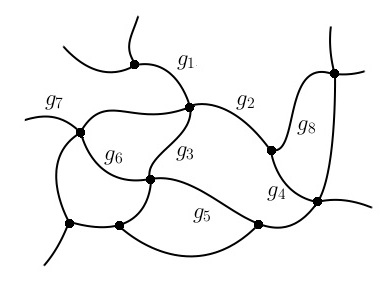}
\caption{States of quantum geometry are given by arbitrary graphs whose edges are labelled by group elements representing the holonomy along each edge. These graphs can take pretty complicated forms.}
\label{fig:graph_state}
\end{center}
\end{figure}
They may also be knotted, and hence classified by knot invariants. And the loops may intersect, creating vertices at which three or more Wilson lines meet. Historically the importance of all these possibilities has been considered, and continues to be assessed. We will simply take the view that a general loop state can have all of the properties listed above. It is therefore valid to consider a loop state to be a graph or network $\Gamma$ with edges $p_i$ labelled by elements of some gauge group (generally $SU(2)$ or $SL(2,\mathbb{C})$ in LQG)
\begin{equation}\label{eqn:cylinder_fn}
\Psi_{\Gamma} = \psi(g_1,g_2,\ldots,g_n)
\end{equation}
where $g_i$ is the holonomy of $A$ along the $i{}^\mathrm{th}$ edge. Pictorially, we can imagine something like figure~\ref{fig:graph_state}. In general we expect there to be an ensemble $\{\Gamma_i\}$ of spin-networks which corresponds to a semiclassical geometry $\{\mc{M},g_{ab}\}$ in the thermodynamic limit\footnote{when the number of degrees of freedom $N \rightarrow \infty$, the volume $V \rightarrow \infty$ and the number density $ N/V \rightarrow n$ where $n$ is bounded above}. 

We now wish to identify operators corresponding to physical observables of the spacetime. These operators should be based upon the physical structure of the graphs under consideration. It is worth noting at this point that in the Hamiltonian approach to quantum gravity that we have pursued there is an ambiguity as to whether we choose the connection or the frame fields as the configuration variables. In fact either choice is permissible, but the physical interpretation of connections as configuration variables and frame fields as conjugate momenta is more straightforward, and as we shall see it allows us to write operators that generate discrete areas and volumes. 

The inner-product of two different states on the \textbf{same} graph can be defined using the Haar measure on the group
\begin{equation}\label{eqn:inner_product}
\langle \Theta_{\Gamma} | \Psi_{\Gamma} \rangle = \int_{\mathcal{G}^n} d\mu_1 \ldots d\mu_n \Theta(g_1,\ldots,g_n) \bar{\Psi} (g_1,\ldots,g_n)
\end{equation}
For e.g. $L^2(\mathcal{G})$ - the space of square integrable functions on the manifold of the group $\mc{G}$ - constitutes the kinematical space of states for a single edge. \todo{Mention that the interpretation of this is uncertain.}

For further details and discussion the reader is referred to \cite{Ashtekar1996Differential,Ashtekar1996Quantum,Ashtekar1997cQuantum,Ashtekar1999Isolated,Ashtekar2000Quantum, Rovelli1994Discreteness}.

\subsection{Area Operator}\label{subsec:areaop}
The area operator in quantum geometry is defined in analogy with the classical definition of the area of a two-dimensional surface $S$ embedded in some higher dimensional manifold $M$. In the simplest case $S$ is a piece of $\mathbb{R}^2$  embedded in $\mathbb{R}^3$, however in general both $S$ and the higher-dimensional manifold may have some curvature. To make use of notation developed above, and without loss of generality, we will presume $S$ is embedded in a three-dimensional manifold $\Sigma$ obtained by foliating four dimensional spacetime (see sec.~\ref{sec:ADMsplit}). To each point $p \in S$ we can associate a triad  or ``frame field'' \ie a set of vectors  $\{ \vec{\mathbf{e}}_1, \vec{\mathbf{e}}_2, \vec{\mathbf{e}}_3 \}$ which form a basis for the tangent space $T_p$ at that point. In abstract index notation this basis can also be written more succinctly as $\{ e_a{}^i \}_p$ where $a,b,c\in\{1,2,3\}$ index the vectors and $i,j,k \ldots$ label the components of each individual vector in the active or ``chosen'' coordinate system. The indices $i,j,k \ldots$ are necessary because if $S$ is curved (\ie the gauge connection $A_a$ is non-zero) the basis at two distinct points in $S$ need not be the same, and hence a given vector $\vec{\mathbf{e}}_a$ will have different components at different points. 

The area of a two-dimensional surface $S$ embedded in $\Sigma$ is given by
\beq
	A_S = \int d^2 x\, \sqrt{{}^2 h}
\eeq
where ${}^2 h_{ab}$ is the metric on $S$, induced by the three-dimensional metric $h_{ab}$ on 
$\Sigma$, and ${}^2 h$ is its determinant, consistent with eq.~(\ref{eqn:area_det_metric}). Given an orthonormal triad field $\{e_a{}^i\}$ on $\Sigma$, we can always apply a local gauge rotation to obtain a new triad basis $\{ e'_a{}^i \}$, such that two of its legs - a ``dyad'' $\{ e'_x{}^i, e'_y{}^j \}$ - are tangent to the surface $S$ and $e'_z{}^k$ is normal to $S$. Then the components of the two-dimensional metric ${}^2 h_{AB}$ ($A,B \in \{x,y\}$ are purely spatial indices) can be written in terms of the dyad basis $\{e_A{}^I\}$\footnote{$I,J \in \{0,1\}$ label generators of the group of rotations $SO(2)$ in two dimensions. They are what is left of the ``internal'' $\mf{su}(2)$ degrees of freedom of the triad when it is projected down to $S$.} as 
\beq
  {}^2 h_{AB} = e_A{}^I e_B{}^J \delta_{IJ}
\eeq
The above expression with all indices shown explicitly becomes
\beq
	{}^2 h_{AB} := \left( \begin{array}{cc}
						h_{xx} & h_{xy} \\
						h_{yx} & h_{yy}
				\end{array} \right) = \left( \begin{array}{cc}
											e_x{}^I e_x{}^J & e_x{}^I e_y{}^J\\
											e_y{}^I e_x{}^J & e_y{}^I e_y{}^J
								\end{array} \right) \delta_{IJ}
\eeq
Now, the determinant of a $2 \times 2$ matrix ${}^2 h_{AB}$ takes the well-known form\footnote{This is a special case of the determinant for an $n \times n$ matrix $A_{ij}$ which can be written as $\mathrm{det}(A) = \sum_{i_1 \ldots i_n \in \mc{P}} A_{1\,i_i} A_{2\,i_2} \ldots A_{n\,i_n} \epsilon^{i_1 i_2 \ldots i_n}$ where the sum is over all elements of the permutation group $\mc{P}$ of the set of indices $\{i_m\}$ and $\epsilon^{i_1 i_2 \ldots i_n}$ is the completely anti-symmetric tensor.}
\beq
	\mathrm{det}({}^2 h_{AB}) = \sum_{i_1,i_2} h_{1\,i_1} h_{2\,i_2} \epsilon^{i_1\,i_2} = h_{11} h_{22} - h_{12} h_{21}
\eeq
For an orthornormal triad $\epsilon^{ij}{}_k e_z{}^k = e_x{}^i e_y{}^j$. Therefore in terms of the dyad basis $\{ e_A{}^I\}$, adapted to the surface $S$, the expression for the determinant becomes
\begin{eqnarray}
	\mathrm{det}({}^2 h_{AB}) & = & \left(e_x{}^i e_x{}^j e_y{}^k e_y{}^l - e_x{}^i e_y{}^j e_y{}^k e_x{}^l \right)\delta_{ij} \delta_{kl} \nonumber \\
			& = & \left( \epsilon^{ik}{}_m \epsilon^{jl}{}_n - \epsilon^{ij}{}_m \epsilon^{kl}{}_n \right) e_z{}^m e_z{}^n \, \delta_{ij} \delta_{kl} \nonumber \\
			& = & \epsilon^{ik}{}_m \, \epsilon_{ikn} \, e_z{}^m e_z{}^n \nonumber \\
			& = & \delta_{mn} \, e_z{}^m e_z{}^n
\end{eqnarray}
where we have used the fact that $\epsilon^{ij}{}_m \, \delta_{ij} = 0$ and also chosen to write the contraction of two completely anti-symmetric tensors in terms of products of Kronecker deltas.

Thus the classical expression\footnote{This is only valid for the case when $\Sigma$ is a three-dimensional manifold. In a general $n$-dimensional manifold, the area is a tensor
\begin{equation}
	A_{\mu \nu}{}^{jk} = e_{[\mu}{}^j e_{\nu]}{}^k
\end{equation}
In order to extract a single number - the ``area'' - from this tensor we project onto a two-dimensional surface spanned by the vectors $\{u^i, v^j\}$
\begin{equation}
	A[S] = e_{[\mu}{}^j e_{\nu]}{}^k u_i v_j
	\label{eqn:Area_of_surface}
\end{equation}
} for the area becomes
\begin{equation}\label{eqn:classical_area}
	A_S = \int_S d^2 x \, \sqrt{\vec{e_z} \cdot \vec{e_z}}
\end{equation}
where $\vec{e_z} \cdot \vec{e_z} \equiv e_z{}^m e_z{}^n \delta_{mn}$. With the classical version in hand it is straightforward to write down the quantum expression for the area operator. In the connection representation, the classical vierbein (tetrad) plays the role of the momenta.
Since the quantum operator for the vierbien is given by $ e_a{}^j \rightarrow -i\hbar \frac{\delta}{\delta A^a{}_j} $ we find that
\begin{equation}\label{eqn:quantum_area}
	\hat A_S = \int_S d^2 x \, \sqrt{\delta_{jk} \frac{\delta}{\delta A^z{}_j} \frac{\delta}{\delta A^z{}_k}}
\end{equation}
In order to determine the action of this operator on a spin-network state, let us recall the form of the state $\Psi_\Gamma$ from eq.~(\ref{eqn:cylinder_fn}), 
$$ \Psi_\Gamma = \psi(g_1,g_2,\ldots,g_n) $$
where $g_l$ is the holonomy along the $l$\supersc{th} edge of the graph. Let the edges of the graph $\Gamma$ intersect the surface $S$ at exactly $m$ locations, $\{P_1,\,P_2,\,\ldots P_m\}$. For the time being let us ignore the cases when an edge is tangent to $S$. We will also ignore the possibility that if the loops intersect, creating vertices, such a vertex happens to lie on $S$. Then, evidently, the action of eq.~(\ref{eqn:quantum_area}) on the state $\Psi_\Gamma$ will give us a non-zero result only in the vicinity of the punctures\footnote{Since the connection is \emph{defined} only along those edges and nowhere else!}. Thus
\begin{equation}
  \hat A_S \Psi_\Gamma \equiv \sum_{p=P_1}^{P_m} \sqrt{\delta_{ij} \frac{\delta}{\delta A^z{}_i(p)} \frac{\delta}{\delta A^z{}_j(p)}} \Psi_\Gamma
\end{equation}
We have written the connections with an explicit dependence on position $p$ to emphasise that at the $l$\supersc{th} puncture, the operator will act only on the holonomy $g_l$. 
We write the holonomies (compare eq.~({\ref{eqn:holonomy_wilsonloop}})) in the form
\begin{equation}\label{eqn:holonomy_b}
g_\lambda [A] = \mathcal{P} \exp\left\{\int_{\lambda_0}^{\lambda_1} dx \, n_a(x) \, A^a{}_I t^I \right\}
\end{equation}
where $\gamma$ is the curve along which the holonomy is evaluated, $x$ is an affine parameter along that curve, the $t^I$ are generators of the appropriate symmetry group as noted in section~\ref{subsec:covariantderiv}, and $n_a$ is the tangent to the curve at $x$. Then recognising that the functional derivative of the holonomy with respect to the connection takes the form
\begin{equation}\label{eqn:connection_deriv}
\frac{\delta}{\delta A^a{}_I} \, g_\lambda [A] = n_a(x) t^I g_\lambda[A]
\end{equation}
it follows easily that
\begin{equation}
  \frac{\delta}{\delta A^a{}_I} \psi(g_1,\ldots,g_k,\ldots,g_n) = n_a t^I \psi(g_1,\ldots,g_k,\ldots,g_n)
\end{equation}
where $n^a$ is the unit vector tangent to the edge at the location of the puncture. Now, recall that the $t^I$ in the above expression is nothing more than the $I$\supersc{th} generator of the Lie group in question. For $SO(3)$, these generators are the same as the angular momentum operators: $t^I \equiv J^I$ Thus the effect of taking the derivative with respect to the connection is to act on the state by the angular momentum operators. This gives us
\begin{equation}
  \frac{\delta}{\delta A^a{}_I} \frac{\delta}{\delta A^b{}_J} \psi = n_a n_b J^I J^J \psi
\end{equation}
Performing the contractions over the spatial and internal indices, noting that $n^a n_a = 1$, we finally obtain
\begin{equation}
  \hat A_S \Psi_\Gamma \equiv \sum_k \sqrt{\delta^{ij}\hat J_i \hat J_j} \Psi_\Gamma = \sum_k \sqrt{\vect{J}^2} \Psi_\Gamma
\end{equation} 
where $\hat J_i$ is the $i$\supersc{th} component of the angular momentum operator acting on the spin assigned to a given edge. $\vect{J}^2$ is the usual Casimir of the rotation group - that is, it is the element $\sum_a X_a X^a$ where the $X_a$ are the basis of the relevant Lie algebra and the $X^a$ are the dual basis defined with respect to some invariant mapping of the basis and dual basis to the scalars. The basic example of a Casimir element encountered at  undergraduate level is the squared angular momentum operator $L^2 = L_x^2 + L_y^2 + L_z^2$. Casimir operators commute with all elements of the Lie algebra. The action of $\vect{J}^2$ upon a given spin state gives us
\begin{equation}
  \vect{J}^2 \fullket{j} = j(j+1) \fullket{j}
\end{equation}
This gives us the final expression for the area of $S$ in terms of the angular momentum label $j_k$ assigned to each edge of $\Gamma$ which happens to intersect $S$,
\begin{equation}
  \hat A_S \Psi_\Gamma = l_p^2 \sum_k \sqrt{j_k (j_k + 1)} \Psi_\Gamma
\end{equation} 
where $l_p^2$ (a unit of area given as the square of the Planck length) is inserted in order for both sides to have the correct dimensions.

\subsection{Volume Operator}
\label{subsec:volumeop}
We have found a way of assigning quantised areas to graph states. It is natural to expect that these areas would lie on the boundaries of volumes, and to search, therefore, for a volume operator analogous to the area operator found in sec.~\ref{subsec:areaop}. Similarly to the two-dimensional case, we find that the volume of a region of space $S$ is given by 
\beq
  V = \int_S d^3 x \, \sqrt{h} = \frac{1}{6}\int_S d^3 x \, \sqrt{\epsilon_{abc}\epsilon^{IJK} e^a_I e^b_J e^c_K}
\eeq
Replacing the tetrads by their operator equivalents gives us the following expression for the volume \emph{operator}:
\beq\label{eqn:volume-classical}
	V = \frac{1}{6}\int_S d^3 x \, \sqrt{\epsilon_{abc}\epsilon^{IJK}\frac{\delta}{\delta A_a{}^I}\frac{\delta}{\delta A_b{}^J}\frac{\delta}{\delta A_c{}^K}}
\eeq

We have already discussed in the previous section that the effect of acting on a spin-network state with the operator corresponding to the tetrad has the effect of multiplying the state by the angular momentum operator:
\beq
	n^a \frac{\delta}{\delta A_a{}^i} \Psi_\Gamma = \hat J^i \psi 
\eeq
Consequently the action of the volume \emph{operator} on a given state can be expressed as
\beq\label{eqn:volume-quantum}
	\hat V \, \Psi_\Gamma = \frac{1}{6}\int_S d^3 x \, \sqrt{\epsilon_{abc}\epsilon^{ijk} n^a n^b n^c \hat J_i \hat J_j \hat J_k}\, \Psi_\Gamma
\eeq
Now, since the operator's action is non-zero only on the vertices $v$ of the graph $\Gamma$, the integral in the above expression reduces to a sum over a finite number of vertices $v\in \Gamma$ which lie in $S \cap \Gamma$:
\beq\label{eqn:volume-quantum-sum}
	\hat V \, \Psi_\Gamma = \frac{1}{6} \sum_{v \in S \cap \Gamma} \sqrt{\epsilon_{abc}\epsilon^{ijk} n^a n^b n^c \hat J_i \hat J_j \hat J_k} \, \Psi_\Gamma
\eeq
\begin{figure}[htbp]
	\centering
	\subfloat[Volume around node in classical geometry. Edges are labelled by vectors of the form $ a \hat x + b \hat y + c \hat z \in \mbb{R}^3 $]{
		\includegraphics[scale=0.4]{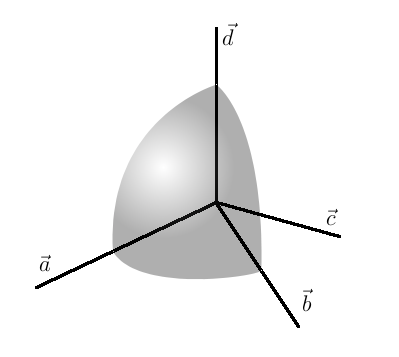}
		\label{fig:volume-op-classical}
	}\hspace{5mm}
	\subfloat[Volume operator in quantum geometry. Edges are labelled by elements of the form $\alpha \sigma_x + \beta \sigma_y + \gamma \sigma_z \in \sltwoc$]{
		\includegraphics[scale=0.4]{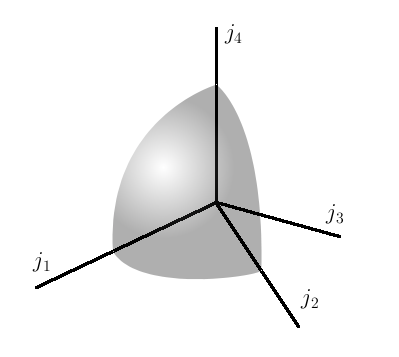}
		\label{fig:volume-op-quantum}
	}
	\caption{In order to calculate the volume around the vertex we must sum over the volume contained in the solid angles between each unique triple of edges. Classically this volume can be determined by the usual prescription $\vec{a}\cdot(\vec{b}\times\vec{c})$, where $\vec{a},\vec{b},\vec{c}$ are the vectors along each edge in the triple. In quantum geometry these vectors are replaced by irreps of $SU(2)$ but the basic idea remains the same.}
\end{figure}

In the literature one finds two forms of the volume operator. These are the Rovelli-Smolin (RS) and Ashtekar-Lewandowski versions. The RS \cite{Rovelli1994Discreteness} version is
\beq\label{eqn:rs-volume}
	\hat V_S^{RS} \Psi_{\Gamma} = \gamma^{3/2} l_p^3 \sum_{v \in S \cap \Gamma} \sum_{i,j,k} \left|\frac{i C_{reg}}{8} \epsilon_{abc} \epsilon^{ijk} n^a n^b n^c \hat J_i \hat J_j \hat J_k \right|^{1/2}  \Psi_{\Gamma}
\eeq
where $\epsilon_{abc}$ is the alternating tensor.

The AL \cite{Ashtekar1997cQuantum} version is
\beq\label{eqn:al-volume}
	\hat V_S^{AL} \Psi_{\Gamma} = \gamma ^{3/2} l_p^3 \sum_{v \in S \cap \Gamma} \left|\frac{i C_{reg}}{8} \epsilon_v (n^a, n^b, n^c) \epsilon_{abc}\epsilon^{ijk} n^a n^b n^c \hat J_i \hat J_j \hat J_k \right | ^{1/2}  \Psi_{\Gamma}
\eeq
Here $ \epsilon_v (n^a, n^b, n^c) \in {-1,1,0} $ is the orientation of the three tangent vectors at $v$ to the three curves/edges meeting at $v$. The key difference between the two version lies in this term. The RS operator does not take into account the orientation of the edges which come into the vertex. This fact is taken into account in the AL version, 
and it allows us to speak of a phase transition from a state of geometry at high-temperature ($T > T_c$)where the volume operator averages to zero for all graphs (which are ``large'' in some suitable sense) and a low-temperature ($T < T_c$) state where a geometric condensate forms and the volume operator gains a non-zero expectation value for states on all graphs. The key point here is that the AL version takes into account the ``sign'' of the volume contribution from any triplet of edges meeting at a vertex. Given any such triplet of edges $e_I, e_J, e_K$, by flipping the orientation of any one edge we flip the sign of the corresponding contribution to $\hat V^{AL}_S$. If we take the orientation of an edge as our random variable for the purposes of constructing a thermal ensemble, then it is clear that in the limit of high-temperature these orientations will flip randomly and the sum over the triplets of edges in $\hat V^{AL}_S$ will give zero for most (if not all) graph states. As we lower the temperature
the system begins to anneal and for some temperature $T = T_c$ the system should reach a critical point where the volume operator spontaneously develops a non-zero expectation on most (if not all) graph states.

Note that;
\begin{enumerate}
 \item Since the result of the volume operator acting on a vertex depends on the signs $\epsilon(e_I,e_J,e_K)$ of each triplet of edges, a simple dynamical system would then consist of a fixed graph with fixed spin assignments ($j_e$) to edges but with orientations that can flip, \ie $j_e \leftrightarrow - j_e$ (much like a spin).
 \item The Hamiltonian must be a hermitian operator. This fixes the various term one can include in it. We must also include all terms consistent with all the allowed symmetries in our model.
 \item The simplest trivalent spin-network has one vertex with three edges, e.g. a vertex of the hexagonal lattice. One can generalize the action of the volume operator on graphs which have vertices with valence $v$ (number of connecting edges) greater than 3. (The volume operator gives zero on vertices with $v \le 2$ so these are excluded). To do so we use the fundamental identity which allows to decompose the state describing a vertex with $v \ge 4$ into a sum over states with $ v = 3$. One example of the decomposition of a four valent vertex into two three-valent vertices is in the following figure;
\begin{figure}[h]
\begin{center}
\includegraphics[scale=0.4]{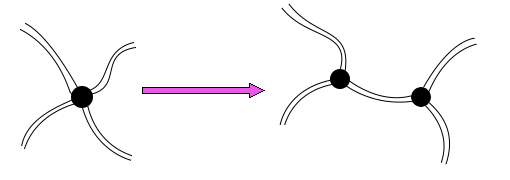}
\caption{}
\label{fig:decompose_vertex}
\end{center}
\end{figure}
 \item This model can help us understand how a macroscopic geometry can emerge from the ``spin'' or many-body system described by a Hamiltonian, which contains terms with the volume and area operators, on a spin-network.
\end{enumerate}

\subsection{Spin Networks}
\label{subsec:spinnets}
This discussion leaves us with a simple mental picture of spin-networks. Briefly, they are graphs with representations (``spins") of some gauge group (generally $SU(2)$ or $SL(2,\mathbb{C})$ in LQG) living on each edge. The links of this network correspond with cross-sectional areas, and the vertices where three or more links meet correspond with discrete volumes. The values of area and volume are determined by the spin labels on the relevant links. Since each link corresponds with the parallel transport of spin from one vertex to another, it is necessary to ensure that angular momentum is conserved at vertices, and so an \emph{intertwiner} is associated with each vertex. For the case of a four-valent vertex we have four spins: $(j_1,j_2,j_3,j_4)$. More generally a polyhedron with $n$ faces represents an intertwiner between the edges piercing each one of the faces. There is a simple visual picture of the intertwiner in the four-valent case.
\begin{figure}[htbp]
\centering
\subfloat[Labelling of edges by group elements]{
\includegraphics[scale=0.4]{network-labelled-g.png}
\label{fig:graph-states-g}
}\hspace{3mm}
\subfloat[Labeling of edges by group representations]{
\includegraphics[scale=0.4]{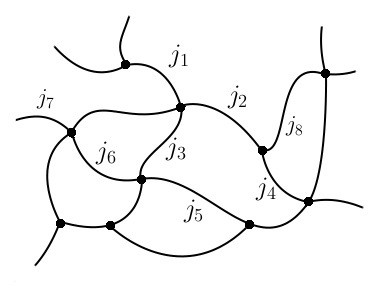}
\label{fig:graph-states-j}
}
\caption{States of quantum geometry are given by arbitrary graphs whose edges are labelled by group elements representing the holonomy along each edge. The Peter-Weyl theorem allows us to decompose these states in terms of \emph{spin-network} states, where edges are now labelled by group representations (angular momenta).}
\end{figure}

Picture a tetrahedron enclosing the given vertex, such that each edge pierces precisely one face of the tetrahedron. Now, the natural prescription for what happens when a surface is punctured by a spin is to associate the Casimir of that spin $ \mathbf{J}^2 $ with the puncture. The Casimir for spin $j$ has eigenvalues $ j (j+1) $. 
These eigenvalues are identified with the area associated with a puncture. 

In order for the edges and vertices to correspond to a consistent geometry it is important that certain constraints be satisfied. For instance, for a triangle we require that the edge lengths satisfy the triangle inequality $ a + b < c $ and the angles should add up to $ \angle a + \angle b + \angle c = \kappa \pi$, with $\kappa = 1$ if the triangle is embedded in a flat space and $\kappa \ne 1$ denoting the deviation of the space from zero curvature (positively or negatively curved).

In a similar manner, for a classical tetrahedron, now it is the sums of the areas of the faces which should satisfy ``closure'' constraints. For a quantum tetrahedron these constraints translate into relations between the operators $j_i$ which endow the faces with area.

For a triangle, giving its three edge lengths $(a,b,c)$ completely fixes the angles and there is no more freedom. However, specifying all four areas of a tetrahedron \emph{does not} fix all the freedom. The tetrahedron can still be bent and distorted in ways that preserve the closure constraints. These are the physical degrees of freedom that an intertwiner possesses - the various shapes that are consistent with a tetrahedron with a given set of face areas. 

\subsection{Spin-Foams}\label{subsec:foams}

\begin{figure}[htbp]
    \centering
    \includegraphics[scale=0.3]{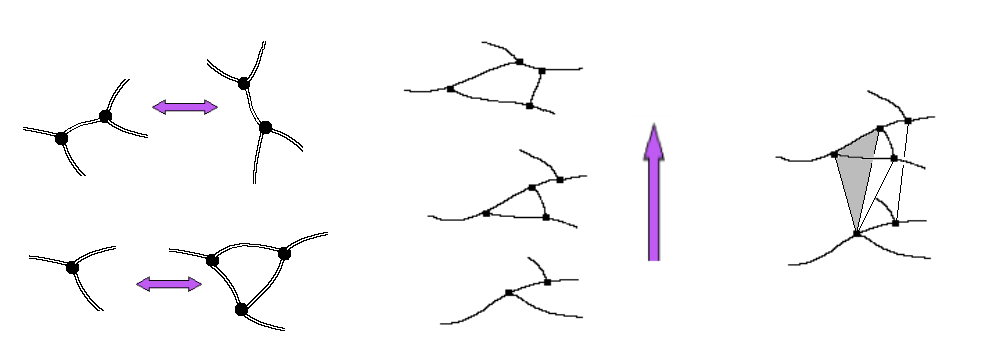}
    \caption{The 2-2 and 1-3 moves (left). The spin networks composing spatial hypersurfaces undergo a succession of such evolution moves as time passes (centre). In spin-foams, the links of a spin-network sweep out sheets in spacetime, such as the shaded region (right).}
    \label{fig:evolution_moves}
\end{figure}

In LQG the kinematical entities describing a given state of quantum geometry are spin-networks. The dynamical entities - \ie those that encode the evolution and history of spin-networks - are known as \emph{spin-foams}. If a spin-network describes a $d$-dimensional spacelike geometry, then a spin-foam describes a possible history which maps this spin-network into another one. In order to determine the transition amplitudes between two different states of quantum-geometry whose initial and final states are given by spin-networks $S_i$ and $S_f$, one must sum over \emph{all} possible spin-foams which interpolate between the two spin-network states. When we perform the sum over all allowed histories we find that the resulting amplitude depends only on the \emph{boundary} configuration of spins. This is holography. The holographic principle boils down to saying that the state of a system is determined by the state of its boundary. Therefore, although the point is not made as often is it possibly should be, LQG embodies the holographic principle in a very fundamental way.  
A spin-foam corresponds to a history which connects two spin-network states. On a given spin-network one can perform certain operations on edges and vertices which leave the state in the kinematical Hilbert space. These involve moves which split or join edges and vertices and those which change the connectivity. There are two basic transformations which the transitions between network states can be built from. These are the ``2-to-2'' move, in which an adjacent pair of trivalent vertices in a network exchange one incoming link each, and the ``1-to-3'' move, in which a single trivalent vertex splits into three vertices, in analogy to the ``star-triangle transformation'' used in the analysis of electrical circuits (see fig.~\ref{fig:evolution_moves}). The inverse moves are also possible, of course, and the reader should easily recognize that the 2-to-2 move is its own inverse, while the inverse of the 1-to-3 move shrinks a trio of vertices down to form a single vertex. 

One can ``formally'' view a spin-foam as a succession of states $\{ \fullket{\Psi(t_i)}\}$ obtained by the repeated action of the scalar constraint
\begin{align}
\fullket{\Psi(t_1)} \sim & \exp{}^{-i\mc{H}_{eha}\delta t} \fullket{\Psi(t_0)} \nonumber \\
\fullket{\Psi(t_2)} \sim & \exp{}^{-i\mc{H}_{eha}\delta t} \fullket{\Psi(t_1)} \ldots
\end{align}
and so on \cite{Reisenberger1994Worldsheet,Reisenberger1997Sum}. It is not our intention to discuss spin-foams in great detail here. Hopefully the preceding material has sufficiently familiarized the reader with the notation and concepts of LQG that they will be able to read other, more specific discussions of spin-foams easily. 

\section{Applications}\label{sec:applications}

Ultimately, the value of any theory is judged by its relevance for the \emph{real} world. Unfortunately, due to the small length scales involved, direct tests of models of quantum gravity are not easy to perform. However one can try to reproduce well-known results from other physical theories as a preliminary consistency test for newer theories. In this section, we will consider how LQG can be applied to the calculation of black hole entropy, and cosmological models. 

While the question of black hole entropy is, as yet, an abstract problem, it is concrete enough to serve as a test-bed for theories of quantum gravity. 
In addition to the Bekenstein area law (mentioned in sec. \ref{sec:intro}), by investigating the behavior of a scalar field in the curved background geometry near a black hole horizon it was determined \cite{Hawking1975Particle} that all black holes behave as almost perfect black bodies radiating at a temperature inversely proportional to the mass of the black hole: $T \propto 1/M_{BH}$. This thermal flux is named Hawking radiation after its discoverer. 
These properties of a black hole turn out to be completely independent of the nature and constitution of the matter which underwent gravitational collapse to form the black hole in the first place. These developments led to the understanding that a macroscopic black hole, at equilibrium, can be described as a thermal system characterized solely by its mass, charge and angular momentum.

Bekenstein's result has a deep implications for any theory of quantum gravity. The ``Bekenstein bound'' refers to the fact that eq.~(\ref{eqn:area-law}) is the \emph{maximum} number of degrees of freedom - of both, geometry and matter - that can lie within \emph{any region of spacetime} of a given volume $V$. The argument is straightforward \cite{Susskind1994The-World}. Consider a region of volume $V$ whose entropy is greater than that of a black hole which would fit inside the given volume. If we add additional matter to the volume, we will eventually trigger gravitational collapse leading to the formation of a black hole, whose entropy will be less than the entropy of the region was initially. However, such a process would violate the second law of thermodynamics and therefore the entropy of a given volume must be at a maximum when that volume is occupied by a black hole. And since the entropy of a black hole is contained entirely on its horizon, one must conclude that the maximum number of degrees of freedom $\mc{N}_{max}$ that would be required to describe the physics in a given region of spacetime $\mc{M}$, in any theory of quantum gravity, scales not as the volume of the region $V({\mc{M}})$, but as the area of its boundary \cite{t-Hooft1993Dimensional,Susskind1994The-World} $\mc{N}_{max} \propto A(\partial \mc{M})$.

In view of the independence of the Bekenstein entropy on the matter content of the black hole, the origin of eq.~(\ref{eqn:area-law}) must be sought in the properties of the horizon geometry. Assuming that at the Planck scale, geometrical observables such as area are quantized such that there is a minimum possible area element $a_0$ that the black hole horizon, or any surface for that matter, can be ``cut up into'', eq.~(\ref{eqn:area-law}) can be seen as arising from the number of ways that one can put (or ``sew'') together $\mc{N}$ quanta of area to form a horizon of area $A = k \mc{N} a_0$, where $k$ is a constant. In this manner, understanding the thermal properties of a black hole leads us to profound conclusions:
\begin{enumerate}
    \item In a theory of quantum gravity the physics within a given volume of spacetime $\mc{M}$ is completely determined by the values of fields on the boundary of that region $\partial \mc{M}$. This is the statement of the \emph{holographic principle}.
	\item At the Planck scale (or at whichever scale quantum gravitational effects become relevant) spacetime ceases to be a smooth and continuous entity, \ie \emph{geometric observables are quantized}.
\end{enumerate}

In LQG, the second feature arises naturally - though not all theorists are convinced that geometry should be ``quantized'' or that LQG is the right way to do so. One can also argue on general grounds that the first feature - holography - is also present in LQG, though this has not been demonstrated in a conclusive manner. Perhaps this paper might motivate some of its readers to close this gap!

Let us now review the black hole entropy calculation in the framework of LQG.

\subsection{Black Hole Entropy}\label{subsec:entropy}

The ideas of \emph{quantum geometry} 
allow us to give a statistical mechanical description of a black hole horizon. This is analogous to the statistical mechanical description of entropy for a gas, or some other system composed of many smaller parts, and can be illustrated by a toy model involving tossed coins. 

Suppose we toss $N$ fair coins in succession, and record both the order of each series of heads and tails (this is called a microstate, since it keeps track of what each ``particle'' in the system does) and the total number of heads and tails that occur, ignoring the order (this is called the macrostate). In general, several microstates will correspond to a given macrostate. For instance, if $N=4$, there is only one way to create a macrostate with zero heads (TTTT), or with four heads (HHHH), for the macrostate with three heads (and hence, one tail), there are four microstates (HHHT, HHTH, HTHH, and THHH), and similarly for the macrostate with one head and three tails. And for the macrostate with two heads and two tails there are six microstates. The number of microstates $y$ as a function of the number of heads in the macrostate, $x$, follows a Gaussian distribution, which has the general form 
\beq
 y = A e^{-B(x-\mu)^2}
\eeq 
where $\mu=N/2$ and $A$ and $B$ are scaling constants. Taking the natural logarithm of both sides we find that 
\beq
(x-\mu)^2 = -\frac{1}{B}\ln (\frac{y}{A})
\eeq
We identify the left-hand side with the entropy of the specified macrostate, remembering that in general different macrostates will have different entropies\footnote{For a hand-waving argument as to why this should be the entropy, consider a simplified gas in which each molecule has just two speeds, $v_{\mathrm min}$ (tails) and $v_{\mathrm max}$ (heads). Then the variable $x$ is related to the average speed of the gas molecules, which when squared is related to their energy, and this may be related to entropy by the equation $dS = dQ/T$ at constant $T$.}. The right-hand side is proportional to the logarithm of the number of microstates allowed in that macrostate. If we were interested in calculating the entropy of a gas, rather than tossing coins, the macrostates would correspond to particular values of pressure and temperature, while the microstates correspond to the positions and momenta of individual molecules.

In general there are two ways to calculate the entropy associated with a given random variable $x$.
\begin{enumerate} \label{enum:entropy_calc}
	\item Using Shannon's formula. Let us say that we sample our random variable from some given ensemble, from which we draw $N$ samples. The variable $x$ takes values in the set $\{x_i\}$ where $i = 1,2, \ldots n$), then the entropy associated with our lack of knowledge of the variable $x$ is given by:
	\begin{equation}\label{eqn:shannon_entropy}
		S(x) = - \sum_{i=1}^{n} p(x_i) \ln p(x_i)
	\end{equation}
	where $p(x_i)$ is the probability that the random variable takes on the value $x_i$. If in the $N$ samples on which the entropy is based, the $i^{\textrm{th}}$ value $x_i$ occurs $k_i$ times (with the constraint that $\sum_{i} k_i = N$), then we have the usual frequentist definition for the probability associated with that value:
	\[
		p(x_i) = \frac{k_i}{N}
	\]
	The definition of the Shannon entropy \eqref{eqn:shannon_entropy} is equivalent to the definition of the Gibbs entropy in statistical mechanics.
	\item Using the statistical mechanics method, or its more general version, Jaynes' formalism  \cite{Jaynes1957Information}. This is based on the \emph{maximum entropy principle}, according to which, in the absence of any prior information about a given random variable the \emph{least unbiased} assumption one can make is that the variable satisfies a probability distribution which possesses the maximum possible entropy. This assumption leads us to the usual Boltzmann form of the probability. For a given value of the random variable $x_i$, the associated probability distribution must satisfy the maximum entropy criterion (wherein \eqref{eqn:shannon_entropy} is maximized) and also the usual axioms of probability theory 
	\begin{subequations}\label{eqn:prob_axioms}
		\begin{align}
			\sum_{i=1}^n p_i & = 1 \label{eqn:prob_axioms_a} \\
			\expect{f(x)} 	 & = \sum_{i=1}^n p_i f(x_i) \label{eqn:prob_axioms_b}
		\end{align}		
	\end{subequations}
	where $f(x)$ is any function of $x$. The unique probability function which satisfies these criteria is found to be (see for e.g. \cite[Sec.~3.2]{Pathria2011Statistical}):
	\begin{equation}\label{eqn:boltzmann_pdf}
		p_i = e^{-\lambda - \mu x_i}
	\end{equation}
	where $\lambda, \mu$ are Lagrange multipliers required for enforcing the constraints given in \eqref{eqn:prob_axioms} \footnote{The quantity being extremized has the form $L = - \sum_{i=1}^{n}\left\{ p(x_i) \ln p(x_i) - \lambda p(x_i) - \mu f(x_i) p(x_i) \right\}$.} $\lambda, \mu$ can be identified with the chemical potential and the inverse temperature, respectively associated with the random variable $x$. Using \eqref{eqn:boltzmann_pdf} we can write down the partition function
	\begin{equation}\label{eqn:partition_fn}
		Z(\mu) = \sum_{i=1}^n e^{-\lambda - \mu x_i}
	\end{equation}
	given which we can evaluate the usual thermodynamic quantities such as expectation values, free energy and the entropy in $x$:
	\begin{subequations}\label{eqn:thermal_quantities}
		\begin{align}
			\expect{f(x)} & = - \frac{\partial \ln Z(\mu)}{\partial \mu} \\
			F(T) & = -k T \ln Z(T) \\
			S & = - \frac{\partial F}{\partial T}
		\end{align}
	\end{subequations}
	where the inverse ``temperature'' is given by $\lambda = 1/kT$.
\end{enumerate}

In the case of quantum geometry, the microstates correspond with the assignments of area to the discrete ``pieces'' of a surface (such as the event horizon of a black hole). Hence for each macroscopic interval of area $[A+\delta A, A-\delta A]$, entropy $S$ is proportional to the log of the number of ways in which we can puncture the sphere to yield an area within that interval.
\begin{figure}[htbp]
    \centering
    \label{fig:bh-entropy}
    \includegraphics[scale=0.3]{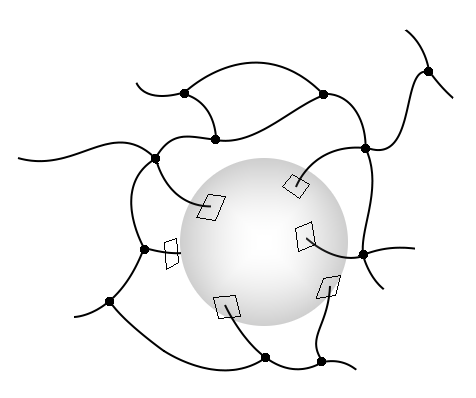}
    \caption{A spin-network corresponding to some state of geometry in the bulk punctures a black-hole horizon at the indicated locations. Each puncture yields a quantum of area $\propto \sqrt{j(j+1)}$ where $j$ is the spin-label on the corresponding edge. The entropy of the black-hole - or, more precisely, of the horizon - can be calculated by counting the number of possible configurations of punctures which add up to give a macroscopic value of the area lying within some finite interval $(A,A+\delta A)$}
\end{figure}

The state of a quantum surface is specified by a sequence of $N$ integers (or half-integers depending on the gauge group) $\{j_i,\ldots,j_N\}$, each of which labels an edge which punctures the given surface. The area of the surface is given by a sum over the Casimir at each puncture:
\begin{equation}\label{eqn:area-formula}
	\mathbf{A} = 8\pi \gamma l_p^2 \sum_{i=1}^{N} \sqrt{\mathbf{j}_i (\mathbf{j}_i + 1)}
\end{equation}
	The eigenvalues of the operator $\mathbf{j}_i$ are of the form $k_i/2$, where $k_i \in \mathbb{Z}$. Thus, the eigenvalues of the area operator are of the form
\begin{equation}\label{eqn:area-eigenvalue}
	A_i = 4\pi \gamma l_p^2 \sqrt{k_i(k_i + 2)} = 4\pi \gamma l_p^2 \sqrt{(k_i+1)^2 -1}
\end{equation}

In addition to \eqref{eqn:area-eigenvalue} the integers $ \{k_I\} $ must also satisfy a so-called \textit{projection constraint}, which is discussed later in this section.

The task at hand is the following; given an interval $[A+\delta A, A-\delta A]$, where $A$ is a macroscopic area value and $\delta A$ is some small interval ($\delta A/A \ll 1$), and the number $N$ of edges which puncture the surface, determine the allowed the number $N(M)$ of sequences of integers $\{k_i,\ldots,k_N\}$, such that the resulting value for the total area falls within the given interval
\begin{equation}\label{eqn:area-interval}
	M = \frac{A}{4 \pi \gamma l_p^2} = \sum_{i}\sqrt{k_i(k_i +2)} \in [A+\delta A, A-\delta A] \,.
\end{equation}

There are various approaches to this problem. We summarize two of these - the simple argument of Rovelli's \cite{Rovelli1996Black} and the number theoretical approach of \cite{Agullo2008Black,Agullo2011Detailed}.

\subsubsection{Rovelli's Counting}\label{subsubsec:rovelli-counting}

We want to compute the number of sequences $N(M)$, where each sequence $\{k_i\}$ satisfies
\begin{equation*}
	M = \frac{A}{4 \pi \gamma l_p^2} = \sum_{i}\sqrt{k_i(k_i +2)}
\end{equation*}

Let us first note the following set of inequalities:
\begin{equation}\label{eqn:bhe_inequality_a}
	\sum_i \sqrt{k_i^2} < \sum_i \sqrt{k_i (k_i + 2)} \equiv \sum_i \sqrt{(k_i + 1)^2 - 1} < \sum_i \sqrt{(k_i + 1)^2}
\end{equation}
Let $N_+(M)$ denote the number of sequences such that $\sum_i k_i = M$ and $N_-(M)$ denote the number of sequences such that $\sum_i (k_i + 1) = M$). Then the above set of inequalities implies that \cite{Rovelli1996Black}
\begin{equation}\label{eqn:bhe_inequality_b}
	N_-(M) < N(M) < N_+(M)
\end{equation}
Computing $ N_+(M) $ boils down to counting the number of partitions of $ M $, \ie the numbers of sets of \textit{ordered}, positive integers whose sum is $ M $. As noted in \cite{Rovelli1996Black}, this can be solved by observing that if $ (k_1,k_2,\ldots,k_n) $ is a partition of $ M $, then $ (k_1,k_2,\ldots,k_n,1) $ and $ (k_1,k_2,\ldots,k_n+1) $ are partitions of $ M+1 $. All partitions of $ M+1 $ can be obtained in this manner and therefore we have $ N_+(M+1) = 2 N_+(M) $, which implies that $ N_+(M) = C 2^M $, where $ C $ is a constant.

\subsubsection{Number Theoretical Approach}\label{subsubsec:number-theory}

This approach consists of two steps;
\begin{enumerate}[label=\textbf{\Alph*.},labelindent=\parindent]
	\item	Determining allowed sequences. This involves solving the BP (Brahmagupta-Pell) equation\footnote{It is well-known that the name of ``Pell's Equation'' was the result of Leonhard Euler's misidentification of John Pell with the mathematician Lord Brouckner. If we gave Euler a second chance to name the equation, he might have called it ``Brouckner's equation''. This equation had previously been intensively studied by the Indian mathematicians Brahmagupta and Bhaskara around the 5\supersc{th} century B.C. and 12 century A.D. respectively. However, Brouckner and Euler are to be forgiven for not having knowledge of the existence of this earlier work. The authors hereby take the liberty of correcting this historical wrong associated with the naming of this equation, by adding the prefix ``Brahmagupta'' to the presently accepted name ``Pell's Equation''.}.
	For now, we will work in units where $4\pi \gamma l_p^2 \equiv 1$. Thus for a given set of $N$ punctures on a quantum horizon, the total area can be written as
	$$ A = \sum_{i=1}^N A_i = \sum_{i=1}^N \sqrt{(k_i+1)^2 -1} $$
	For each possible value of $k$, let $g_k$ be the number of punctures which have that eigenvalue. So, we can write
	$$ A = \sum_{k=1}^{k_{max}} g_k \sqrt{(k+1)^2 -1} $$
	with $g_k = 0$ if no puncture has spin $k/2$. As shown in Appendix \ref{app-sec:square-free}, the square root of any integer can be written as the product of an integer and the square-root of a square-free integer. Since $k\in\mathbb{Z} \Rightarrow (k + 1)^2 -1\in \mathbb{Z}$, therefore we can write
	$$ \sqrt{(k+1)^2 -1} = y_k \sqrt{p_k} $$
	for some $ y_k \in \mathbb{Z}$ and $ p_k \in \mathbb{A}$, where $\mathbb{A}$ is the set of square-free integers. This implies that the area eigenvalue can be written as an integer linear combination of square-roots of square-free numbers:
	$$ A = \sum_{i=1}^{i_{max}} y_i \sqrt{p_i} $$
	leading us to the condition that
	$$ \sum_{k=1}^{k_{max}} g_k \sqrt{(k+1)^2 -1} = \sum_{i=1}^{i_{max}} y_i \sqrt{p_i} $$
	As a first step towards solving the general case, let us first try to determine the solution of the above equation for a single area eigenvalue $k_i/2$:
	$$ \sqrt{(k_i + 1)^2 - 1} = y_i \sqrt{p_i} $$
	knowing which we will be able to solve the general equation. Here the unknown variables are $k_i, y_i$. The $p_i$ are the known square-free numbers.
	Setting $x_i = k_i + 1$ and squaring both sides we obtain
	$$ x_i^2 - p_i y_i^2 = 1$$
	This is commonly known as Pell's equation, or perhaps more appropriately as the Brahmagupta-Pell equation. A method for obtaining its solutions is given in Appendix \ref{app-sec:pells-eqn}.
\end{enumerate}

\begin{enumerate}[label=\textbf{\Alph*.},labelindent=\parindent,resume]
	\item Determining the number of valid ways of \emph{sprinkling} labels from an allowed sequence onto the edges. This can be mapped to one of the simpler examples of NP-complete problems in the field of computational complexity - the Number Partitioning Problem \cite{Mertens1998Phase,Mertens2000A-physicists}.
	
	The relevance of the NPP for black hole entropy arises as follows. The counting of states of a horizon for a non-rotating black hole boils down to determining the number of ways in which we can choose spin-labels $k_i$ from a given sequence $\{k_1,...,k_N\}$ (where the allowed sequences are determined by solving the Brahmagupta-Pell equation) to each of the $i=1...N$ edges puncturing the horizon, such that $\sum k_i = 0$.
	
	More generally the case where $\sum k_i = m \,(m > 0)$, corresponds to a horizon with angular momentum $m$. This is equivalent to the statement of the NPP, where given an arbitrary but fixed sequence of (positive) integers $A=\{a_i,...,a_N\}$, one asks for the number of ways $N_A$ in which we can partition $A$ into two subsets $A_+$ and $A_-$, such that the difference of the sum of the elements of each subset is minimized: $\sum_{A_+} a_i - \sum_{A_-} a_i = m$. For the BHE problem $m$ is given by $\sum k_i^+ - \sum k_i^- = m$.
	
	As shown in \cite{Agullo2008Black} this problem can be mapped to a non-interacting spin-system \cite{De-Raedt2001Number} as follows. Consider a chain of $N$ spins each of which can be in an up $\fullket{\uparrow}$ state or a down $\fullket{\downarrow}$ state. If $a_i$ belongs to $A_+$ ($A_-$) then we set the $i^{\textrm{th}}$ spin to up (down). Consequently the constraint $A_+ - A_- = m$ can be expressed as the condition that
	\begin{equation}
		m - \sum_{i=1}^{N} a_i S_i = 0
	\end{equation}
where $S_i \in \{+1,-1\}$ are the possible eigenvalues of $\sigma_z$. The problem of partitioning $A$ is then equivalent \cite{De-Raedt2001Number} to determining the ground state of the Hamiltonian
	\begin{equation}\label{eqn:npp-hamiltonian}
		H = m - \sum_{i=1}^{N} a_j \sigma^j_z
	\end{equation}
where $\sigma^j_z$ is the Pauli spin operator for the $j^{\textrm{th}}$ spin. Any eigenstate of $H$ with zero energy corresponds to a solution of the NPP for the set $A$.
\end{enumerate}

\subsection{Loop Quantum Cosmology}\label{subsec:lqc}

One of the first avenues to follow when approaching old problems with new tools is to select the simplest possible scenarios for study, in the hope that the understanding gained in this arena would ultimately lead to a better understanding of more complex systems and processes. In classical GR this corresponds to studying the symmetry reduced solutions\footnote{That is, the solutions of the EFEs possessing strong global symmetries which reduces the effective local degrees of freedom to a small number.} of Einstein's equations, such as the FLRW cosmologies and their anisotropic counterparts, and various other exact solutions such as deSitter, anti-deSitter, Schwarzschild, Kerr-Newman etc.\footnote{We refer the reader to the extremely comprehensive and well-researched catalog of solutions to Einstein's field equations, in both metric and connection
variables, presented in \cite{Mueller2009Catalogue}. A somewhat older, but still valuable, catalog of exact solutions is given in \cite{Stephani2003Exact}.} 
which correspond respectively to a ``universe'' (in this very restricted sense) with positive cosmological constant ($\Lambda > 0$), a universe with $\Lambda < 0$, a non-rotating black hole and a rotating black hole (both in asymptotically flat spacetimes\footnote{A metric with a radial dependence is considered asymptotically flat if it approaches (in a well-defined sense) a flat Minkowski metric as $r \rightarrow \infty$.}). In each of these cases the metric has a very small number of local degrees of freedom and hence provides only a ``toy model''.
Of course, in the \emph{real world}, the cosmos is a many-body system and reducing its study to a model such as the FLRW universe is a gross simplification. However, via such models, one can obtain a qualitative grasp of the behavior of the cosmos on the largest scales.

\subsubsection{FLRW Models}\label{subsubsec:flrw}

The simplest quantum cosmological model is that which corresponds to the Friedmann metric whose line-element is given by\footnote{The following discussion is taken from \cite[Section 4]{Bojowald2006aLoop}.}
\begin{equation}\label{eqn:flrw-metric}
	ds^2 = -N(t)^2 dt^2 + a(t)^2 \left( \frac{1}{1-kr^2} dr^2 + r^2 d\Omega^2 \right)
\end{equation}
where the only dynamical variable is the scale factor $a(t)$ which depends only on the time parameter; $r = \sqrt{x^2 + y^2 + z^2}$ is the radial dimension of the spatial slices; $d\Omega^2 = d\theta^2 + \sin^2 \theta d\phi^2	$ is the angular volume element and $ k = -1, 0, +1 $ determines whether our spatial slices are open ($k=-1$), flat ($k=0$) or closed ($k=1$). For this metric we can perform the $ 3+1 $ decomposition and write down the action in terms of the various constraints. By comparing this metric with the general form given in eq.~\eqref{eqn:line-element}, we see that $N(t)$ is the lapse function and the shift vanishes $N^a = 0$. This implies that the diffeomorphism constraint $ D_a \pi^{ab} $ must also vanish.

Inserting this metric into the the EFE \eqref{eqn:einstein-eqn} gives us the vacuum Friedmann-LeMaitre-Robertson-Walker equations which describe the dynamics of homogenous, isotropic spacetimes:
\begin{equation}\label{eqn:flrw-equation}
	\left(\frac{\dot{a}}{a}\right)^2 + \frac{k}{a^2} = \frac{8 \pi G}{3 a^2} H_{matter}(a) 
\end{equation}
where $ H_{matter} $ is the Hamiltonian for any matter fields that might be present. This equation gives us the Hamiltonian constraint for the FLRW metric. This can be seen by starting from the Lagrangian formulation where
\begin{equation}
	S_{EH} = \frac{1}{16 \pi G} \int dt\, d^3 x \, \sqrt{-g} R[g]
\end{equation}
The Ricci scalar $ R[g] $ for the FLRW line-element \eqref{eqn:flrw-metric} is
\begin{equation}\label{eqn:flrw-ricci}
	R = 6 \left( \frac{\ddot{a}}{N^2 a} + \frac{\dot{a}^2}{N^2 a^2} + \frac{k}{a^2} - \frac{\dot{a} \dot N}{a N^3}\right)
\end{equation}
Substituting the above into the $ S_{EH} $ we obtain
\begin{equation}
	S = \frac{V_0}{16 \pi G } \int dt \, N a(t)^2 R = \frac{3 V_0}{8 \pi G} \int dt \, N \left( - \frac{a \dot{a}^2}{N^2} + ka  \right)
\end{equation}
From this equation we can identify the momentum $ p_a $ conjugate to the (only) degree of freedom - the scale factor $ a(t) $:
\begin{equation}
	p_a = \frac{\partial L}{\partial \dot a} = - \frac{3 V_0}{4 \pi G} \frac{a \dot{a}}{N^2}
\end{equation}
Since the action does not contain any terms depending on $ \dot{N} $, we have $ p_N = 0 $, implying that the lapse function $ N(t) $ is not a dynamical degree of freedom. We can now write down the Hamiltonian for the system in the usual manner: $ H = \sum_i p_i \dot{q}_i - L = p_a \dot{a} - L$, which gives:
\begin{equation}\label{eqn:flrw-hamiltonian}
	H = N \left[ \frac{4 \pi G}{3 V_0} \frac{p_a^2}{2 a} - \frac{3 V_0 k a}{8 \pi G} \right]
\end{equation}
It is clear from the form of this expression that this Hamiltonian will become divergent as $ a \rightarrow 0 $. Changing from metric to connection variables will allow us to alleviate this problem.

\subsubsection{Connection Variables}\label{subsubsec:connection-variables}

For isotropic spacetimes, the triad and extrinsic curvature take the form \cite[Section IV]{Alexander2006Gravity}:
\begin{equation}
	e^a_i = a(t)^2 \delta^a_i \qquad K_a^i = \dot{a} a^2 \delta_a^i
\end{equation}
In variables adapted to the particular form of the metric \eqref{eqn:flrw-metric}, the connection $ |\tilde{p}| $ and triad $ \tilde{c} $ are expressed as:
\begin{equation}
	|\tilde{p}| = \tilde{a}^2 = \frac{a^2}{4}; \qquad	\tilde{c} = \tilde{\Gamma} + \gamma \dot{\tilde{a}} = \frac{1}{2}(k + \gamma \dot{a})
\end{equation}
where $ \gamma $ is the Immirzi parameter. These components satisfy the commutation relations:
\begin{equation}
	\left\{ \tilde{c}, \tilde{p} \right\} = \frac{8 \pi \gamma G}{3} V_0
\end{equation}
The factors of $ V_0 $ can be absorbed into the definition of the variables to give us:
\begin{equation}
	c = V_0^{1/3} \tilde{c} \qquad p = V_0^{2/3} \tilde{p}
\end{equation}
In terms of these the Hamiltonian constraint \eqref{eqn:flrw-hamiltonian} becomes:
\begin{equation}
	H = - \frac{3}{8 \pi G} \left( \frac{(c-\Gamma)^2}{\gamma^2} + \Gamma^2 \right)\sqrt{|p|} + H_{matter}(p) = 0
\end{equation}
where $ \Gamma = V_0^{1/3} \tilde{\Gamma} $

\subsection{Semiclassical Limit}

The graviton propogator has a robust quantum version in these models. Its long-distance limit yields the $1/r^2$ behavior \cite{Rovelli2005Graviton} expected for gravity and an effective coarse-grained action given by the usual one consisting of the Ricci scalar plus terms containing quantum corrections.



\section{Discussion and recent developments}\label{sec:discussion}

Any fair and balanced review paper on LQG should also mention at least a few of the many objections its critics have presented. A list of a few of the more important points of weakness in the framework and brief responses to them follows:
\begin{enumerate}
\item \emph{LQG admits a volume extensive entropy and therefore does not respect the Holographic principle}:  This criticism hinges upon the description of states of quantum gravity as spin-networks which are essentially spin-systems on arbitrary graphs. However, spin-networks only constitute the \emph{kinematical} Hilbert space of LQG. They are solutions of the spatial diffeomorphism and the gauss constraints but \emph{not} of the Hamiltonian constraint which generates time-evolution. This criticism is therefore due to a (perhaps understandable) failure to grasp the difference between the kinematical and the dynamical phase space of LQG.

In order to solve the Hamiltonian constraint we are forced to enlarge the set of states to include \emph{spin-foams} which are histories of spin-networks. In a nutshell then, as we mentioned in sec.~\ref{subsec:foams}, the kinematical states of LQG are the spin-networks, while the dynamical states are the spin foams. The amplitudes associated with a given spin-foam are determined completely by the specification of its boundary state. Physical observables do not depend on the possible internal configurations of a spin-foam but only on its boundary state. In this sense LQG satisfies a stronger and cleaner version of holography than string theory, where this picture emerges from considerations involving graviton scattering from certain extremal black hole solutions.

%
%

\item \emph{LQG violates the principle of local Lorentz invariance/picks out a preferred frame of reference}: Lorentz invariance is obeyed in LQG but obviously not in the exact manner as for a continuum geometry. As has been shown by Rovelli and Speziale \cite{Rovelli2010Lorentz} the kinematical phase space of LQG can be cast into a manifestly Lorentz covariant form. A spin-network/spin-foam state transforms in a well-defined way under boosts and rotations. Similarly in quantum mechanics one finds that a quantum rotor transforms under discrete representations of the rotation group $SO(3)$.
\item \emph{LQG does not have stable semiclassical geometries as solutions - geometry ``crumbles''} - CDT simulations e.g. \cite{CDT2005} show how a stable geometry emerges. As mentioned in sec.~\ref{subsec:eh-action}, this involves calculating a sum over histories for the geometry of spacetime, between some initial and final state. The stability of the spacetimes studied in such simulations appears to be dependent on causality - that is, spacetime geometries develop unphysical structures in the Euclidean case, which are controlled when there is a well-defined past and future, as is the case in LQG. The question of exactly how similar CDT and LQG are to each other is a matter of continuing investigation. 
\item \emph{LQG does not contain fermionic and bosonic excitations that could be identified with members of the Standard Model}: The area and volume operators do not describe the entirety of the structures that can occur within spin networks. LQG or a suitably modified version which allows braiding between various edges will exhibit invariant topological structures. Recent work \cite{SBT1,SBT2} has been able to identify some such structures with SM particles. In addition, in any spin-system - such as LQG - there are effective (emergent) low-energy degrees of freedom which satisfy the equations of motion for Dirac and gauge fields. Xiao-Gang Wen and Michael Levin \cite{Levin2004String-net,Levin2007Detecting} have investigated so-called ``string-nets'' and find that the appropriate physical framework is the so-called ``tensor category'' or ``tensor network'' theory \cite{Biamonte2010Categorical,Evenbly2011Tensor,Haegeman2011Entanglement}. In fact string-nets are very similar to spin-networks so Wen and Levin's work - showing that gauge bosons and fermions are quasiparticles of string-net condensates - should carry over into LQG without much modification.
\item \emph{LQG does not exhibit dualities in the manner String Theory does}: Any spin-system exhibits dualities. A graph based model like LQG even more so. One example of a duality is to consider the dual of a spin-network which is a so-called 2-skeleton or simplicial cell-complex. Another is the star-triangle transformation, which can be applied to spin-networks which have certain symmetries, and which leads to a duality between the low and high temperature versions of a theory on a hexagonal and triangular lattice respectively \cite{Baxter2008Exactly}.
\item \emph{LQG doesn't admit supersymmetry, wants to avoid extra dimensions, strings, extended objects, etc}: Extra dimensions and supersymmetry are precisely that - ``extra''. Occam's razor dictates that a successful physical theory should be founded on the \emph{minimum} number of ingredients. 
It is worth noting that at the time of writing of this paper, results from the LHC appear to have ruled out many supersymmetric extensions of the standard model. By avoiding the inclusion of extra dimensions and supersymmetry, LQG represents a perfectly valid attempt to create a theory that is consistent with observations. 
\item \emph{LQG has a proliferation of models and lacks robustness}: Again a lack of extra baggage implies the opposite. LQG is a tightly constrained framework. There are various uniqueness theorems which underlie its foundations and were rigorously proven in the 1990s by Ashtekar, Lewandowski and others. There are questions about the role of the Immirzi parameter and the ambiguity it introduces however these are part and parcel of the broader question of the emergence of semi-classicality from LQG (see Simone Mercuri's papers in this regard).
\item \emph{LQG does not contain any well-defined observables and does not allow us to calculate graviton scattering amplitudes}: Several calculations of two-point correlation functions in spin-foams exist in the literature \cite{Rovelli2005Graviton} These demonstrate the emergence of an inverse-square law.
\end{enumerate}

As well as discussing criticisms of LQG, it is also fair to consider what role this theory may have in the future. We would not have written a paper reviewing the formulation and current status of LQG if we did not consider it an important and interesting theory - one which we feel is probably a good representation of the nature of spacetime. However it is wise to remember that most physical theories are ultimately found to be flawed or inadequate representations of reality, and it would be unrealistic to think that the same might not be true of LQG. Questions linger about the nature of time and the interpretation of the hamiltonian constraint, among other things. What is the value then, in studying LQG? Perhaps LQG will eventually be shown to be untenable, or perhaps it will be entirely vindicated. As authors of this paper, we feel that the truth will probably lie somewhere in the middle, and that however much of our current theories of LQG survive over the next few decades, this research program does provide strong indications about what some future (and, we hope, experimentally validated) theory of ``Quantum Gravity'' will look like.

\begin{acknowledgments}
DV would like to thank SBT for invitations to visit the Perimeter Institute in Fall 2009, where this collaboration was born, and to visit the University of Adelaide in August, 2011 where this project was continued. DV also thanks the Perimeter Institute and the University of Adelaide for their hospitality during these visits. SBT would like to thank the Ramsay family for their support through the Ramsay Postdoctoral Fellowship. Finally, a special note of thanks is due to Martin Bojowald, without whose help this work would not have stayed true to its claim of being ``self-contained''.
\end{acknowledgments}

\appendix

%
%
\section{Conventions}\label{app-sec:index_conventions}

Uppercase letters $I,J,K,\ldots \in \{0,1,2,3\}$ are ``internal'' indices which take values in the $\mf{sl}(2,\mathbb{C})$ Lorentz lie-algebra. Greek letters $\mu,\nu,\alpha,\beta \in \{0,1,2,3\}$ are four-dimensional spacetime indices. Lowercase letters $i,j,k,\ldots $ from the middle of the alphabet will be used for indices in a space of $N$ dimensions, $\mf{su}(2)$ lie-algebra indices, etc. Lowercase letters $a,b,c,\ldots \in \{1,2,3\}$ from the start of the alphabet are three-dimensional spatial indices.

\subsection{Lorentz Lie-Algebra}\label{app-subsec:gamma-matrices}

The generators of the $n$-dimensional representation of the Lorentz Lie algebra can be written in terms of the ($n\times n$) Dirac gamma matrices $\{\gamma^I\}$, which satisfy the anticommutation relations
\begin{equation}
		\left\{ \gamma^I,\gamma^J \right\} = 2 g^{IJ} \times \mb{1}_{n\times n}
		\label{eqn:dirac_anticommut}
\end{equation}
where $g^{IJ}$ is the metric tensor and $\mb{1}_{n\times n}$ is the identity matrix. 

For the case of $n=4$, the matrices are given by
\begin{align}\label{eqn:gamma_matrices}
    \gamma^{0}=\left(\begin{array}{cc}0&1\\-1&0\end{array}\right), \quad
    \gamma^{i}=\left(\begin{array}{cc}0&\sigma^{i}\\\sigma^{i}&0\end{array}\right)
\end{align}
where $i,j,k \in \{1,2,3\}$ and $\sigma^i$ are the usual Pauli matrices, and in this case $g^{IJ}$ is equivalent to $\eta^{IJ} = \mathrm{diag}(-1,1,1,1)$, the usual Minkowski metric.

In terms of the $\{\gamma^\mu\}$, the generators of the Lorentz group $SO(3,1)$ can be written as \cite{PeskinSchroeder}
\begin{equation}
	T^{IJ} = \frac{i}{4}\left[ \gamma^I, \gamma^J \right]
\end{equation}
Note that, whereas in the above we have restricted ourselves to the case of $3+1$ dimensions, the expression for the generators of the Lorentz group goes through in any dimension, with either Lorentzian or Euclidean metric \cite[Section 3.2]{PeskinSchroeder}. An $\mf{so}(3,1)$-valued connection can then be written as
\begin{equation}
	\mb{A}_\mu = A_\mu{}^{IJ} T_{IJ} = \frac{i}{4} A_\mu{}^{IJ} \left[ \gamma_I, \gamma_J \right]
\end{equation}
but by the antisymmetry of the gamma matrices, the above expression can be shortened to $\mb{A}_\mu = \frac{i}{2} A_\mu{}^{IJ} \gamma_I \gamma_J $, where we remember that the connection is antisymmetric in the internal indices $A^{IJ} = -A^{JI}$. 

\section{Lie Derivative}\label{app-sec:lie-derivative}

The Lie derivative $\pounds_X$ of a tensor $T$ is the change in $T$ evaluated along the flow generated by the vector field $\vec{X}$ on a manifold. When $T$ is simply a function $T \equiv f$ on the manifold, the Lie derivative reduces to the directional derivative of $f$ along $X$\footnote{this fact is related to the interpretation of the differential $dx$ as a component of a 1-form, and the derivative operator $\partial_x = \partial/\partial x$ as a component of a vector field}:
\[ \pounds_X T \equiv X^a \partial_a f = \frac{\partial}{\partial \gamma} f(\gamma) \]
where $\gamma$ co-ordinatizes the points along the curve genereted by $X$. When the connection is torsion-free, we may replace $\partial_\alpha$ with $\nabla_\alpha$.

It can be shown ~\cite{Wald1984General} that:
\beq
    \pounds_{X} T^{\mu_1 \ldots \mu_n}_{\,\,\nu_1\ldots\nu_m} = X^\alpha \nabla_\alpha T^{\mu_1 \ldots \mu_n}_{\,\,\nu_1\ldots\nu_m} - \sum_{i=1}^{n} T^{\ldots \alpha\ldots}_{\,\,\nu_1\ldots\nu_m} \nabla_\alpha X^{\mu_i} + \sum_{i=1}^{m} T^{\mu_1 \ldots \mu_n}_{\,\,\ldots\alpha\ldots} \nabla_{\nu_i} X^\alpha 
    \label{eqn:LieDeriv1}
\eeq
where $\ldots\alpha\ldots$ is shorthand for an expression with $\alpha$ in the $i^\th$ position and $\mu$(s) or $\nu$(s) elsewhere, \eg $\mu_1 \ldots \mu_{i-1} \, \alpha \, \mu_{i+1}\ldots \mu_n$. In particular the Lie derivative of a vector field $T^\mu$ along a vector field $X^\nu$ reduces to the commutator of the two vector fields:
\beq 
\quad \pounds_X T^\mu = X^\alpha \nabla_\alpha T^\mu - T^\alpha \nabla_\alpha X^\mu \equiv \left[ X, T \right] \label{eqn:LieDeriv2}
\eeq

In the case of a rank-2 tensor $T^{\mu\nu}$:
\beq 
    \pounds_X T_{\mu\nu} = X^\alpha \nabla_\alpha T_{\mu\nu} + T_{\alpha\nu} \nabla_\mu  X^\alpha + T_{\mu\alpha} \nabla_\nu X^\alpha \label{eqn:LieDeriv3}
\eeq
Applying this to the metric tensor $g_{\mu\nu}$ we find the relation:
\beq
\pounds_X g_{\mu\nu} = \nabla_{\mu} X_{\nu} + \nabla_{\nu} X_{\mu}\label{eqn:LieDeriv4} 
\eeq
since the covariant derivative of the metric vanishes.

\section{ADM Variables}\label{app-sec:adm-variables}

One would like to be able to determine the data required to embed the spatial hypersurfaces $\Sigma$ within the 4-manifold $\mc{M}$, given the spacetime metric $g_{ab}$ \& the unit time-like vector field $n^a$ normal to $\Sigma$. This data consists of the intrinsic \& extrinsic curvature tensors $(h_{ab},k_{ab})$. As explained in the main text the object defined by \ref{eqn:intrinsic_curv} plays the role of the intrinsic metric (or ``curvature'') of $\Sigma$. The quantity $k_{ab}$ is the \emph{extrinsic curvature} of $\Sigma$ determined by the particular form its embedding in $\mc{M}$. In order to define $k_{ab}$ we first need to determine the form of the covariant spatial derivative.

\subsection{Covariant Spatial Derivative}\label{app-subsec:spatial-deriv}

The covariant spatial derivative $D_a$ on $\Sigma$ acting on purely spatial object is given by \cite[Sec.~3.2.2.2]{Bojowald2011Canonical}:
\begin{equation}\label{eqn:spatial_deriv1}
 D_a T_{b_1 \ldots b_i}{}^{c_1 \ldots c_j} = h_a^{a'} h_{b_1}{}^{b'_1} \ldots h_{b_i}{}^{b'_i} h^{c_1}{}_{c'_1} \ldots h^{c_j}{}_{c'_j} \nabla_{a'} T_{b'_1 \ldots b'_i}{}^{c'_1 \ldots c'_j}
\end{equation}
where $T_{b_1 \ldots b_i}{}^{c_1 \ldots c_j}$ is an arbitrary spacetime tensor. The \emph{spatial} derivative of an \emph{arbitrary} vector field $n_a$ can then be written as:
\beq\label{eqn:spatial_deriv2}
D_a n_b = h_a{}^c h_b{}^d \nabla_c n_d = (g_b{}^d + n_b n^d) h_a{}^c \nabla_c n_d = h_a{}^c \nabla_c n_b
\eeq
using the fact that $n^d \nabla_c n_d = 1/2 \nabla_c (n_d n^d) = (1/2) \nabla_c (-1) = 0$ because $n_a$ is a unit vector $n^a n_a = -1$.

There is nothing mysterious about \ref{eqn:spatial_deriv2}. It simply measures how the vector field $n^a$ changes from point to point as we move around the spatial manifold $\Sigma$. To help visualize this one can think of an arbitrary configuration of the electric field $\vect{E}$ in three-dimensional space ${}^3\Sigma$. For simplicity, if ${}^3\Sigma$ is $\mbb{R}^3$ and ${}^2\Sigma \subset {}^3\Sigma$ is the surface $z=0$, then the three-dimensional derivative operator $\vect{\nabla} = (\partial_x, \partial_y, \partial_z)$ on $\mbb{R}^3$ reduces to the two-dimensional derivative $\vect{D} = (\partial_x, \partial_y)$ on the $xy$ plane. $D_a E_b$ tells us how $\vect{E}$ changes as we move from one point to another in ${}^2\Sigma$.

\subsection{Extrinsic Curvature}\label{app-subsec:extrinsic-curv}

The extrinsic curvature of a given manifold is a mathematical measure of the manner in which it is \emph{embedded} in a manifold of higher dimension. As illustrated in \ref{fig:extrinsic-curv}, a two-dimensional cylinder embedded in $\mbb{R}^3$ has zero instrinsic curvature, but non-zero extrinsic curvature. The normal at each point of the cylinder is a three-dimensional vector $n_b$ and this vector changes as one moves around the cylindrical surface if the extrinsic curvature of the surface is non-zero. Thus, the simplest definition for a tensorial quantity which measures this change is given by:
\beq\label{eqn:extrinsic-curv-a}
k_{ab} = D_a n_b = h_a{}^c h_b{}^d \nabla_c n_d
\eeq
where $D_a$ is the covariant spatial derivative defined in \ref{app-subsec:spatial-deriv}. This quantity turns out to be symmetric. In order to see this (\cite[Sec.~3.2.2.2]{Bojowald2011Canonical}), note that given two spatial vector fields $Y^a$ and $Z^a$, their commutator $[Y,Z]^a = Y^b \nabla_b Z_a - Z^b \nabla_b Y_a$ will also be spatial, \ie $n_a [Y,Z]^a = 0$. This implies:
\[ n_a [Y,Z]^a = n_a (Y^b \nabla_b Z^a - Z^b \nabla_b Y^a) = - Z^a Y^b \nabla_b n_a + Z^b Y^a \nabla_b n_a = Y^a Z^b (\nabla_b n_a - \nabla_a n_b) = 0 \]
using the fact that since $n_a Y^a = 0$, $n_a \nabla_b Y^a = - Y^a \nabla_b n_a $ and similarly for the remaining term. And since $Y^a, Z^a$ are purely spatial, this implies that (the spatial projection of) $\nabla_a n_b = \nabla_b n_a$.

Thus the extrinsic curvature of ${}^3 \Sigma$ can be written as:
\beq\label{eqn:extrinsic-curv-b}
k_{ab} = \frac{1}{2} (D_a n_b + D_b n_a)
\eeq

\subsection{Canonical Momentum in ADM Formulation}\label{app-subsec:canonical-momentum}

Recall that the time vector field is written in terms of the lapse $N$, shift $N^{\mu}$ and the normal to the hypersurface $n^\mu$, so that $t^\mu = N n^\mu + N^\mu$ (eq.~(\ref{eqn:time-evolution})). We wish to write down the explicit form of the Lie-derivative of a one-index $X_a$ and two-index object $h_{ab}$, with respect to a vector field $v_a$. Conveniently this is already present in equations~(\ref{eqn:LieDeriv2})...(\ref{eqn:LieDeriv4})! As we may expect, when a vector field is a sum of two (or more) vector fields (as for the time-evolution field above), the Lie derivative with respect to that field decomposes into the sum of Lie derivatives with respect to each of the components fields. So if $X_a = u_a + v_a + w_a$, then $\pounds_{X} [T] = \pounds_{u} [T] + \pounds_{v} [T] + \pounds_w [T]$, where $T$ is the arbitrary tensor whose Lie derivative we want to find. You can see this directly from eq.~(\ref{eqn:LieDeriv1}) by writing the field X as a sum of other vector fields. When, $T$ is a vector, then $\pounds_{X} T = [X,T] = [u,T] + [v,T] + [w,T]$ and so on ($[A,B]$ is the commutator of two vector fields as in eq.~(\ref{eqn:LieDeriv2})).

There are two steps involved in deriving the form of the canonical momentum. First is to prove the identity \eqref{eqn:dot_h_ab}. The second is to use that result to perform the functional derivative of the Einstein-Hilbert Lagrangian $L_{EH}$ w.r.t. the $\dot h_{ab}$ to obtain eq.~\eqref{eqn:canonical-momentum}.

First, we wish to show that $\pounds_{\vec{t}} h_{ab} = 2 N k_{ab} + \pounds_{\vec{N}} h_{ab}$, which we can do by finding a suitable expression for $\pounds_{\vec{t}} h_{\mu\nu}$, and then restricting the indices to the range $\mu,\,\nu \rightarrow a,\,b \in \{1,2,3\}$. So, since $t_\mu = N n_\mu + N_\mu$, using the above mentioned additive property of Lie derivatives, we have $\pounds_{\vec{t}} h_{\mu\nu} = \pounds_{N \vec{n}} h_{\mu\nu} + \pounds_{\vec{N}} h_{\mu\nu}$. The second term is present unchanged in eq.~(\ref{eqn:dot_h_ab}). Now it remains to be shown that $2N k_{ab} = \pounds_{N \vec{n}} h_{ab}$.

In the following we follow the treatment of \cite[Sec.~3.2.2.2]{Bojowald2011Canonical}. First we show that $\pounds_{\vec{n}} h_{ab} = 2 k_{ab}$. Using the definition of the Lie derivative, we have:

\begin{align}\label{eqn:Lie_deriv_h}
 \pounds_{\vec{n}} h_{ab} & = n^c \nabla_c h_{ab} + h_a{}^c \nabla_b n_c + h_b{}^c \nabla_a n_c \nonumber \\
 & = n^c \nabla_c (g_{ab} + n_a n_b) + h_a{}^c \nabla_b n_c + h_b{}^c \nabla_a n_c \nonumber \\ 
			  & = n^c \nabla_c (n_a n_b) + (g_a{}^c + n_a n^c) \nabla_b n_c + (g_b{}^c + n_b n^c) \nabla_a n_c \nonumber \\
			  & = n^c \nabla_c (n_a n_b) + \nabla_b n_a + \nabla_a n_b \nonumber \\
			  & = n_a n^c \nabla_c n_b + n_b n^c \nabla_c n_a + g_a{}^c \nabla_c n_b + g_b{}^c \nabla_c n_a \nonumber \\
			  & = (g_a{}^c + n_a n^c) \nabla_c n_b + (g_b{}^c + n_b n^c) \nabla_c n_a \nonumber \\
			  & = h_a{}^c \nabla_c n_b + h_b{}^c \nabla_c n_a \nonumber \\
			  & = D_a n_b + D_b n_a = 2 k_{ab}
\end{align}
where in the second and third lines we have used the expression for the spatial metric in terms of the four-metric and the unit normal to $\Sigma$. In the third line the term containing $n^c \nabla_a n_c$ is zero. In the third and fourth lines metric compatibility ($\nabla_a g_{bc} = 0$) is used to commute the 4-metric through the spacetime derivative. In the fifth line we have again used metric compatibility to write $\nabla_a n_b$ as $g_a{}^c \nabla_c n_b$, \etc.

For the last step we have need 
which finally yields:
\beq\label{eqn:CanMomentum1}
\pounds_{\vec{n}} h_{ab} = D_a n_b + D_b n_a = 2 k_{ab}
\eeq

All this algebra is not necessary if one notes that the Lie derivative of a \emph{metric tensor} is given by \ref{eqn:LieDeriv4}, which we restate for convenience:
\beq
	\pounds_{\vec{n}} h_{ab} = \nabla_{\mu} n_{\nu} + \nabla_{\nu} n_{\mu} \nonumber
\eeq
Now, note that the above equation holds true only when the derivative operator $\nabla_\mu$ is compatible ($\nabla_{\mu} h_{\mu\nu} = 0$) with the metric $h_{ab}$ whose Lie derivative we wish to determine. Hence we should use the correct notation $D$ for the spatial derivative operator instead of $\nabla$. Then we have:
\beq
	\pounds_{\vec{n}} h_{ab} = D_{a} n_{b} + D_{b} n_{a} = 2 k_{ab}
\eeq
which, by definition \ref{eqn:extrinsic-curv-b}, is twice the extrinsic curvature of ${}^3 \Sigma$. Given this expression we proceed as follows:
\begin{align}\label{eqn:CanMomentum2}
	2 k_{ab} = \pounds_{\vec{n}}{h_{ab}} & = n^c \nabla_c h_{ab} + h_{ac} \nabla_b n^c + h_{bc} \nabla_a n^c \nonumber \\
	& = \frac{1}{N} \left( N n^c \nabla_c h_{ab} + N h_{ac} \nabla_b n^c + N h_{bc} \nabla_a n^c \right) \nonumber \\
	& = \frac{1}{N} \left( N n^c \nabla_c h_{ab} + h_{ac} \nabla_b (N n^c) + h_{bc} \nabla_a (N n^c) \right) \nonumber \\
	& = \frac{1}{N} \pounds_{\vec{t}-\vec{N}} h_{ab} = \frac{1}{N} \left( \pounds_{\vec{t}} h_{ab} - \pounds_{\vec{N}} h_{ab} \right) \nonumber \\
	& = \frac{1}{N} h_a{}^c h_b{}^d \left( \pounds_{\vec{t}} h_{cd} - \pounds_{\vec{N}} h_{cd} \right) \nonumber \\
	& = \frac{1}{N} \left( \dot h_{ab} - D_a N_b - D_b N_a \right)
\end{align}
where in the first line we have used \ref{eqn:extrinsic-curv-b} alongwith the definition \ref{eqn:LieDeriv3} of the Lie derivative. In the second we have multiplied \& divided by the scale factor $N$. In the third we have used the fact that $n^c h_{ac} = 0$ to move $N$ inside the derivative operator. In going from the third to the fourth, we have used the relationship between the lapse, shift and time-evolution fields: $N n^a = t^a - N^a$, followed by \ref{eqn:LieDeriv3} (in reverse) and then used the linearity of the Lie derivative. In the fifth we have, in the words of Bojowald \cite{Bojowald2011Canonical}, ``smuggled in'' two factors of $h$ knowing that $k_{ab}$ is spatial to begin with. In the sixth, the spatial projection $h_a{}^c h_b{}^d \pounds_t h_{cd} = \dot h_{ab}$ is identified as the ``time-derivative'' of the spatial metric. We leave the remaining step (to show that $h_a{}^c h_b{}^d \pounds_{\vec{N}} h_{cd} = D_a N_b + D_b N_a$ ) as an exercise for the reader.

To summarize, we have:
\begin{equation}\label{eqn:extrinsic_inverse}
	k_{ab} = \frac{1}{2N}\left[ \pounds_{\vec{t}} h_{ab} - D_{(a} N_{b)} \right] = \frac{1}{2N}\left[ \dot h_{ab} - D_{(a} N_{b)} \right]
\end{equation}
Now, the Einstein-Hilbert Lagrangian is given by:
\[
	L_{EH} = N\sqrt{h} \left[ {}^{(3)}R + k^{ab} k_{ab} - k^2 \right]
\]
The first term does not contain any dependence on $k_{ab}$ or $N_a$ and so its derivative w.r.t. $\dot h_{ab}$ vanishes. For the remaining two terms we have:
\begin{equation}
	\frac{\delta L_{EH}}{\delta \dot h_{ef}} = N \sqrt{h} \left[ k^{ab} \frac{\delta k_{ab}}{\delta \dot h_{ef}} + k_{ab} \frac{\delta k^{ab}}{\delta \dot h_{ef}} - 2 k \frac{\delta k}{\delta \dot h_{ef}} \right] \nonumber
\end{equation}
here $k = h^{ab} k_{ab}$. $k^{ab}$ can be written as $h^{ac} h^{ac} k_{bd}$:
\begin{equation}
	\frac{\delta L_{EH}}{\delta \dot h_{ef}} = N \sqrt{h} \left[ k^{ab} \frac{\delta k_{ab}}{\delta \dot h_{ef}} + k_{ab} h^{ac} h^{bd} \frac{\delta k_{cd}}{\delta \dot h_{ef}} - 2 \, k \,h^{ab}\frac{\delta k_{ab}}{\delta \dot h_{ef}} \right]
\end{equation}
From \eqref{eqn:extrinsic_inverse} we have:
\begin{equation}
	\frac{\delta k_{ab}}{\delta \dot h_{ef}} = \frac{1}{2N} \delta^e_a \delta^f_b
\end{equation}
Inserting this into the previous expression we have:
\begin{align}
	\frac{\delta L_{EH}}{\delta \dot h_{ef}} & = N \sqrt{h} \left[ k^{ab} \frac{1}{2N} \delta^e_a \delta^f_b + k_{ab} h^{ac} h^{bd} \frac{1}{2N} \delta^e_c \delta^f_d - 2 \, k \,h^{ab} \frac{1}{2N} \delta^e_a \delta^f_b \right] \nonumber \\
	& = \sqrt{h} \left[ k^{ef} - k h^{ef} \right ] = \pi^{ef}
\end{align}
which is identical to \eqref{eqn:canonical-momentum} as desired.

\section{Duality}\label{app-sec:duality}

The notion of self-/anti-self-duality of the gauge field $F_{\alpha\beta}$ is central to understanding both the topological sector of Yang-Mills theory and the solutions of Einstein's equations in the connection formulation. As discussed in sec.~\ref{sec:k-vectors-k-forms}, the use of multivectors and $k$-forms can be very helpful for understanding duality. Let us review these concepts. 

\subsection{Multivectors and Differential Forms}\label{app-subsec:diff-forms}
A vector is a directed line segment with a magnitude which is interpreted as a length. One way to form the product of two vectors $u$ and $v$ is the dot product $u\cdot v$, which is a scalar that is maximised when the vectors are parallel. We can also form the wedge product, $u\wedge v$, which is a directed surface spanned by $u$ and $v$ (the direction being both an orientation in space and a preferred direction of rotation around the boundary of the surface), with a magnitude interpreted as the area of the surface. This area is called a bivector, and its magnitude is maximised when $u$ and $v$ are perpendicular (and zero when they are parallel). The wedge product of three non-coplanar vectors is a trivector, which is a parallelipiped with a direction (a preferred directed path around the edges of the parallelipiped) and a magnitude interpreted as its volume. The wedge product of $k$ vectors (assuming they are not parallel, coplanar, etc.) will in general be called a multivector, being an oriented parallelipiped in $k$ dimensions, with a magnitude given by its enclosed volume. A scalar may be regarded as a $0$-vector. We can define the Clifford product of multivectors as 
\beq
  uv = u\cdot v + u \wedge v
\eeq 
If $u$ and $v$ are ordinary vectors, then if $u$ and $v$ are perpendicular $uv=u\wedge v$ since in this case $u\cdot v=0$. Hence when dealing with orthonormal basis vectors we may adopt the notation $e_i e_j = e_i\wedge e_j = e_{ij}$, and likewise $e_i e_j e_k = e_i\wedge e_j\wedge e_k = e_{ijk}$, etc. Conversely if $u$ and $v$ are parallel then $uv=u\cdot v$ since in this case $u\wedge v=0$, hence $e_i e_i = e_i\cdot e_i = 1$. 

The importance of multivector quantities in physics can be seen if we consider the case of four-dimensional Minkowski spacetime, where the scalar product is taken using the metric $\eta_{\mu\nu}$. Hence $e_0 e_0 = -1$, and $e_1 e_1 = e_2 e_2 = e_3 e_3 = +1$. In this case the basis vectors are isomorphic to the Dirac gamma matrices, $\gamma_\mu$, and the reader can verify that they satisfy $\{e_\mu,\,e_\nu\}=2\eta_{\mu\nu}$, the defining relation of the Dirac matrices (see eq.~(\ref{eqn:dirac_anticommut})). Since this anticommutator is formed by taking Clifford products of the $e_\mu$, the gamma matrices are said to generate a representation of a Clifford algebra.

Differential forms and multivectors can be seen to correspond closely. A bivector and a 2-form both define a plane. A trivector and a 3-form both define a volume, etc. However as mentioned in sec.~\ref{sec:k-vectors-k-forms}, multivectors can be easier to visualise, as the magnitude of a $k$-form is a density, while the magnitude of a multivector is a $k$-dimensional volume. It can therefore often be easier to think of how the wedge products and duals of $k$-forms behave by visualising them as multivectors. 

Duality is a notion that emerges naturally from the construction of the space of multivectors, and likewise from the construction of the space $ \oplus_{k=0}^n {}^n \Lambda_k$ of differential forms on a $n$-dimensional manifold $M$. Let ${}^n\Lambda_k$ denote the subspace consisting only of forms of order $k$ in $n$ dimensions \emph{e.g.} in three dimensions the space of two-forms ${}^3\Lambda_2$ is spanned by the basis $\left\{ dx^1 \wedge dx^2, dx^2 \wedge dx^3, dx^3 \wedge dx^1 \right\}$ where $\{ x^1, x^2, x^3 \}$ is some local coordinate patch - \ie a mapping from a portion of the given manifold to a region around the origin in $\mathbb{R}^3$.

Now one can show \cite{Baez1994Gauge,Wald1984General} that $ {}^n \Lambda_k = {}^n \Lambda_{n-k} $, \ie the space of $k$-forms is the same as the space of ($n-k$)-forms. Thus any $k$-form $F_{a_1 a_2 \ldots a_k}$, defined on an $n$ dimensional manifold, can be mapped to an ($n-k$)-form $ (\star F_{a_1 a_2 \ldots a_{n-k}})$. This is accomplished with the completely antisymmetric tensor $\epsilon_{x_1 \ldots x_n}$ on $M$:
\begin{equation}
	(\star F)^{a_1 \ldots a_{n-k}} = \frac{1}{(n-k)!} \epsilon^{a_1 \ldots a_{n-k}}{}_{a_{n-k+1} \ldots a_n} F^{a_{n-k+1}\ldots a_n}
	\label{eqn:k-form_dual}
\end{equation}

This expression may appear daunting, but as suggested we can make its meaning clearer by examining duality with multivectors.  Consider the case of three dimensions. The antisymmetry of the wedge product means that the unit trivector $e_{ijk} = e_i e_j e_k$ picks up a factor of $-1$ each time the order of any two of its factors is swapped, hence $e_{ijk} = -e_{ikj}$, etc. and so the unit trivector is a geometrical representation of the antisymmetric tensor $\epsilon_{ijk}$. 
\begin{figure}[t]
    \centering
    \includegraphics[scale=0.3]{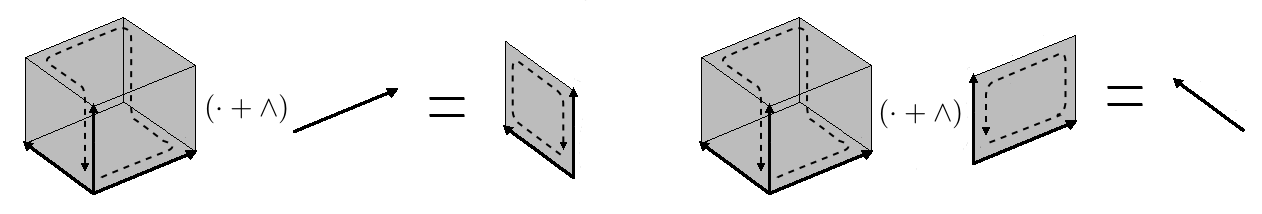}
    \caption{The unit trivector $e_{123}$ allows us to explore duality in three dimensions. When we take the Clifford product, indicated here by $(\cdot+\wedge)$, of the unit trivector with a vector, the part of $e_{123}$ parallel to the vector yields a scalar factor via the dot product, and a factor of zero via the wedge product part. This leaves us with a bivector perpendicular to the original vector (left). Likewise the Clifford product of $e_{123}$ with a bivector yields a vector (right). In each case the bivector and vector are dual to each other, since each spans the directions the other doesn't. Duality is therefore an extension of the concept of orthogonality. For a four-dimensional object, the dual would be taken with $e_{1234}$, the dual of a vector would be a trivector, and the dual of a bivector would be another bivector.}
        \label{fig:multivector_dual}
\end{figure}
Multiplying a vector by the unit trivector yields a bivector, and multiplying a bivector by the unit trivector yields a vector (Figure~\ref{fig:multivector_dual}). To see why, consider the familiar cross product. Any two, non-degenerate, vectors $a, b \in \mathbb{R}^3$ span a two-dimensional subspace of $\mathbb{R}^3$. Using these two vectors we construct a third vector $c=\vect{a} \times \vect{b}$, where the components of $c$ are given by $c^i = \epsilon^i{}_{jk} a^j b^k$. This construction is taught to us in elementary algebra courses, but never quite seemed to make complete sense because it seemed to be peculiar to three-dimensions. The product $a \times b$ is a vector which is perpendicular to the plane defined by the vectors $a$ and $b$. But this plane is the same one that the wedge product $a\wedge b$ lies in. If we take the Clifford product of $a\wedge b$ with the unit 3-vector, $e_1\wedge e_2\wedge e_3=e_{123}$ we are left with a vector that is perpendicular to the plane of $a\wedge b$, and which equals $-(a\times b)$. Why? Because the components of the vector (bivector) parallel with components of the unit trivector yield scalars, leaving only the components perpendicular to $a\wedge b$, as we can see by expanding the Clifford product in full 
\beqar
 (e_1\wedge e_2\wedge e_3)(a\wedge b) & = & (e_{123})((a^1 e_1+a^2 e_2+a^3 e_3)\wedge (b^1 e_1+b^2 e_2+b^3 e_3)) \nonumber \\
  & = & (e_{123})((a^1 b^2 - a^2 b^1) e_{12} + (a^1 b^3 - a^3 b^1) e_{13} + (a^2 b^3 - a^3 b^2) e_{23}) \nonumber \\ 
  & = & (a^1 b^2 - a^2 b^1) e_{12312} + (a^1 b^3 - a^3 b^1) e_{12313} + (a^2 b^3 - a^3 b^2) e_{12323} \nonumber \\
  & = & (a^1 b^2 - a^2 b^1) (-e_3) + (a^1 b^3 - a^3 b^1) e_2 + (a^2 b^3 - a^3 b^2) (-e_1) \nonumber \\
  & = & - a\times b 
\eeqar
where in the second-last line we have used $e_i\wedge e_j = - e_j\wedge e_i$ to rearrange the basis vector terms, so as to eliminate extra terms by using $e_i e_i = e_i\cdot e_i = 1$. We also find that the wedge product of $a$ and $b$ has components $(a \wedge b)_{ij} = a_{[i}b_{j]}$.



This allows us to view the cross product as a three-dimensional special case of a procedure that can be performed in any number of dimensions. This procedure is ``forming the dual''. We can say that the cross product of two vectors is the dual of the wedge product, $(a\times b) = \star(a\wedge b)$. In the language of differential forms this procedure is described by eq.~(\ref{eqn:k-form_dual}), and utilises the antisymmetric tensor $\epsilon^{a_1 a_2\ldots a_n}$. In the language of multivectors, it involves taking the Clifford product with the unit multivector $e_1 e_2\ldots e_n = e_{12\ldots n}$. 

From now on we will speak of $k$-forms, rather than $k$-forms and/or multivectors. But their equivalence, and the geometric interpretation arising from this, should be kept in mind.

\subsection{Spacetime Duality}\label{app-subsec:spacetime}
From the discussion in sec.~\ref{app-subsec:diff-forms}, it should be obvious that in four dimensions the dual of any two-form is another two-form
\begin{equation}
{\star F}_{\alpha \beta} = \frac{1}{2} \epsilon_{\alpha \beta}{}^{\mu \nu} F_{\mu\nu} \end{equation}
(compare this with eq.~(\ref{eqn:dualfmunu_components}), and as noted there, the quantity defined on the plane between any pair of spacetime axes is associated to the quantity defined on the plane between the other two spacetime axes). It is due to this property of even-dimensional manifolds that we can define \emph{self-dual} and \emph{anti-self-dual} $k$-forms, where a form is self-/anti-self-dual if:
\beq\label{eqn:euclideandual}
	{\star F} = \pm F
\eeq

Given an arbitrary 2-form $G_{\mu\nu}$ its self-dual part $G^+$ and anti-self-dual part $G^-$ are given by 
$$ G^{+} = \frac{G + \star G}{2 \alpha} \qquad G^- = \frac{G - \star G}{2 \beta} $$
where $\alpha$ and $\beta$ are constants we have introduced for later convenience. We can check that
\beq
 \star(G\pm\star G) = \pm(G\pm\star G)
\eeq 
because $ \star \star = \mathbf{1}$ in a Euclidean background. In other words $\star G^+ = G^+$ and $\star G^- = - G^- $, which is precisely the definition of (anti-)self-duality. Thus any 2-form can always be written as a linear-sum of a self-dual and an anti-self-dual piece
$$ G = \alpha G^+ + \beta G^- \qquad \star G = \alpha G^+ - \beta G^- $$

The above results hold for a Euclidean spacetime. For a Lorentzian background we would instead have $ \star \star = \mathbf{-1}$ and the dual of a two-form is given by:
\begin{equation}
 \star F_{\alpha\beta} = \frac{i}{2}\epsilon_{\alpha\beta}{}^{\gamma\delta} F_{\gamma \delta}
\end{equation}
and the statement of (anti-)self-duality becomes
\beq\label{eqn:lorentziandual}
	\star F = \pm i F
\eeq
with the self-dual and anti-self-dual pieces of a two-form $G$ given by $G^+ = (G + \star i G )/2\alpha$ and $G^- = (G - \star i G )/2\beta$


\subsection{Lie-algebra duality}\label{app-subsec:lie-duality}

The previous section discussed self-duality in the context of tensors with spacetime indices $T^{\alpha\beta\ldots}{}_{\gamma\delta\ldots}$. In gauge theories based on non-trivial Lie-algebras we also have tensors with lie-algebra indices, such as the curvature $F_{\mu\nu}{}^{IJ}$ of the gauge connection $A_\mu{}^{IJ}$ where $I,J$ label generators of the relevant Lie algebra.. The dual of the connection can then be defined using the completely antisymmetric tensor acting on the Lie algebra indices, as in:
\begin{equation}
	\star A_\mu{}^{IJ} = \frac{1}{2} \epsilon^{IJ}{}_{KL} A_\mu{}^{KL}
\end{equation}

\subsection{Yang-Mills}\label{app-subsec:yangmills}

Let us illustrate the utility of the notion of self-duality by examining the classical Yang-Mills action:
$$ S_{YM} = \int_{R^4} {\mathrm{Tr}} \left[ F \wedge \star F \right] $$
Varying this action with respect to the connection \href{http://en.wikipedia.org/wiki/Instanton#Yang.E2.80.93Mills_theory}{gives us} the equations of motion:
$$ d F = 0 \,\,; \quad d \star F = 0 $$
which are satisfied if $ F = \pm \star F$, \ie if the gauge curvature is self-dual or anti-self-dual. Thus for self-/anti-self-dual solutions the Yang-Mills action reduces to:
$$ S^{\pm}_{YM} = \pm \int_{R^4} {\mathrm{Tr}} \left[ F \wedge F \right] $$
which is a topological invariant of the given manifold and is known as the \emph{Pontryagin index}. Here the $\pm$ superscript on the r.h.s. denotes whether the field is self-dual or anti-self-dual. 

\subsection{Geometrical interpretation}

Given any (Lie-algebra valued) two-form $F^{I}_{ab}$ (where $I, J, K \ldots$ are Lie-algebra indices) we can obtain an element of the Lie-algebra by contracting it with a member of the basis of the space of two-forms: $\{dx^i \wedge dx^j \}$ where $x^i$ denotes the $i^\textrm{th}$ vector and \textbf{not} the components of a vector. The components are suppressed in the differential form notation as explained in the preceding sections. The resulting lie-algebra element is
$$ \Phi^I{} = F^I{}_{ab}\, dx^a \wedge dx^b $$ 
and $ \Phi^I{}$ is the \emph{flux} of the field strength through the two-dimensional surface spanned by $\{dx^a, dx^b\}$.

We can also define
$$ \star\Phi^I = \star F^I{}_{ab}\, dx^a \wedge dx^b = \frac{1}{2}\epsilon_{ab}{}^{cd} F^I{}_{cd}\, dx^a \wedge dx^b$$
which implies that $\star\Phi^I{}_{ab} = \frac{1}{2}\epsilon_{ab}{}^{cd} \Phi^I{}_{cd} $, \ie the flux of the field strength through the $ab$ plane is equal to the flux of the \emph{dual} field through the $cd$ plane.

\subsection{(Anti) Self-dual connections}
When we say that the connection is (anti-)self-dual, explicitly this means that
\beq
A_\mu^{IJ} = \pm\star{A_\mu}^{IJ} = \pm\frac{i}{2}\epsilon^{IJ}{}_{KL}A_\mu{}^{KL}
\eeq 
Let us now show the relation between the (anti-)self-dual four-dimensional connection and its restriction to the spatial hypersurface $\Sigma$. We begin by writing the full connection in terms of the generators $\{\gamma^I\}$ of the Lorentz lie-algebra: ${}^\pm\vect{A} := A_\mu^{IJ}\gamma_I\gamma_J$ and expanding the sum (see \cite[Section 2]{Alexander2006Gravity} and \ref{app-subsec:gamma-matrices}):
\begin{eqnarray}\label{eqn:connection_decomp}
	A_\mu^{IJ}\gamma_I\gamma_J & = & A_\mu^{i0}\gamma_i\gamma_0 +
    A_\mu^{0i}\gamma_0\gamma_i + A_\mu^{ij} \gamma_i\gamma_j \nonumber \\
    & = & 2A_\mu^{0i}\gamma_0\gamma_i+ A_\mu^{ij}\gamma_i\gamma_j \nonumber  \\
    & = & 2A_\mu^{0i}\left(\begin{array}{cc}\sigma_i&0\\0&-\sigma_i\end{array}\right)
    + iA_\mu^{jk}\epsilon^{ijk}\left(\begin{array}{cc}\sigma_i&0\\0&\sigma_i\end{array}\right)
\end{eqnarray}
In the second line we have used the fact that $A_\mu^{IJ}$ is antisymmetric in the internal indices and that the gamma matrices anticommute. In the third we have used the expressions for the gamma matrices given in Appendix \ref{app-subsec:gamma-matrices} to expand out the matrix products. This allows us to write the last line in the above expression in the form
\begin{align}
	\vect{A} = A_\mu^{IJ}\gamma_I\gamma_J = 2i\left(\begin{array}{cc}A_\mu^{i+}\sigma_i&0\\0&A_\mu^{i-}\sigma_i\end{array}\right)
\end{align}
where:
\begin{subequations}
	\begin{align}
	    A_\mu^{i+} & = \frac{1}{2}\epsilon^{ijk}A_\mu^{jk} - iA_\mu^{0i} \\
	    A_\mu^{i-} & = \frac{1}{2}\epsilon^{ijk}A_\mu^{jk} + iA_\mu^{0i}
	\end{align}
\end{subequations}
For $I=0, J \in \{1,2,3\}$, using the definition of the dual connection, we find that
$$ A_\mu{}^{0i} = \frac{i}{2} \epsilon^{0i}{}_{jk} A_\mu{}^{jk} $$
and so we may rewrite these expressions as 
\begin{subequations}
    \begin{align}
	    A_\mu^{i+} & = \frac{1}{2}\left(\epsilon^{ijk} + \epsilon^{0i}{}_{jk}\right)A_\mu^{jk} \\
	    A_\mu^{i-} & = \frac{1}{2}\left(\epsilon^{ijk} - \epsilon^{0i}{}_{jk}\right)A_\mu^{jk}
	\end{align}
\end{subequations}

\section{Path Ordered Exponential}
\label{app:pathordered}
From eq.~(\ref{eqn:schwinger_line_integral}) we see that the effect of a holonomy of a connection along a path $\lambda$ (for either an open or closed path) in a manifold $M$ is defined as
\beq
	\psi_{|(\tau=1)} = \mc{P} \left\{ e^{ \int_\lambda ig d\tau' A_\mu n^\mu } \right\} \psi_{|(\tau=0)} = U_\lambda \, \psi_{|(\tau=0)}
\eeq

The exponential can be formally expressed in terms of a Taylor series expansion:
\beq\label{eqn:pathordering}
	e^{ - \int_\gamma d\tau' A_\mu n^\mu } = \mathbf{1} + \sum_{n=1}^{\infty} \frac{1}{n!} \left\{ \int_{\sigma_0 = 0}^{\sigma_1} \int_{0}^{\sigma_2} \ldots \int_{0}^{\sigma_{n}=1} d\tau_1 d\tau_2 \ldots d\tau_{n} \,\, A(\sigma_n) A(\sigma_{n-1}) \ldots A(\sigma_1) \right\}
\eeq
where for the $n$\supersc{th} term in the sum, the path $\lambda$ is broken up into $n$ intervals parametrized by the variables $\{\tau_1,\tau_2,\ldots,\tau_{n}\}$ over which the integrals are performed. The path ordering enforces the condition that the effect of traversing each interval is applied the order that the intervals occur. The interested reader is referred to pgs. 66 - 68 of \cite{Carroll1997Lecture}.

\section{Peter-Weyl Theorem}
\label{app:peter-weyl}
The crucial step involved in going from graph states with edges labelled by holonomies to graph states with edge labelled by group representations (angular momenta) is the Peter-Weyl theorem . This theorem allows the generalization of the notion of Fourier transforms to functions defined on a group manifold for compact, semi-simple Lie groups.

Given a group $\mc{G}$, let $D^j(g)_{mn}$ be the matrix representation of any group element $g \in \mc{G}$. Then we have (see Chapter 8 of \cite{Tung1985Group}):
\begin{theorem}
\label{thm:orthonormality}
The irreducible represenation matrices $D^j(g)$ for the group $SU(2)$ satisfy the following orthonormality condition
\beq\label{eqn:group-orthonormality}
	\int d\mu(g) D_j^\dag(g)^m{}_n D^{j'}(g)^{n'}{}_{m'} = \frac{n_G}{n_j} \delta^{j'}{}_j \delta^{n'}{}_n \delta^{m'}{}_m
\eeq
\end{theorem}

Here $n_j$ is the dimensionality of the $j$\supersc{th} representation of $G$ and $n_G$ is the \emph{order} of the group. For a finite group this is simply the number of elements of the group. For example, for $\mbb{Z}_2$, $n_G=2$. However a continuous or Lie group such as $SU(2)$ has an uncountable infinity of group elements. In such cases $n_G$ corresponds to the ``volume'' of the group manifold.

This property allows us to decompose any square-integrable function $f(g):\mc{G}\rightarrow \mbb{C}$ in terms of its components with respect to the matrix coefficients of the group representations:
\begin{theorem}
The irreducible representation functions $D^j(g)^m{}_n$ form a complete basis of (Lebesgue) square-integrable functions defined on the group manifold.
\end{theorem}
Any such function $f(g)$ can then be expanded as
\beq\label{eqn:peterweyl-expansion}
	f(g) = \sum_{j;mn} f_j{}^{mn} D^j(g){}_{mn}
\eeq
where $f_j{}^{mn}$ are constants which can be determined by inserting the above expression for $f(g)$ in \ref{eqn:group-orthonormality} and integrating over the group manifold. Thus we obtain
\begin{eqnarray}
  \int d\mu(g) f(g) D^{\dag}_j(g)^{mn} & = & \sum_{j';m'n'} \int d\mu(g) f_{j'}{}^{m'n'} D^{j'}(g){}_{m'n'} D^{\dag}_j(g)^{mn} \nonumber \\
 & = & \sum_{j';m'n'} f_{j'}{}^{m'n'} \frac{n_G}{n_j} \delta^{j'}{}_j \delta^{n'}{}_n \delta^{m'}{}_m
\end{eqnarray}
which gives us
\begin{equation}\label{eqn:peterweyl-coefficents}
  f_j{}^{mn} = \sqrt{\frac{n_j}{n_G}} \int d\mu(g) f(g) D^{\dag}_j(g)^{mn}
\end{equation}

\section{Square-Free Numbers}\label{app-sec:square-free}

According to the fundamental theorem of arithmetic, any integer $d \in \mathbb{Z}$, has a unique factorization in term of prime numbers:
$$ d = \prod_{i=1}^{N} p_i ^ {m_i} $$
where $ \{p_1,p_2,\ldots,p_N\}$ are the $N$ prime-numbers which divide $d$, one or more times. $m_i$ is the number of times the prime number $p_i$ occurs in the factorization of $d$. Thus, we have:
$$ \sqrt{d} = \prod_{i=1}^{N} p_i^{m_i/2} $$
We can partition the set $\{m_i\}$ into two sets containing only the even or odd elements respectively
$$ \{m_i\} \equiv \{m^e_j\} \cup \{m^o_k\} $$
where $j \in 1\ldots n_e$, $k \in 1\ldots n_o$, and $n_e + n_o = N$. This gives:
$$ \sqrt{d} =\left( \prod_{i=1}^{n_e} p_{i}^{\frac{m^e_i}{2}} \right) \left( \prod_{j=1}^{n_o} p_{j} ^{\frac{m^o_j}{2}} \right) $$
Since each of the $m_i^e = 2 a_i^e$ and $m_j^o = 2 b_j + 1$, for some $a_i, b_j \in \mathbb{Z}$, we have:
$$ \sqrt{d} = \left( \prod_{i=1}^{n_e} p_{i}^{a_i^e} \right) \left( \prod_{j=1}^{n_o} p_{j} ^{m^o_j} \right) \sqrt{\prod_{k=1}^{n_o} p_k } = A \sqrt {B} $$
It is evident that since the third term in the product has no repeating elements, its square-root $\sqrt{B}$ cannot be an integer (\ie the presence of repeating elements would lead to an expression like $\sqrt{X\cdot X}$). Such an integer $B$, in therefore known as a square-free integer. Thus any integer $d$ can be written as the product of a square-free integer ($B$) and another (non square-free) integer $C = A^2$ such that $d = C \times B $.

\section{Brahmagupta-Pell Equation}\label{app-sec:pells-eqn}

Around the 7\supersc{th} century A.D. the Indian mathematician \href{http://en.wikipedia.org/wiki/Brahmagupta}{Brahmagupta}, demonstrated the \href{http://en.wikipedia.org/wiki/Brahmagupta%27s_identity}{Brahmagupta-Fibonacci Identity},
\begin{eqnarray}
	(a^2 + nb^2) (c^2 + n d^2) & = & (ac)^2 + n^2 (bd)^2 + n[(ad)^2 + (bc)^2] + 2acbdn - 2acbdn \\
	& = & (ac + nbd)^2 + n(ad - bc)^2
\end{eqnarray}
where we have added and subtracted $2acbdn$ from the \emph{l.h.s.} on the first line. The above goes through for all $n \in \mathbb{Z}$. Given any pair of triples of the form $(x_i, y_i, k_i)$, where $i = 1,2$, which are solutions of the Diophantine equation $x_i^2 - n y_i^2 = k_i^2$, we can construct a third triple $(x_3,y_3,k_3)$, which is also a solution of the same equation, by applying the Brahmagupta-Fibonacci identity to the first two pairs
\begin{equation}
	(x_1^2 - n y_1^2) (x_2^2 - n y_2^2) = (x_1 x_2 - n y_1 y_2)^2 - n (x_1 y_2 - x_2 y_1)^2
\end{equation}
which tells us that $x_3 = x_1 x_2 - n y_1 y_2$, $y_3 = x_1 y_2 - x_2 y_1$ and $k_3 = k_1 k_2$. One can easily check that the triple $\{x_3, y_3, k_3\}$ is also a solution of the Diophantine equation.

When we apply the restriction that $k_i = 1$, the Diophantine equation $x_i^2 - n y_i^2 = k_i^2$ reduces to the Brahmagupta-Pell equation,
$$x_i^2 - n y_i^2 = 1$$
and given two pairs of solutions $\{(x_i,y_i) , (x_j, y_j) \}$ to the BP equation (for the same fixed value of $n$), we can generate a third solution given by $ (x_k, y_k) = ((x_i x_j - n y_i y_j),(x_i y_j - x_j y_i))$. More generally given any solution $(x_0,y_0;n)$ to the BP equation, one can generate an infinite set of solutions $(x_i,y_i;n)$ by repeatedly applying the BF identity to the starting solution
\begin{eqnarray}
(x_1,y_1) & = & (x_0, y_0)^2 \nonumber \\ (x_2,y_2) & = & (x_0, y_0) (x_1,y_1) \nonumber \\ \vdots \nonumber \\ (x_n,y_n) & = & (x_0, y_0) (x_{n-1},y_{n-1})
\end{eqnarray}
Here, the pair $(x_0,y_0;n)$ is referred to as the fundamental solution.

\subsection{Quadratic Integers and the BP Equation}\label{app-subsec:quadratic-int}

We are familiar with solutions of equations of the form
$$ x^2 + Bx + c = 0 $$
This is the quadratic equation from beginning algebra courses, which has as solutions
$$ x_\pm = \frac{-B \pm \sqrt{B^2 - 4c}}{2} $$
when the \emph{discriminant} $ B^2 - 4c $ is negative, the roots of the equation are imaginary or complex numbers
$$ x_\pm = \frac{-B \pm i d}{2} \in \mathbb{C} $$
where $d = |B^2 - 4c|$ and $i = \sqrt{-1}$.
When $\{B,c\} \in \mathbb{Z}$, the solutions of the quadratic equations can be characterized as elements of the field of quadratic integers $\mathbb{Q}(\sqrt d)$, which is an extension of the familiar field of rational numbers $\mathbb{Q}$. Such numbers have the form
$$ z = a + \omega b$$
where $\{a,b\} \in \mathbb{Z}$, $\omega = \sqrt{d} $ if $d\mod 4 \equiv 2,3 $ and $\omega = \frac{1+\sqrt{D}}{2}$ otherwise (if $d\mod 4 \equiv 1$). $d \in \mathbb{A}$, where $\mathbb{A}$ is the set of square-free integers.

It is at this point that one makes a connection to the square-free quadratic extension of the field of rationals $\mathbb{Q}(\sqrt{n})$ and its integral subset $\mathbb{Z}(\sqrt{n})$, by noting that any solution $(x_i,y_i;n)$ of the BP equation can be represented as a quadratic integer:
$$ (x_i,y_i;n) \Rightarrow z_i^n = x_i + y_i \sqrt n \in \mathbb{Z}(\sqrt{n})$$

The consistency of this representation is enforced by the fact that the multiplication law for two quadratic integers $z_i, z_j \in \mathbb{Z}(\sqrt{n})$ is the same condition satisfied when multiplying two pairs of solutions of the BP equation to obtain a third pair, \ie, if $ z_i = x_i + y_i \sqrt{n}$ and $z_j = x_j + y_j \sqrt{n}$ are two members of $\mathbb{Z}(\sqrt{n})$, then their product $z_k = z_i \times z_j = x_k + y_k \sqrt{n}$ is given by:
\begin{align} x_k & = x_i x_j + n y_i y_j \\ y_k & = x_i y_j + x_j y_i \end{align}
which is identical to the multiplication law satisfied by pairs of solutions of the BP equation.

\section{Kodama State}
\label{app:kodama}

The Kodama state is an exact solution of the Hamiltonian constraint for LQG with positive cosmological constant $ \Lambda > 0 $ and hence is of great importance for the theory. It is given by
\beq \Psi_K (A) = \mc{N} e^{\int S_{CS}} \eeq
where $\mc{N}$ is a normalization constant. The action $ S_{CS}[A] $ is the Chern-Simons action for the connection $ A_{\mu}^I $ on the spatial 3-manifold $M$, given by
$$ S_{CS} = \frac{2}{3\Lambda} \int Y_{CS} $$
where
$$ Y_{CS} = \frac{1}{2} Tr \left[ A \wedge \mathbf{d} A + \frac{2}{3} A \wedge A \wedge A \right] $$
with $\mathbf{d}A \simeq \partial_{[\mu}A_{\nu]}^I $ being the exterior derivative. Consistent with our discussion of bivectors and $k$-forms in sec.~\ref{app-subsec:diff-forms} the wedge product $\wedge$ between two 1-forms $P$ and $Q$ is:
$$ P \wedge Q \simeq P_{[a} Q_{b]} $$

For identical one-forms the wedge product gives zero. That is why for the Chern-Simons action to have a non-zero cubic term the connection must be non-abelian. Let us write the various terms in the Chern-Simons density explicitly;
$$ A \wedge \mathbf{d} A \equiv A_{[p}^{i} \partial_{q}^{} A_{r]}^{j} T_{i} T_{j} \qquad A \wedge A \wedge A \equiv A_{[p}^i A_{q}^j A_{r]}^k T_i T_j T_k $$
where $ p,q,r\ldots$ are worldvolume (``spacetime'') indices, $ i,j,k \ldots $ are worldsheet (``internal'') indices and $T_i$ are the basis vectors of the lie-algebra/internal space.

Taking the trace over these terms gives us
$$ Y_{CS} = \frac{1}{2} Tr \left[ A_{[p}^{i} \partial_{q}^{} A_{r]}^{j} T_{i} T_{j} + \frac{2}{3} A_{[p}^i A_{q}^j A_{r]}^k T_i T_j T_k \right] $$

The trace over the lie-algebra elements gives us:
$$ Tr \left[ T_i T_j \right] = \delta_{ij} \qquad Tr \left[ T_i T_j T_k \right] = f_{ijk} $$
where $f_{ijk}$ are the structure constants of the gauge group.

\section{3j-symbols}

The \href{http://en.wikipedia.org/wiki/3-jm_symbol}{Wigner 3j-symbol} is related to the Clebsch-Gordan coefficients through:
$$ \left( \begin{array}{ccc} j_1 & j_2 & j_3 \\ m_1 & m_2 & m_3 \end{array} \right) \equiv \frac{(-1)^{j_1 - j_2 - m_3}}{\sqrt{2 j_3 + 1}} \langle j_1,m_1;j_2,m_2 | j_3,m_3 \rangle $$

where the $(j_i,m_i)$ are the orbital and magnetic quantum numbers of the $i^{th}$ particle. $ \mid j_1,m_1;j_2,m_2 \rangle $ is the state representing two particles (or systems) each with their separate angular momentum numbers. $ \mid j_3,m_3 \rangle $ represents the \emph{total} angular momentum of the system. Classically we have two systems with angular momentum $\vec{L}_1$ and $\vec{L}_2$, then the angular momentum of the combined system is: $\vec{L}_3 = \vec{L}_1 + \vec{L}_2 $.

In quantum mechanics, however, the angular momentum of the composite system can be any one of a set of possible allowed choices. Whether or not the angular momentum of the composite system can be specified by quantum numbers $j_3,m_3$ is determined by whether or not the Clebsch-Gordan coefficient is non-zero.

\section{Regge Calculus}
Regge showed in 1961 that one could obtain the continuum action of general relativity ``in 2+1 dimensions'' from a discrete version thereof given by decomposing the spacetime manifold into a collection of tetrahedral simplices \cite{Regge1961General,Iwasaki1994A-reformulation}. When many such tetrahedra are joined together, curvature of the resulting discrete manifold is represented by positive or negative deficit angles (for instance, a plane 2D surface can be tiled with equilateral triangles, with six such triangles meeting at each vertex. If one attempted to increase the number of degrees round a given vertex by fitting a seventh triangle in, the only way it could be accommodated would be by curving the resulting surface).  
\beq 
S_{i} = \sum_{a=1}^{6} l_{i,a} \theta_{i,a} 
\eeq 
is the Regge action for the $i^{\textrm{th}}$ tetrahedron. Here the sum over $a$ is the sum over the edges of the tetrahedron. $l_{i,a}$ and $\theta_{i,a}$ are the length of the edge and the dihedral deficit angle, respectively, \emph{around} the $a^{\textrm{th}}$ edge of the $i^{\textrm{th}}$ tetrahedron.

The Regge action for a manifold built up by gluing such simplices together is simply the sum of the above expression over all $N$ simplices
$$ S_{Regge} = \sum_{i=1}^N S_{i} $$

It was later shown by Ponzano and Regge \cite{Ponzano1968Semiclassical} that in the  $j_i \gg 1$ limit, the 6-j symbol corresponds to the cosine of the Regge action \cite{Regge2000Discrete}
$$ \left\{ \begin{array}{ccc} j_1 & j_2 & j_3 \\ j_4 & j_5 & j_6 \end{array} \right\} \sim \frac{1}{12\pi V} \cos\left( \sum_i j_i \theta_i + \frac{\pi}{4} \right) $$




%

\printbibliography
\end{document}